\documentclass[11pt, oneside]{article}   	
\usepackage{geometry}                		
\usepackage{amssymb,bm,amsmath,amsfonts}
\usepackage{multicol,lipsum,float}
\usepackage{graphicx}

\usepackage{subcaption}
\title{\textbf{Nonlinear dynamics of the fishbone-induced alpha transport on ITER}}
\author{\textbf{G. Brochard$^{1,2}$, R. Dumont$^{1}$, H. L\"utjens$^{2}$, X. Garbet$^{1}$, T. Nicolas$^2$, P. Maget$^{1}$}}
\date{}							
\begin{document}
\maketitle
\begin{center}
$^1$\emph{ CEA, IRFM, F-13108 Saint-Paul-lez-Durance, France} \\ $^2$ \emph{CPHT, CNRS, \'Ecole Polytechnique, Institut Polytechnique de Paris, Route de Saclay, 91128 PALAISEAU} \\ 
E-mail : guillaume.brochard@polytechnique.edu
\end{center}
\begin{abstract}
\noindent
The fishbone-induced transport of alpha particles is computed for the ITER 15 MA baseline scenario \cite{ITERB}, using the nonlinear hybrid Kinetic-MHD code XTOR-K. Two limit cases have been studied, in order to analyse the characteristic regimes of the fishbone instability : the weak kinetic drive limit \cite{Berk1999} and the strong kinetic drive limit \cite{Zonca2015}. In both those regimes, characteristic features of the $n=m=1$ fishbone instability are recovered, such as a strong up/down-chirping of the mode frequency, associated with a resonant transport of trapped and passing alpha particles.  The effects of the $n=m=0$ sheared poloidal and toroidal plasma rotation are taken into account in the simulations. The shear is not negligible, which implies that the fishbone mode frequency has a radial dependency, impacting the wave-particle resonance condition. Phase space hole and clump structures are observed in both nonlinear regimes, centered around the precessional and passing resonances. These structures remains attached to the resonances as the different mode frequencies chirp up and down. In the nonlinear phase, the transport of individual resonant trapped particles is identified to be linked to mode-particle synchronization. On this basis, a partial mechanism for the nonlinear coupling between particle transport and mode dominant down-chirping is proposed. The overall transport of alpha particles inside out the $q=1 $ surface is of order 2-5\% of the initial population between the simulations.  The loss of alpha power is found to be directly equal to the loss of alpha particles.
\end{abstract}
\section{Introduction}
The nonlinear interplay between macroscopic MHD modes and energetic particles is a subject of crucial interest for burning plasma experiments such as ITER. In these plasmas, the resonant interaction between the characteristic frequencies of the supra-thermal particles and MHD modes frequencies can lead to the destabilization of Alfv\'en eigenmodes (AE), or the emergence of new Energetic Particle Modes (EPM). Such instabilities tend to degrade the confinement of energetic particles, decreasing the fusion performances of burning plasmas due to a net power transfer from the fast particles towards the macroscopic Kinetic-MHD modes. Among EPM's, the fishbone instability was first observed on the PDX tokamak \cite{McGuire1983}, and named after the characteristic fishbone-like signal obtained on the Mirnov coils. The instability was then reproduced in a large number of experiments on various fusion devices \cite{Campbell1988}\cite{Heidbrink1990}\cite{Mantsinen2000}, some of which featuring a complete stabilization of the internal kink mode \cite{Guenter1999}, or a co-existence of the fishbone and the internal kink modes \cite{Nave1991}. The fishbone instability results, as it was demonstrated in linear theory, from the resonant interaction between the $n=m=1$ internal kink mode frequency and the precessional and/or transit frequencies of energetic particles \cite{Chen1984}\cite{Coppi1986}\cite{Betti1993}\cite{Porcelli1994}. The fishbone is triggered past a threshold in fast particle beta $\beta_{h,thres}$. Below this threshold, the internal kink is partially or entirely stabilized.
\\ \\
Recently, it was shown that the alpha fishbone instability is likely to be triggered for Kinetic-MHD equilibria relevant to the ITER 15 MA baseline scenario \cite{Brochard2020a}. These results stress further the necessity of studying the nonlinear phases of the fishbone instability in ITER like conditions, in order to assess the amount of alpha particles transported out of the plasma core, hence reducing the alpha heating power $P_{\alpha}$. Reduced kinetic/hybrid models \cite{Berk1999}\cite{Candy1999}\cite{Odblom2002}\cite{Idouakass2016}\cite{Chen2013}\cite{Zonca2014}
were initially considered to analyse the nonlinear phase of the instability. Besides recovering characteristic features of the instability such as mode up/down chirping coupled to particle transport, these models identified two nonlinear regimes. A weak kinetic drive regime near marginal stability \cite{Berk1999} dominated by fluid nonlinearities \cite{Odblom2002}, where the bouncing time  $\tau_B$ of resonant kinetic particles in a phase space island is much lower than the characteristic time evolution of phase space zonal structures $\tau_{NL}$; and a strong kinetic drive regime \cite{Zonca2015}\cite{Chen2016} far from marginal stability, dominated by kinetic nonlinearities with $\tau_B \sim \tau_{NL}$. In the past decade, as computational capabilities increased, a number of global nonlinear hybrid Kinetic-MHD simulations were performed to study the physics of the alphas \cite{Fu2006}, NBI induced \cite{Wang2013}\cite{Wang2016}\cite{Pei2017}\cite{Shen2017} or electronic fishbones \cite{Vlad2013}\cite{Vlad2016} in either one of these regimes. In order to assess the alpha transport in ITER as generally as possible, both these regimes need to be explored. \\ \\
In this paper, the first ITER relevant nonlinear hybrid simulations of the alpha fishbone instability are performed in these two nonlinear regimes, with the code XTOR-K \cite{Luetjens2010}\cite{Leblond2011}\cite{Brochard_PhD}. This code solves the nonlinear extended two-fluid MHD equations in toroidal geometry, while advancing self-consistently kinetic thermal and/or supra-thermal particles in 6D with a full-f method. The code takes into account all fluid and kinetic nonlinearities, which implies that it is an ideal tool for the study of the fishbone instability in any nonlinear regime. XTOR-K was linearly verified for the alpha fishbone instability \cite{Brochard2020a} against a comprehensive linear fishbone model developed in \cite{Brochard2018}. It ensures a correct physical description of Kinetic-MHD instabilities with the code. In the present work, bulk diamagnetic effects are neglected and one kinetic population of alpha particles is considered. The fluid set of equations solved by XTOR-K in that case reduces to the single fluid resistive MHD equations, with a kinetic coupling in the perpendicular equation of motion \cite{Brochard2020a}. \\ The nonlinear evolution of the fishbone modes is analyzed in section 2, highlighting the mode saturation in the weak and strong drive regime. The nonlinear evolution of phase space resonant structures is presented in section 3. In section 4, the individual behavior of resonant trapped particles is studied, to gain a better understanding of the nonlinear relationship between mode chirping and particle transport. To conclude, the overall transport of alpha particles during the entire nonlinear fishbone phases is analyzed, discriminating between trapped and passing particles. The loss of alpha heating power due to the alpha transport over one fishbone burst is evaluated.
\section{Saturation dynamics of the alpha fishbone instability}
\subsection{Characteristics of the XTOR-K hybrid nonlinear simulations}
\begin{figure}[h!]
\begin{subfigure}{.33\textwidth} 
   \centering
   \includegraphics[scale=0.22]{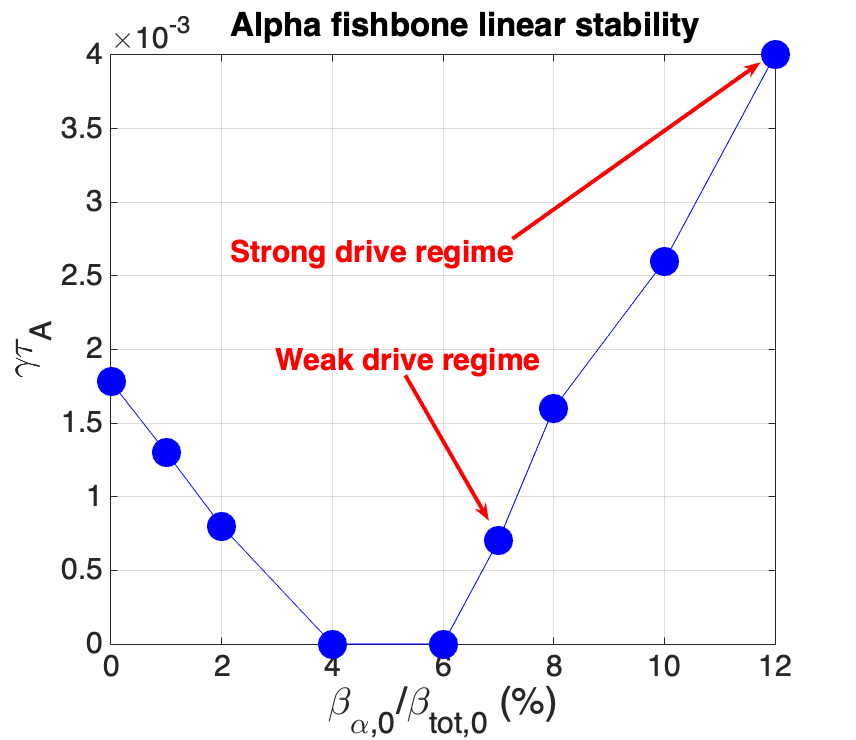}
   \caption{}
\end{subfigure}     
\begin{subfigure}{.33\textwidth} 
   \centering
   \includegraphics[scale=0.22]{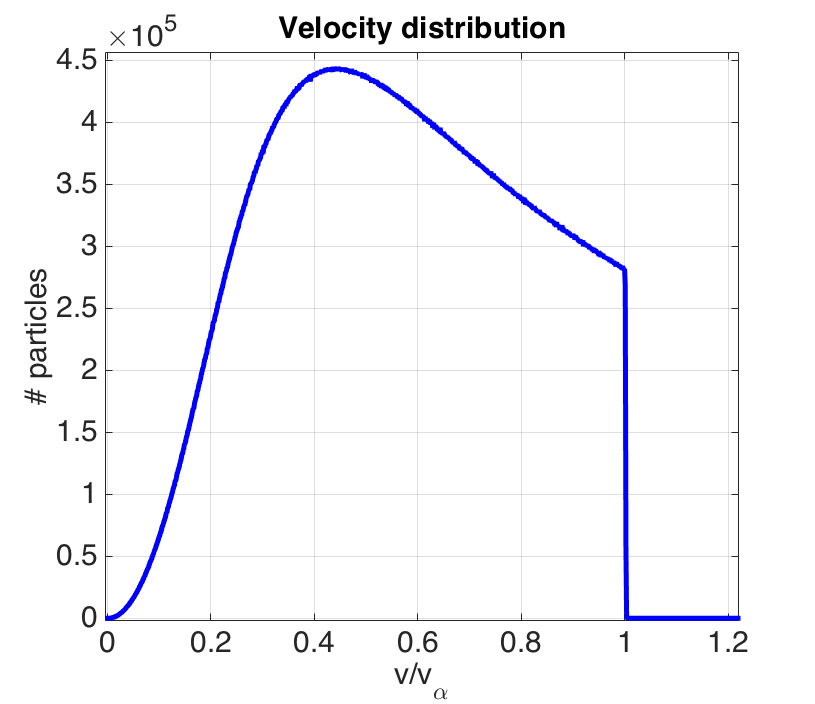}
   \caption{}
\end{subfigure}
\begin{subfigure}{.33\textwidth} 
   \centering
      \includegraphics[scale=0.22]{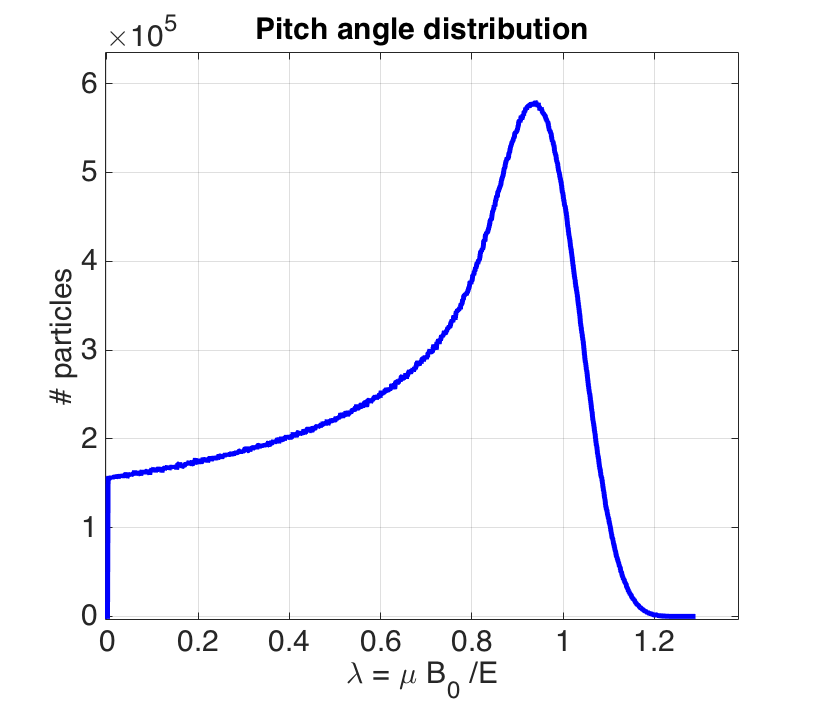}
   \caption{}
\end{subfigure}  
\caption{Characteristics of the nonlinear simulations. (a) Linear growth rates for different on-axis beta ratio $\beta_{\alpha,0}/\beta_{tot,0}$. (b) Alpha particles velocity distribution. (c) Alpha particles pitch-angle distribution.}
\label{chara}
\end{figure}

The Kinetic-MHD equilibria considered for XTOR-K nonlinear simulations are taken from Ref. \cite{Brochard2020a}, Section 4.1. Profiles and geometry are relevant for the ITER 15 MA baseline scenario. The safety factors considered are parabolic, with a $q=1$ surface located at $r=\sqrt{\psi/\psi_{edge}} = 0.35$, where $r$ is the normalized radial coordinate, $\psi$ the local poloidal flux and $\psi_{edge}$ the poloidal flux at the plasma edge. In \cite{Brochard2020a}, the stability of the alpha fishbone in ITER is investigated with a series of linear XTOR-K simulations, in which the on-axis safety factor $q_0$ and beta ratio $\beta_{\alpha,0}/\beta_{tot,0}$ are varied. $\beta_{\alpha,0}$ stands for the on-axis beta of alpha particles and $\beta_{tot,0}$ the on-axis total beta. In the present work, the cases $q_0= 0.95, \ \beta_{\alpha,0}/\beta_{tot,0} = 12 \%$ and $q_0 = 0.95, \ \beta_{\alpha,0}/\beta_{tot,0} = 7\%$ are chosen respectively to study the strong and the weak kinetic drive regimes. \\ \\ The linear growth rates for different on-axis beta ratio with $q_0=0.95$ are displayed in Figure \ref{chara} (a), highlighting the nonlinear regimes is which each of the equilibria lie. Cases with null linear growth rates $\gamma\tau_A$ indicate that no unstable modes grew out of the PIC noise, introduced by XTOR-K's kinetic module, after $5000 \tau_A$ of simulation. $\tau_A = 5.4\times10^{-7}$ s is the Alfv\'en time in both simulations. For these cases, $\gamma\tau_A < \gamma_L\tau_A = 6\times10^{-4}$, where $\gamma_L$ is the  linear growth rate at the fishbone threshold $\beta_{\alpha,thres}/\beta_{tot} \sim 6\%$, estimated by quadratic extrapolation. The weak drive regime arises near marginal stability, when the mode linear growth rate verifies $|\gamma-\gamma_L|/\gamma_L \ll 1$. As it can be seen in Figure \ref{chara} (a), for $\beta_{\alpha}/\beta_{tot} = 7\%$, $\gamma\tau_A = 7\times 10^{-4}$. This nonlinear simulation is then in the weak drive regime, whereas the other simulation with $\gamma\tau_A = 4\times10^{-3}$ is in the strong drive regime. A comparison between the conditions $|\gamma-\gamma_L|/\gamma_L \ll$ and $\tau_B \ll \tau_{NL}$ characterizing the weak drive regime will be provided in Section 4. \\ \\
To describe the kinetic population of alpha particles, an isotropic slowing-down distribution function is imposed with
\begin{equation}
F_{\alpha}(r,v) = \frac{n_{\alpha}(r)}{C}\frac{\sigma_H(v_{\alpha}-v)}{v^3+v^3_c(r)}, \ \ \ \ \ v^3_c(r) = \frac{3\sqrt{\pi}}{4}\frac{m_e}{m_{\alpha}} \bigg(\frac{2T_e(r)}{m_e}\bigg)^{3/2} 
\end{equation}
where $v$ the particle norm velocity, $\sigma_H$ the Heaviside function, $C$ a normalization constant, $v_{\alpha} = \sqrt{2E_{\alpha}/m_{\alpha}}$ the birth velocity of alpha particles, $n_{\alpha}(r)$ the alpha density profile, $T_e(r)$ the electron temperature profile, $m_e$ the electron mass, $m_{\alpha}$ the alpha mass. $v_c$ is the critical velocity at which alpha particles give as much energy to both bulk electrons and ions through Coulombian collisions. By imposing such an initial distribution, it is assumed that the alpha particle resonant transport time $\tau_{trans}$ is much lower that the time $\tau_{th,part}$ it takes for a particle to exit a phase space resonant structure through thermalization. Such a time is a fraction of the thermalization time $\tau_{th}$ \cite{ITERB}, which is equal to $\tau_{th} = 4.6\times 10^5 \tau_A$ at the plasma core for the present parameters. For $\tau_{trans}\sim\tau_{th,part}$, collisions between alpha particles and the bulk plasma would need to be taken into account, as well as a source of alpha particles. It will be shown in section 5 that the ordering $\tau_{trans} \ll \tau_{th,part}$ is relatively well respected in both simulations. \\ \\
In these full-f simulations, 240 millions of macro-particles are used in XTOR-K PIC module to describe the kinetic population of alpha particles. Such a number enables a good resolution of resonant structures in phase space and ensures low PIC noise level. The velocity and pitch-angle distributions in both simulations are displayed respectively in Figure \ref{chara} (b) and (c).
\subsection{Nonlinear dynamics of the fishbone modes}
The nonlinear dynamics of the fishbone modes in both simulations is analyzed here comprehensively with a set of fluid diagnostics. The time evolution of the toroidal modes kinetic energy is displayed in Figure \ref{KE}, the time evolution of the $n=m=1$ mode frequency in Figure \ref{omega}, the mode structure associated to polar Poincar\'e plots at specific times in Figure \ref{modeSD} and Figure \ref{modeWD}. The strong and the weak drive cases are now discussed separately.
\subsubsection{Strong drive case}
\begin{figure}[h!]
\begin{subfigure}{.49\textwidth} 
   \centering
   \includegraphics[scale=0.25]{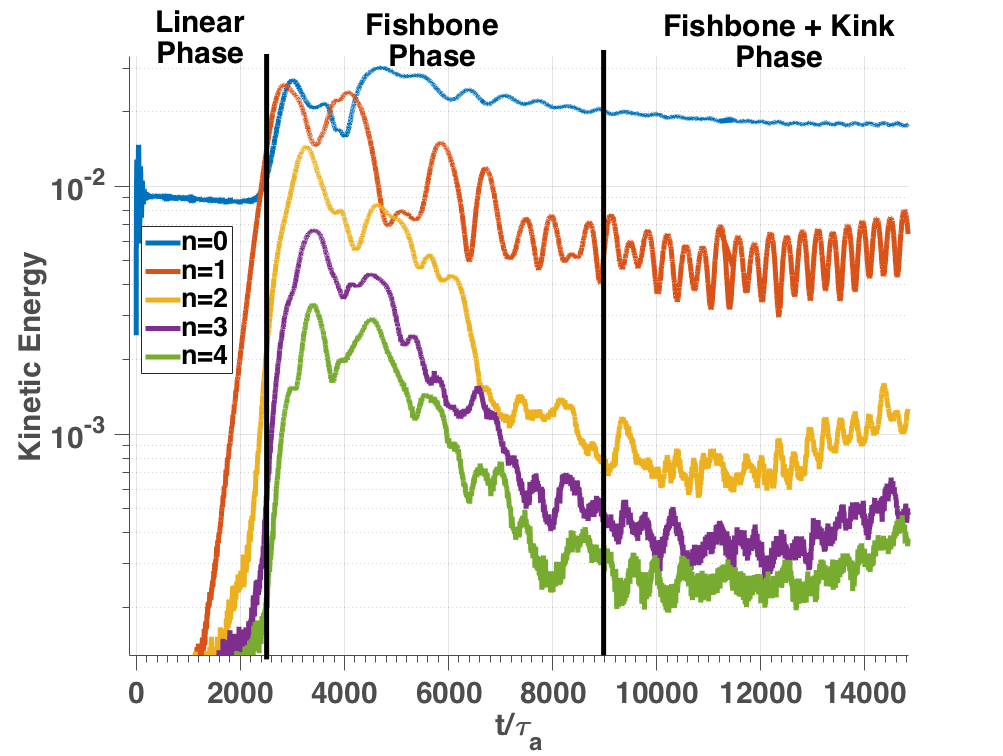}
   \caption{}
\end{subfigure}
\begin{subfigure}{.49\textwidth} 
   \centering
   \includegraphics[scale=0.25]{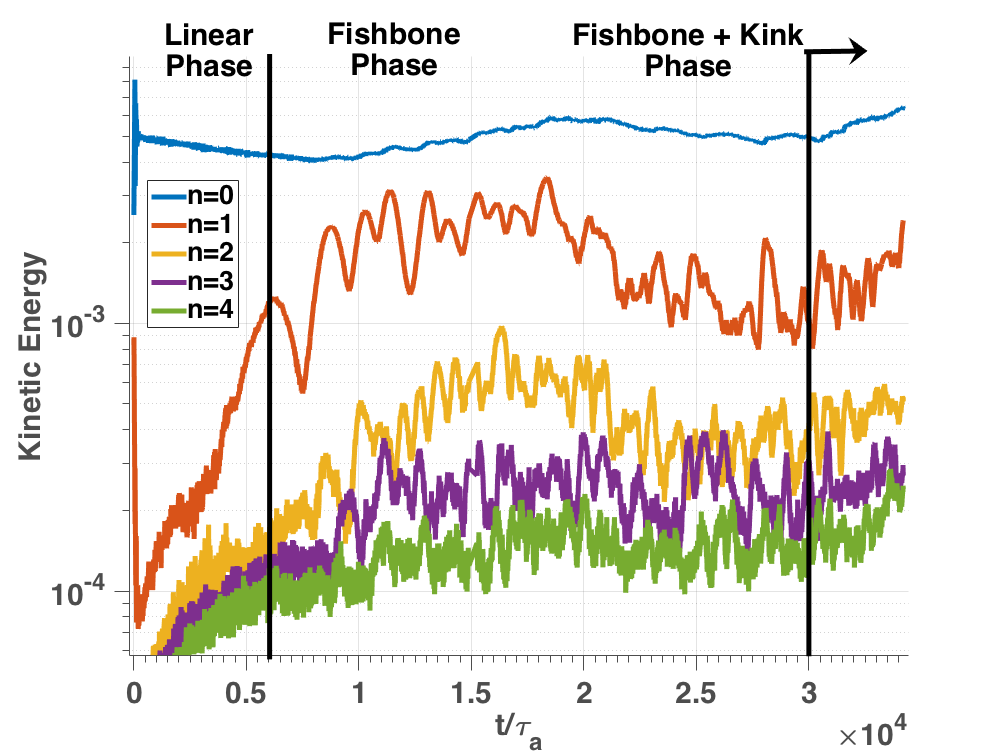}
   \caption{}
\end{subfigure}    
\caption{Time evolution of the toroidal $n$ mode kinetic energies in (a) the strong drive regime (b) the weak drive regime}
\label{KE}
\end{figure}

\begin{figure}[h!]  
\begin{subfigure}{.49\textwidth} 
   \centering
   \includegraphics[scale=0.22]{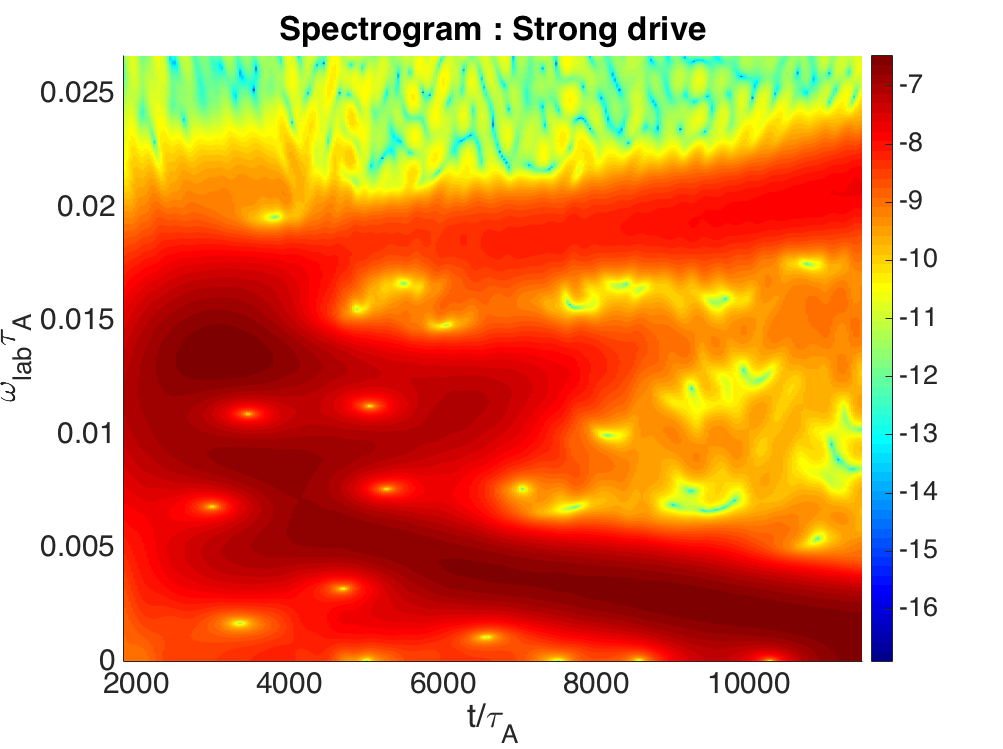}
   \caption{}
\end{subfigure}
\begin{subfigure}{.49\textwidth} 
   \centering
   \includegraphics[scale=0.22]{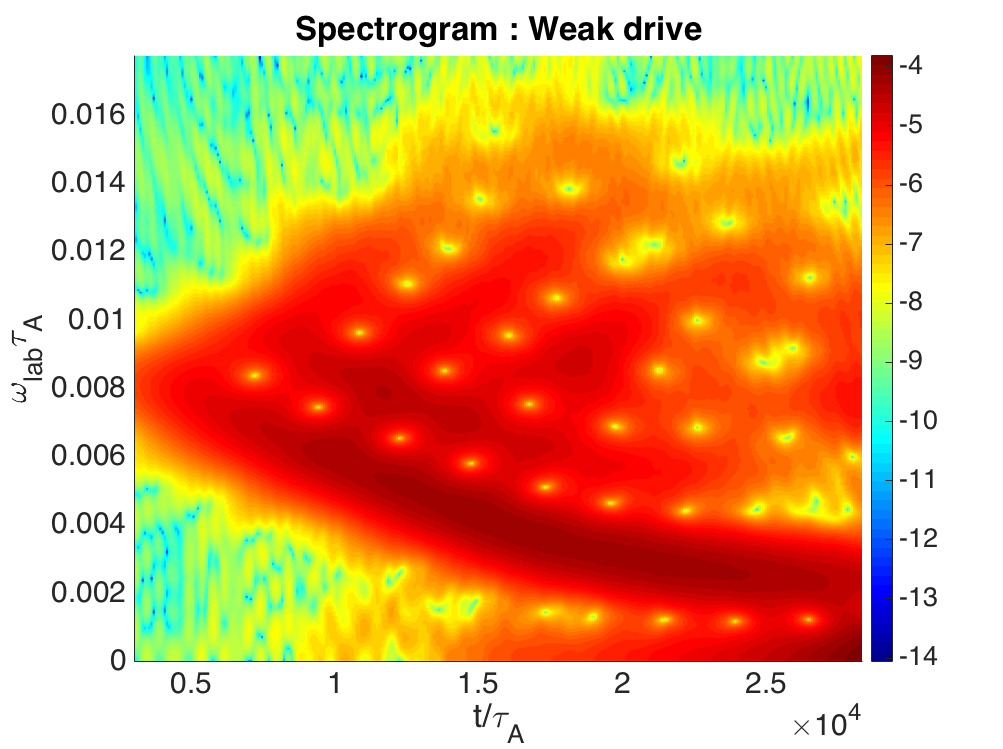}
   \caption{}
\end{subfigure}
\begin{subfigure}{.49\textwidth} 
   \centering
   \includegraphics[scale=0.22]{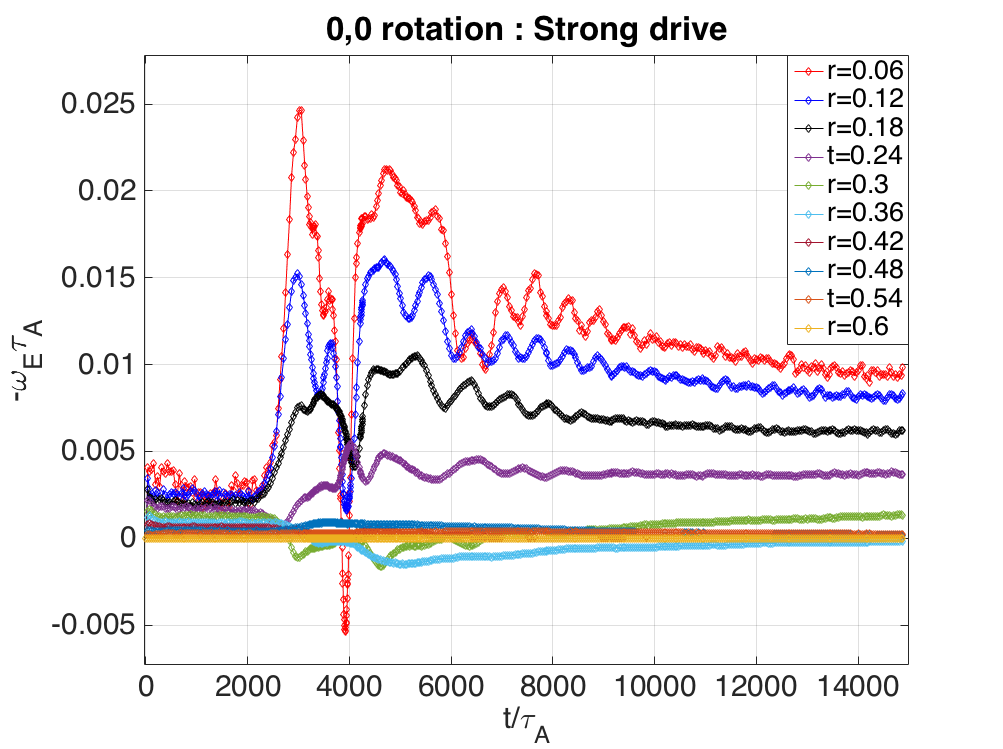}
   \caption{}
\end{subfigure}
\begin{subfigure}{.49\textwidth} 
   \centering
   \includegraphics[scale=0.22]{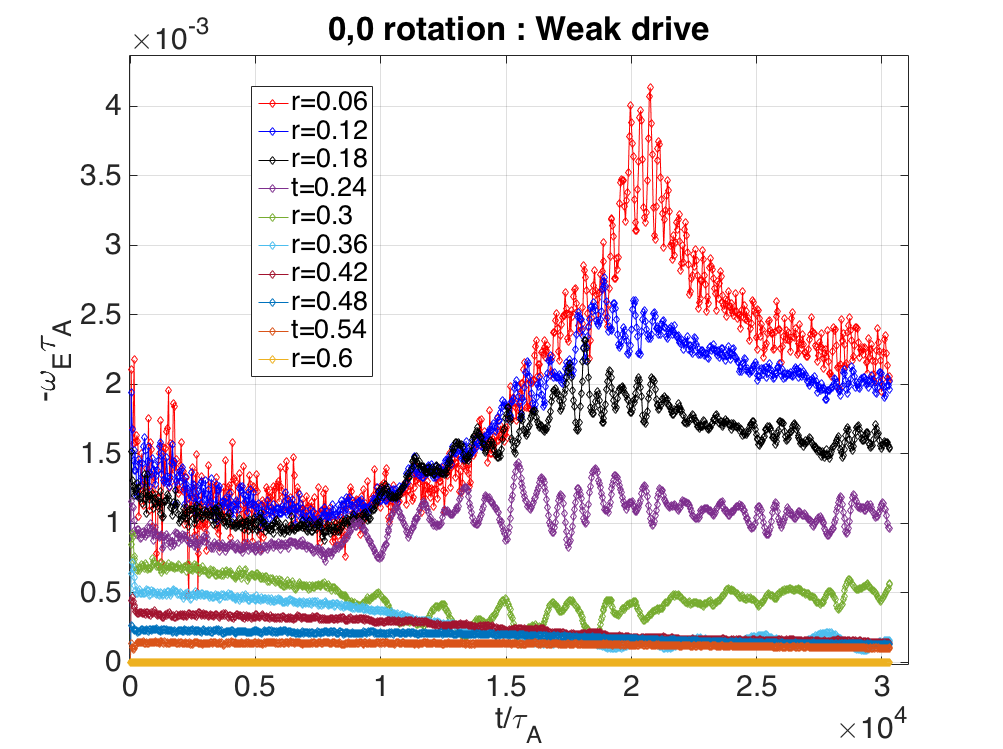}
   \caption{}
\end{subfigure} 
\begin{subfigure}{.49\textwidth} 
   \centering
   \includegraphics[scale=0.22]{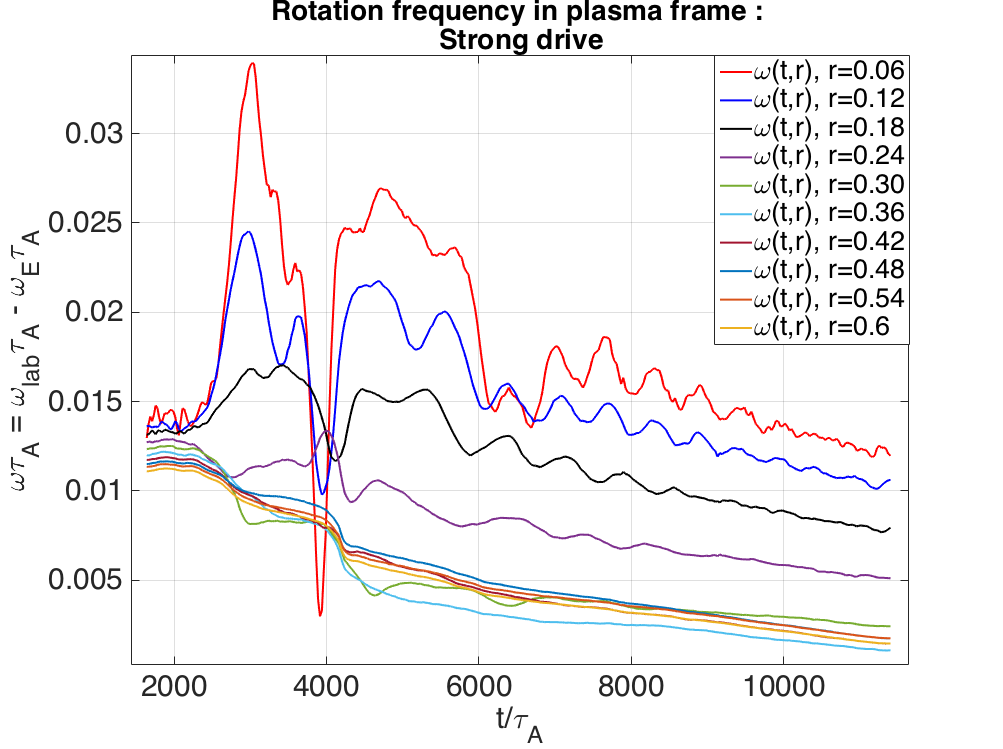}
   \caption{}
\end{subfigure}
\begin{subfigure}{.49\textwidth} 
   \centering
   \includegraphics[scale=0.22]{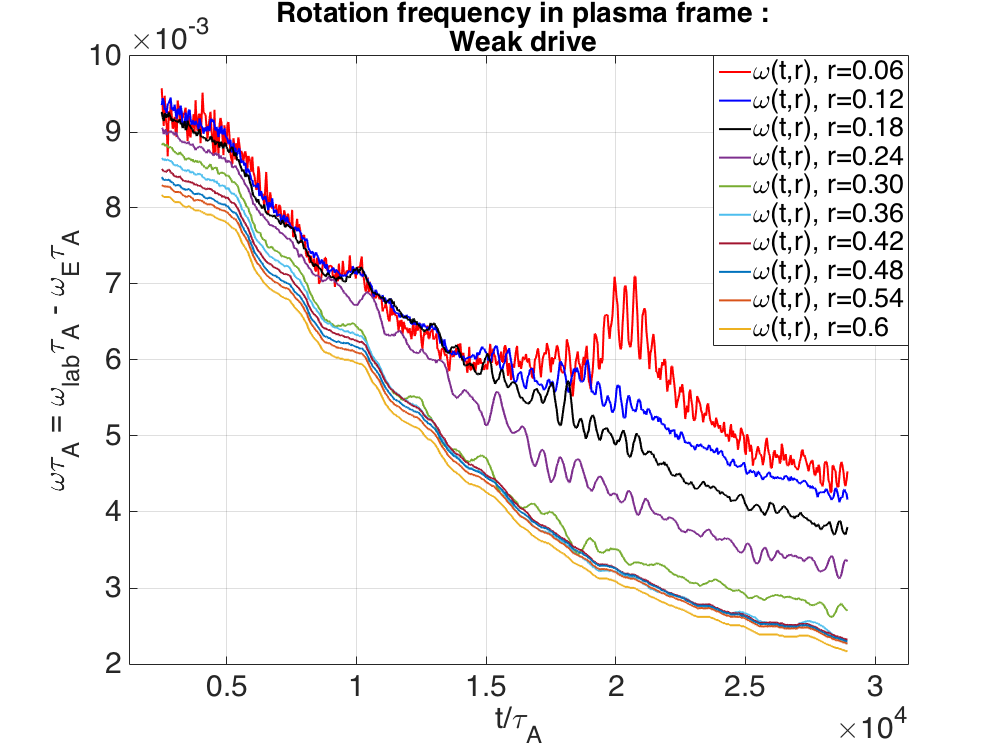}
   \caption{}
\end{subfigure}           
\caption{Time evolution of the 1,1 frequency in different frames. Figures on the left correspond to the strong drive regime, and to the weak drive regime on the right. (a,b) Time evolution of the 1,1 mode frequency in the laboratory frame (c,d) Time evolution of the sheared toroidal and poloidal plasma rotation (e,f) Time evolution of the 1,1 mode frequency down-chirping branch in the plasma frame.}
\label{omega}
\end{figure}

The strong drive simulation can be separated in three distincts phases as depicted in Figure \ref{KE} (a). First a linear phase occurs, where the 1,1 fishbone mode grows exponentially. It ends around $t\sim2500\tau_A$ when the mode saturates. A nonlinear fishbone phase then arises, ranging between $t\in[2500\tau_A,9000\tau_A]$, followed by a mixed phase combining at the same time a nonlinear fishbone instability and an internal kink instability still in its linear growth phase.
The nonlinear fishbone phase is characterized by several features. An up-down chirping of the mode frequency can be observed in Figure \ref{omega}. The time evolution of the 1,1 mode frequency in laboratory frame $\omega_{lab}$ (Figure \ref{omega} (a)) is obtained by performing a spectrogram analysis on the poloidal magnetic field component $B_{\theta,1,1,cos}$, cosine part of the projection on the $n=m=1$
 harmonics. The linear fishbone frequency first splits into one up-chirping and one down-chirping branches at the beginning of the fishbone phase. Then the down-chirping branch also splits in two around $t=5000\tau_A$. The lower down-chirping branch is dominant after $t=7000\tau_A$. The effects of the rigid $n=m=0$ plasma rotation $\omega_E$ need also to be taken into account in this analysis, since the wave-particle resonance condition is defined in the plasma frame, where the mode frequency reads $\omega = \omega_{lab} - \omega_E $. This rigid sheared rotation $\omega_E(r) = \omega_{\theta}(r) - \omega_{\varphi}(r)$ is computed with both poloidal $\omega_{\theta}$ and toroidal $\omega_{\varphi}$ contributions \cite{Maget2016}
\begin{equation}
\omega_{\theta}(r) = 2\pi/\oint_{\mathcal{C}}\frac{\mathcal{J}(r,\theta)}{R(r,\theta)V_{\theta,n=0}(r,\theta)}d\theta
\end{equation}
\begin{equation}
\omega_{\varphi}(r) = \oint_{\mathcal{C}}d\theta \ V_{\varphi,n=0}/(2\pi R(r,\theta))
\end{equation}
where $\mathcal{J}$ is the Jacobian of the coordinate system $(r,\theta,\varphi)$, $V_{\theta/\varphi}$ the covariant poloidal/toroidal MHD velocity, $R$ the major radius and $\mathcal{C}$ the poloidal contour of the magnetic surface $\psi(r)$ associated to the radial coordinate $r$. It has been shown \cite{Graves2000}\cite{Graves2003} that the toroidal shear of the plasma rotation can be non-negligible for the internal kink instability. In recent nonlinear hybrid simulations of the fishbone instability, the effects of $n=m=0$ rotation \cite{Shen2017} or of sheared $n=m=0$ rotation \cite{Shen2015}\cite{Wang2016} have been discarded. However in the present work, it is found that in the fishbone nonlinear phase, both effects are non-negligible, even dominant at some stages. \\ \\ The time evolution of the plasma rotation at different radii is displayed in Figure \ref{omega} (c). The plasma rotation is mostly dominated by poloidal rotation. During the linear phase, the rotation is weakly sheared, with an overall amplitude an order of magnitude below the 1,1 rotation in laboratory frame. 
However, after the fishbone saturation at $t=2500\tau_A$, the poloidal $0,0$ rotation in the core plasma increases drastically and becomes strongly sheared, with an amplitude similar or higher than the $1,1$ rotation in laboratory frame. This is visible in Figure \ref{omega} (e), where the mode rotation in plasma frame $\omega = \omega_{lab} - \omega_E$ of the down-chirping branch has been plotted at different radii. This radial dependency of the mode rotation is taken into account in section 3 when investigating the evolution of resonant structures in phase space. 
\\ 
\\ It is noticed that this sudden evolution of the core plasma rotation is associated with an increase of the $n=0$ mode kinetic energy (Figure \ref{KE} (a)) at the beginning of the nonlinear fishbone phase. Therefore, when the fishbone saturates in the strong drive case, kinetic energy is nonlinearly transferred from the $n=1$ mode to the $n=0$ mode. It is however not possible to differentiate if such nonlinearities are fluid or kinetic, since all nonlinear effects are accounted for in XTOR-K \cite{Chen2016}.
\begin{figure}[h!]
\begin{subfigure}{.24\textwidth} 
   \centering
   \includegraphics[scale=0.22]{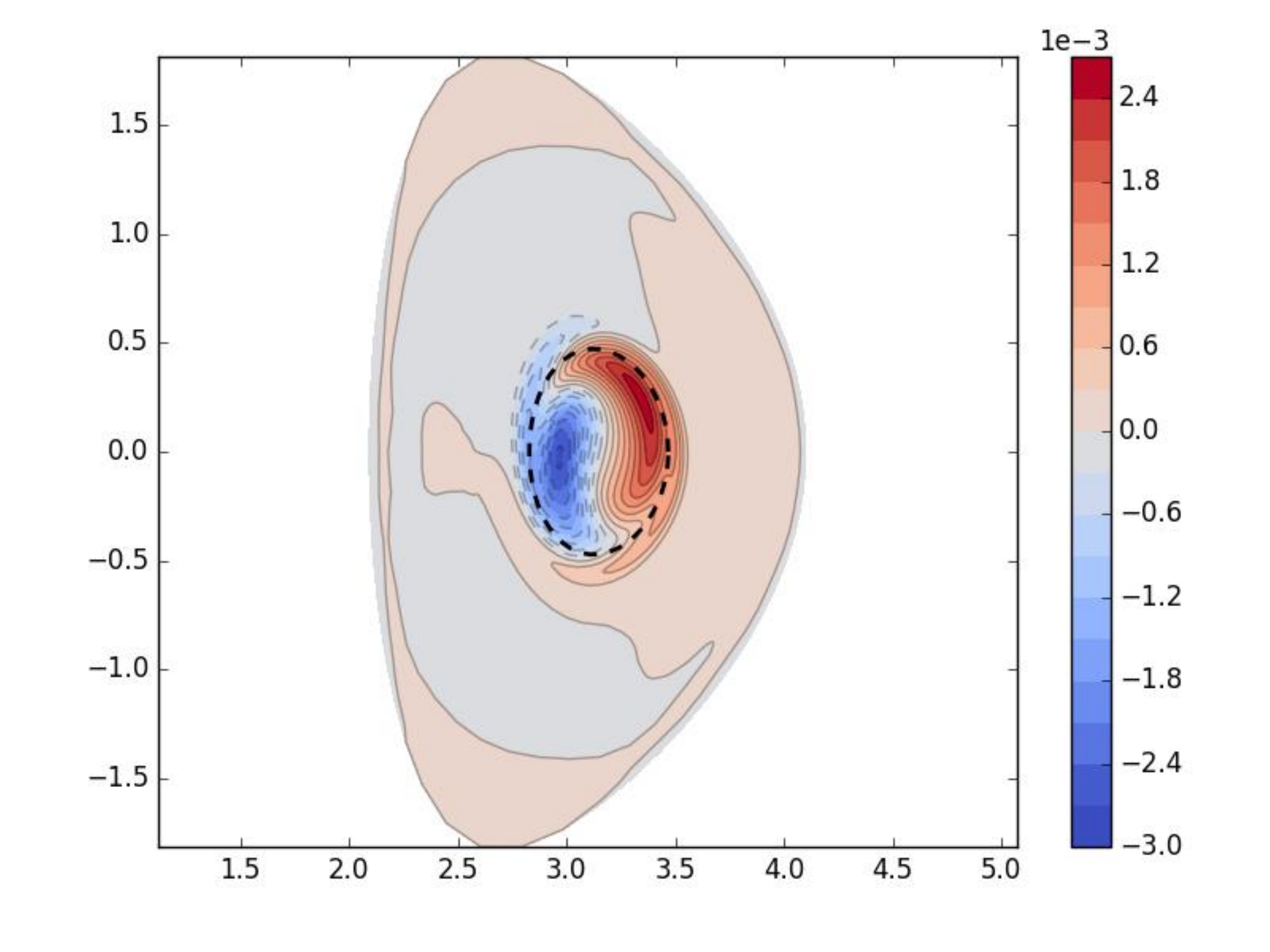}
   \caption{$t/\tau_A = 2639$}
\end{subfigure}
\begin{subfigure}{.24\textwidth} 
   \centering
   \includegraphics[scale=0.22]{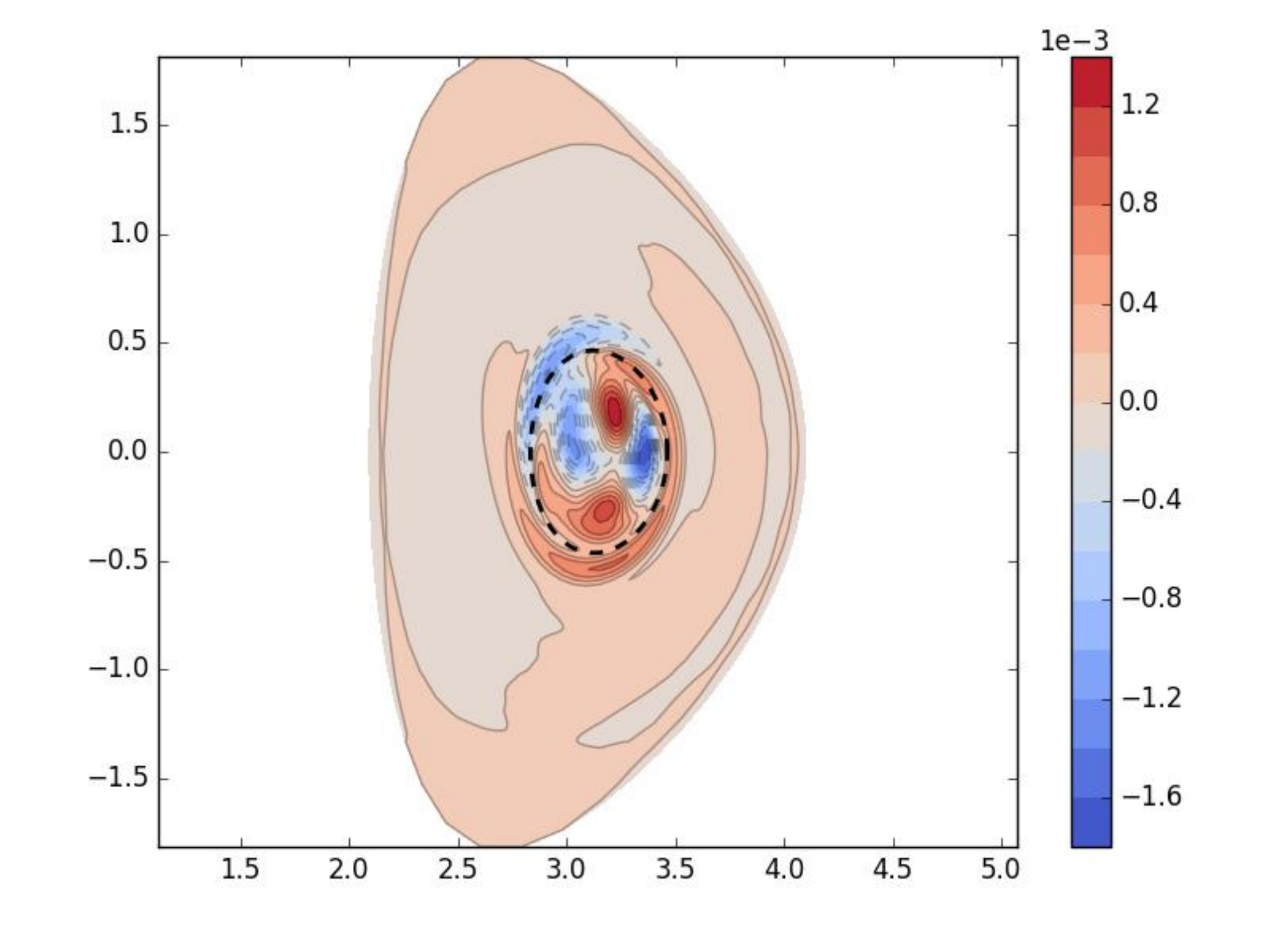}
   \caption{$t/\tau_A = 3564$}
\end{subfigure}
\begin{subfigure}{.24\textwidth} 
   \centering
   \includegraphics[scale=0.22]{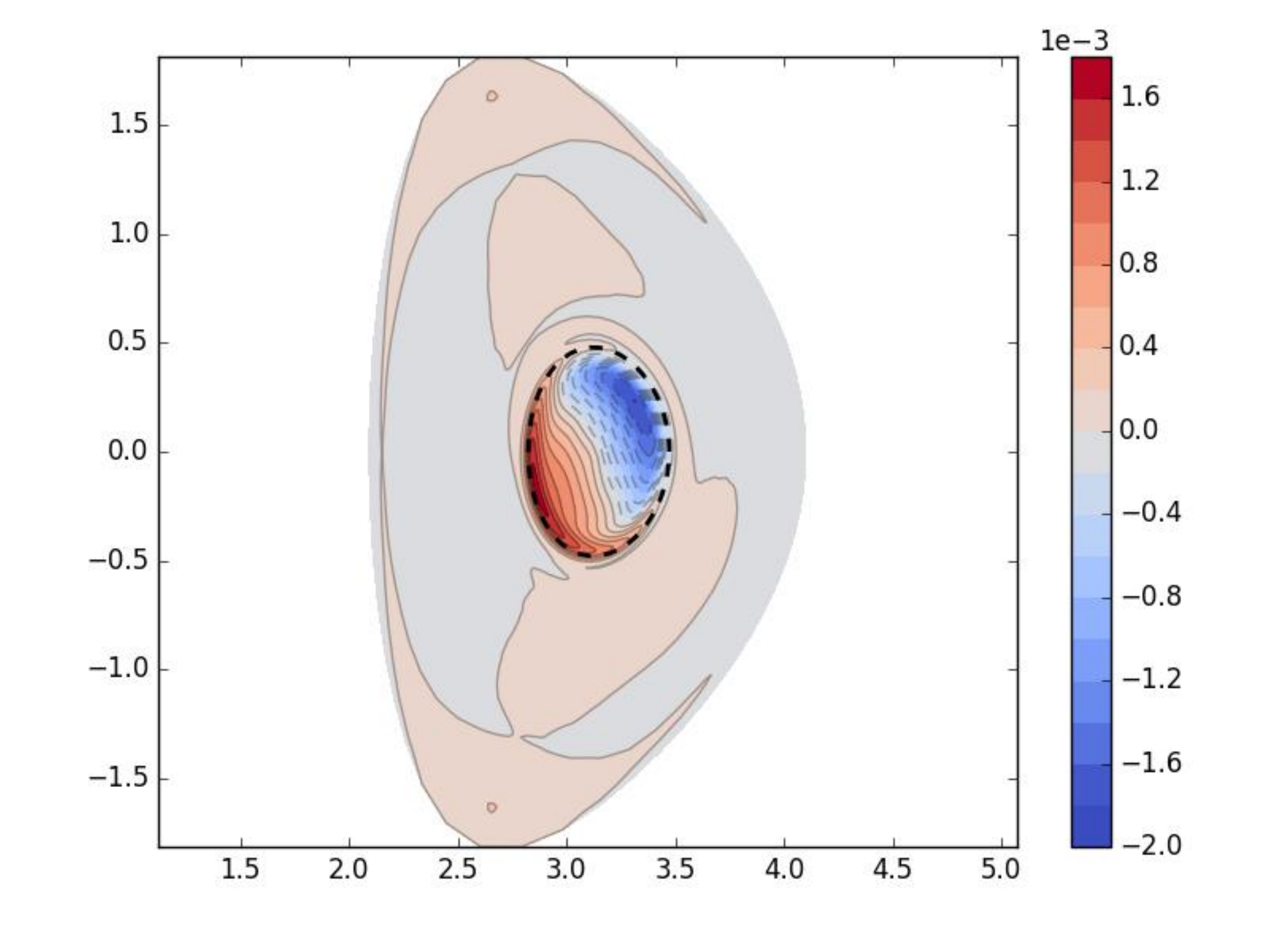}
   \caption{$t/\tau_A = 9265$}
\end{subfigure}
\begin{subfigure}{.24\textwidth} 
   \centering
   \includegraphics[scale=0.22]{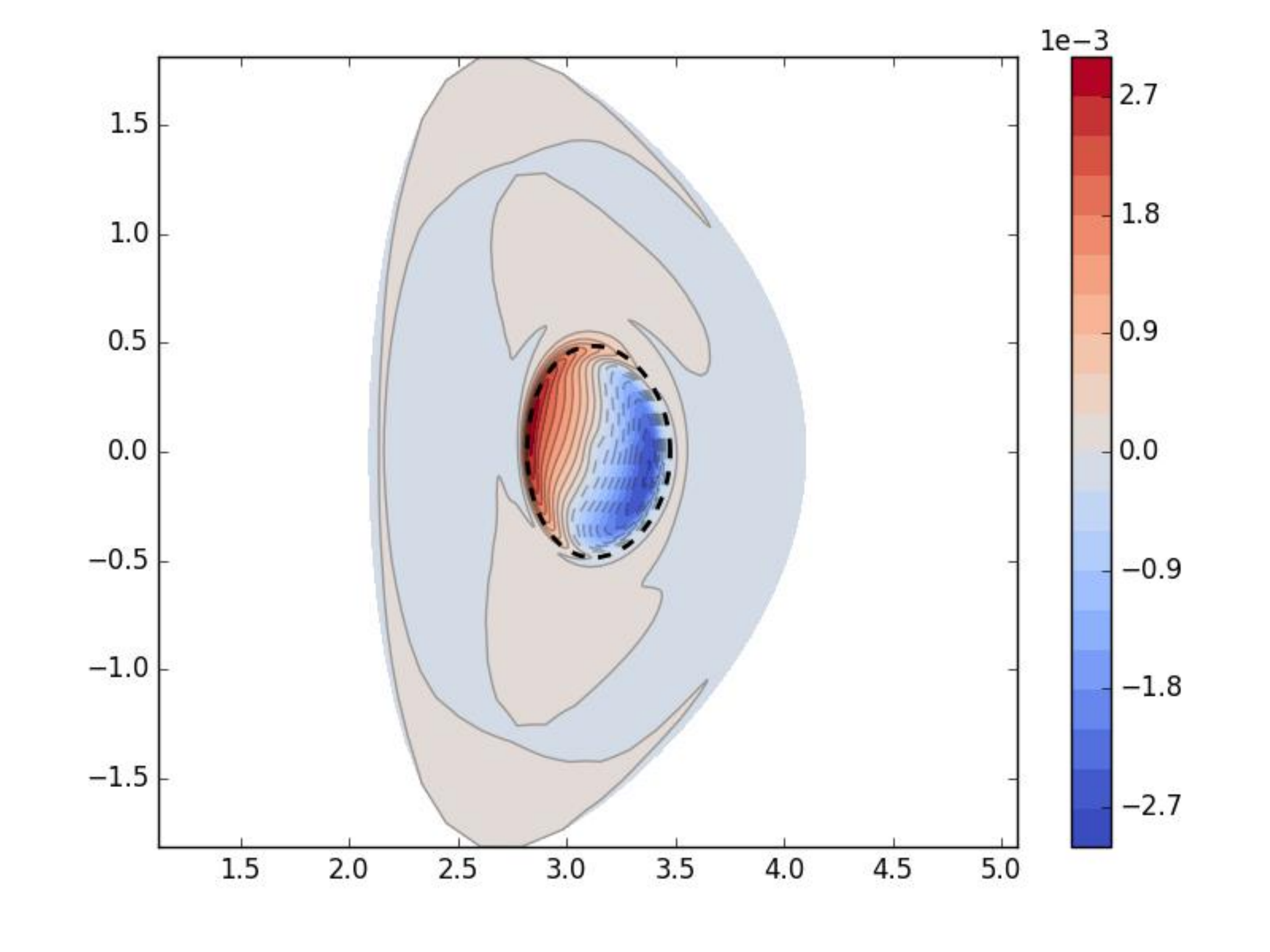}
   \caption{$t/\tau_A = 14878$}
\end{subfigure}
\begin{subfigure}{.24\textwidth} 
   \centering
   \includegraphics[scale=0.31]{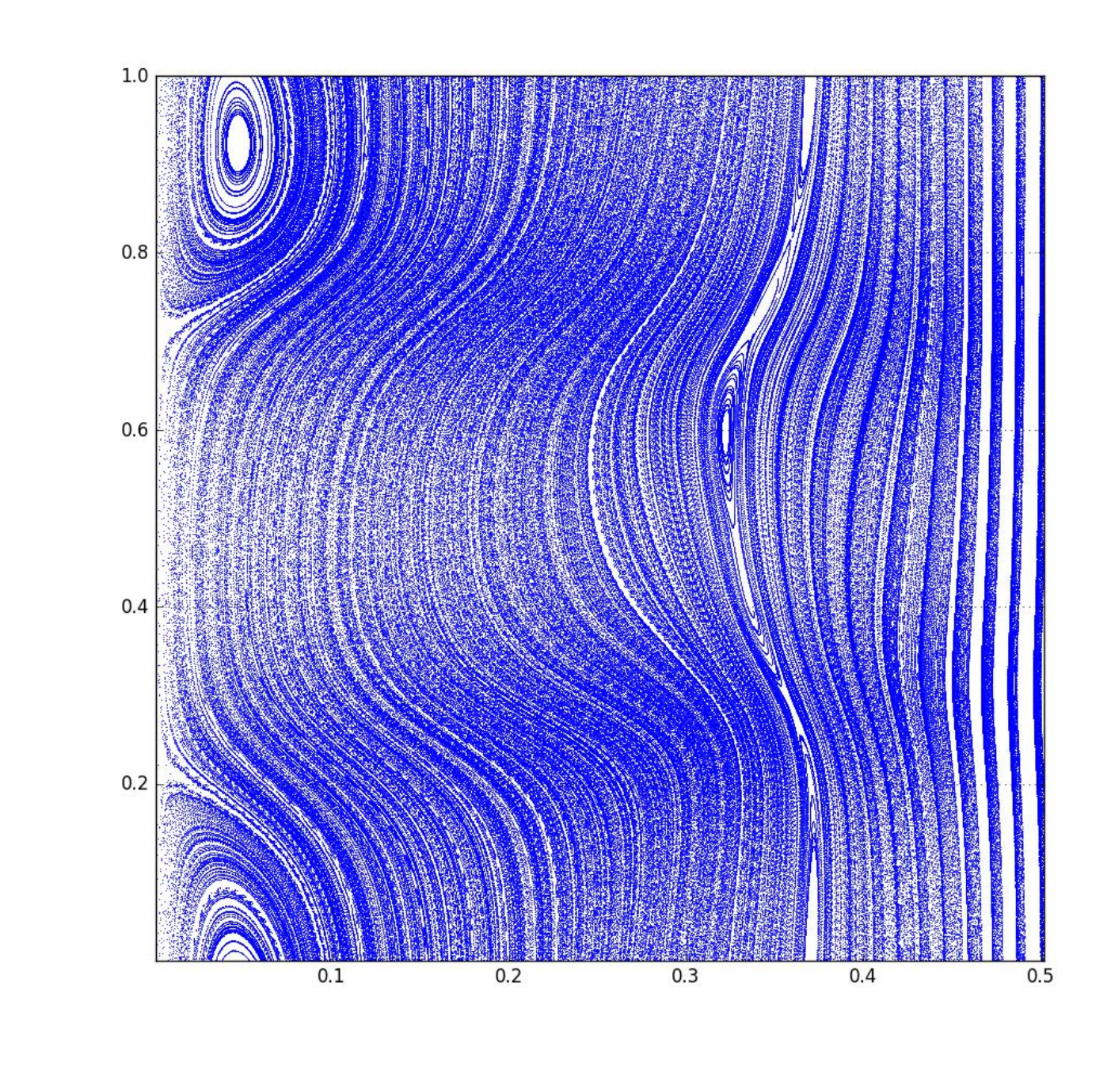}
   \caption{$t/\tau_A = 2639$}
\end{subfigure}
\begin{subfigure}{.24\textwidth} 
   \centering
   \includegraphics[scale=0.31]{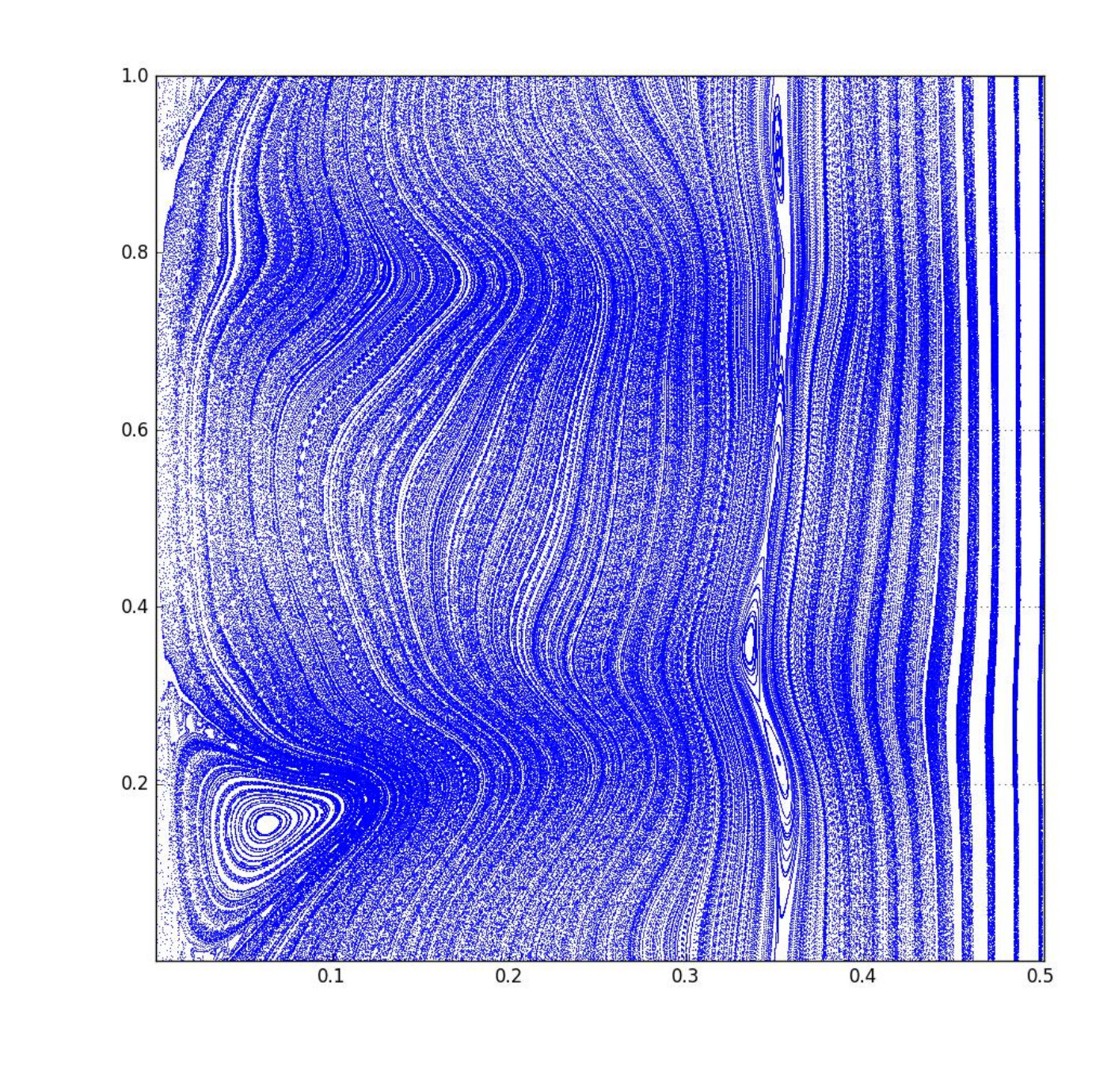}
   \caption{$t/\tau_A = 3564$}
\end{subfigure}
\begin{subfigure}{.24\textwidth} 
   \centering
   \includegraphics[scale=0.31]{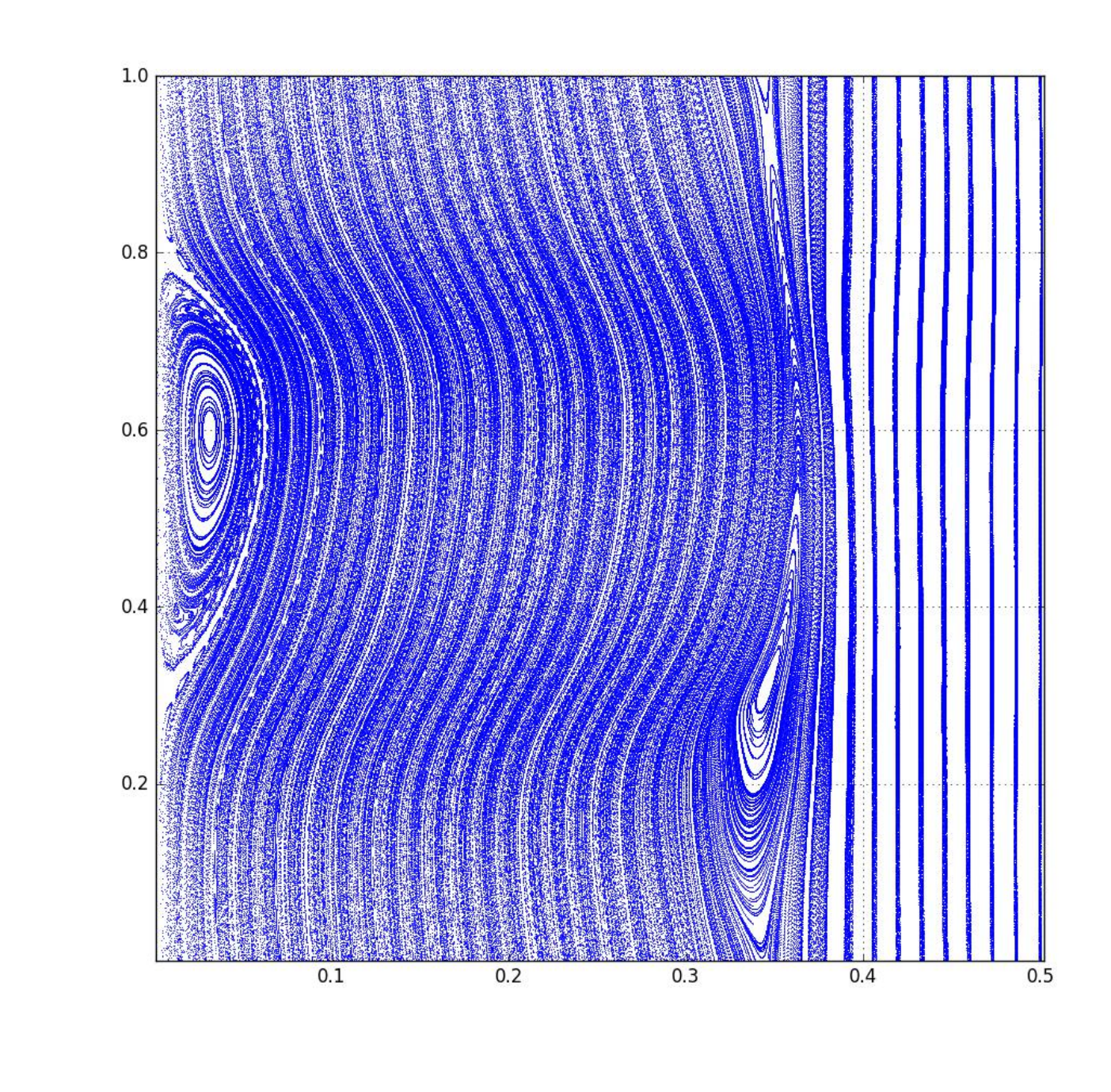}
   \caption{$t/\tau_A = 9265$}
\end{subfigure}
\begin{subfigure}{.24\textwidth} 
   \centering
   \includegraphics[scale=0.31]{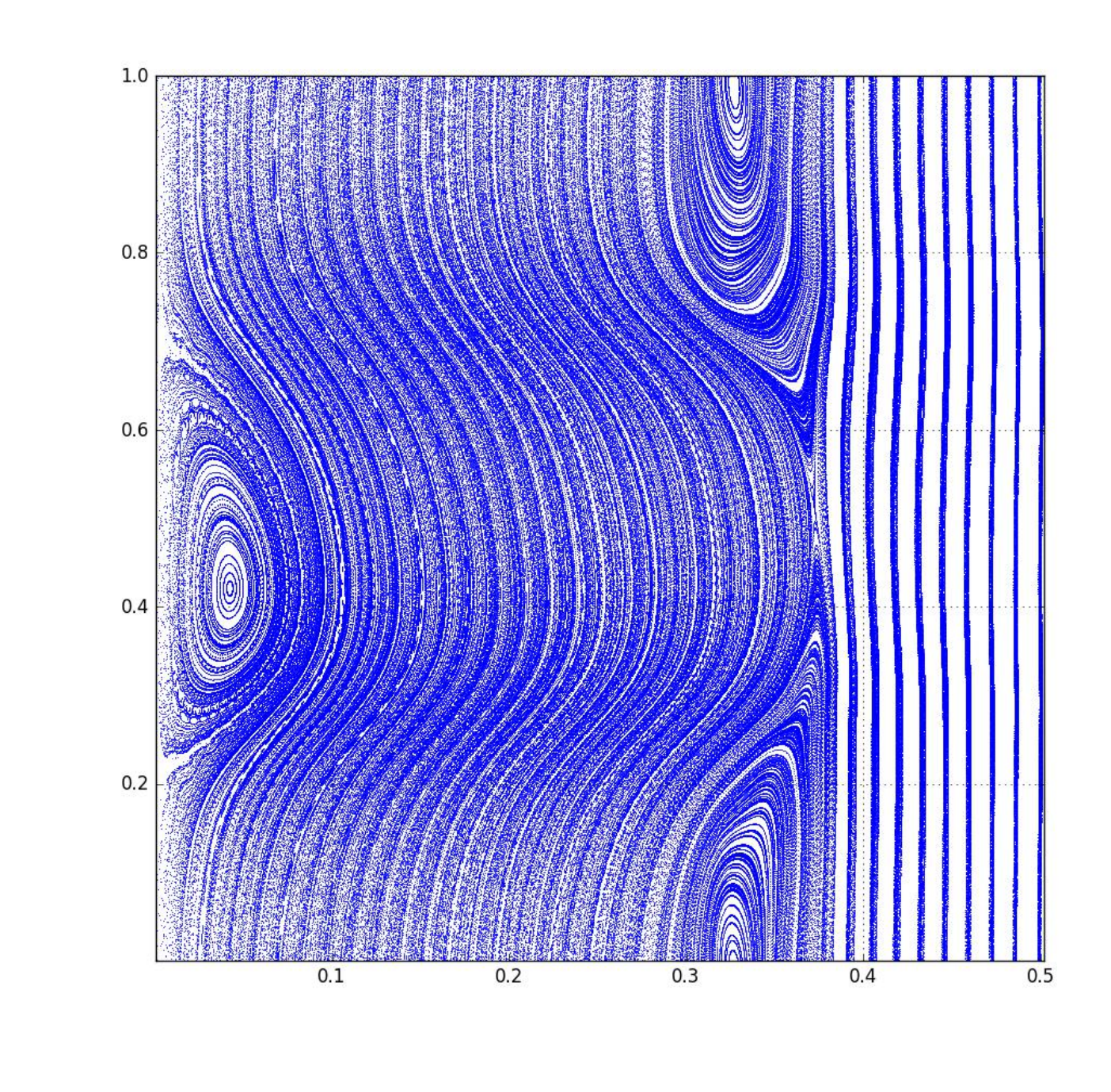}
   \caption{$t/\tau_A = 14878$}
\end{subfigure}   
\caption{Time evolution of the mode amplitude and of the Poincar\'e plots in the strong drive regime. Top figures correspond to the perturbation of the electron temperature in the poloidal plane (R/a,Z/a). Bottom ones correspond to the Poincar\'e plots in polar $(r,\theta)$ coordinates. The dashed line represents the position of the initial $q=1$ surface.}
\label{modeSD}
\end{figure}
The nonlinear fishbone phase is also characterized by a strong evolution of the mode structure and of the magnetic islands, as can be observed in Figure \ref{modeSD} (a) (b) (e) (f). At the beginning of the fishbone phase (Figure \ref{modeSD} (a) (d)), as expected from Figure \ref{KE} (a) at $t = 2640\tau_A$, the mode structure is of $m/n=1/1$ type, with a twisted feature that is due to the weakly $m/n=0/0$ sheared poloidal rotation. It is noted that the mode expands somewhat beyond the $q=1$ radius. Then, until the end of the fishbone phase around $t/\tau_A \sim 9000 \tau_A$, the mode oscillates between 1,1 and  2,2 structures as displayed in Figure \ref{modeSD} (b) at $t=3560\tau_A$. In this figure, it can be observed that the 1,1 and 2,2 modes are superposed, with the 2,2 mode located inside the $q=1$ surface, and the 1,1 mode located just outside of it. These oscillations of the mode structure are consistent with the time evolution of the $n$ modes kinetic energy during the fishbone phase in Figure \ref{KE} (a), where the $n=1$ and $n=2$ modes have at times comparable amplitude, due to the pumping of the $n=1$ mode energy towards the $n=0$. During the nonlinear fishbone phase, in Figure \ref{modeSD} (e) (f), the polar $(r,\theta)$ Poincar\'e maps show the formation of successives 2,2 and 3,3 magnetic islands of small size on the $q=1$ surface. The structure observed for $r<0.1$ in Figure \ref{modeSD} (e)-(h) is due to a shift of the flux tubes from the center of the simulation grid.
\\ \\
The nonlinear phase that occurs after $t\sim 9000\tau_A$ is defined as a mixed phase because it captures characteristic features of both the fishbone and the internal kink instability. Such mixed phases have been observed experimentally \cite{Guenter1999}. During this phase, as it can be noticed in Figure \ref{omega} (a), the 1,1 mode frequency in laboratory frame keeps on chirping down, typical of a nonlinear fishbone phase. Features of an linear internal kink phase can also be noted. As observed in Figure \ref{modeSD} (c) (d), the dominant mode structure is of 1,1 type after $t\sim9000\tau_A$, and is located entirely inside the $q=1$ surface. During this phase, it can been seen in Figure \ref{KE} (a) that the $n=1$ kinetic mode energy grows exponentially in time, and in Figure \ref{modeSD} (g) (h) that a dominant 1,1 magnetic island with increasing size in time appears at the $q=1$ surface. The simulation is stopped once the internal kink magnetic island reaches a significant size around $t=15000\tau_A$.
\subsubsection{Weak drive case}
The weak drive simulation can be decomposed in the same three phases as highlighted in Figure \ref{KE} (b). It can be noted that for this regime, the $n=1$ mode kinetic energy saturates at an amplitude one order of magnitude lower than in the strong drive scenario. At the beginning of the fishbone phase, the $n=1$ mode saturates with increasing oscillations ending near $t\sim 20000\tau_A$. This behavior corresponds to one of the four nonlinear regimes for near-threshold instabilities \cite{Breizman2011} : the periodic amplitude modulation regime. Such a behavior is also illustrated in Ref. \cite{Berk1999}\cite{Idouakass2016}. Moreover in that case, the nonlinear fishbone phase lasts 4 four times longer than in the strong drive scenario, indicating a slower dynamics.
\begin{figure}[h!]
\begin{subfigure}{.24\textwidth} 
   \centering
   \includegraphics[scale=0.22]{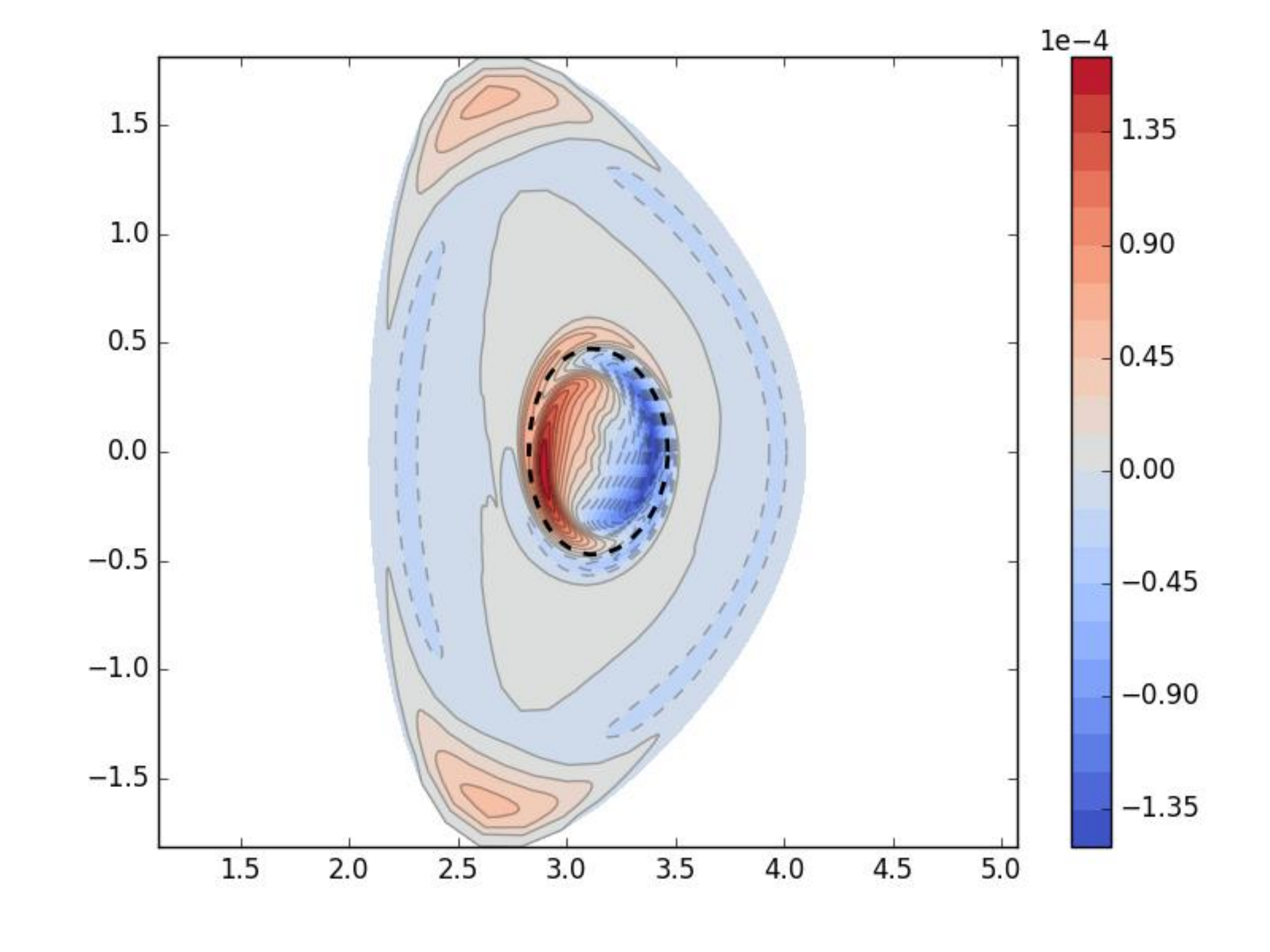}
   \caption{$t/\tau_A = 6281$}
\end{subfigure}
\begin{subfigure}{.24\textwidth} 
   \centering
   \includegraphics[scale=0.22]{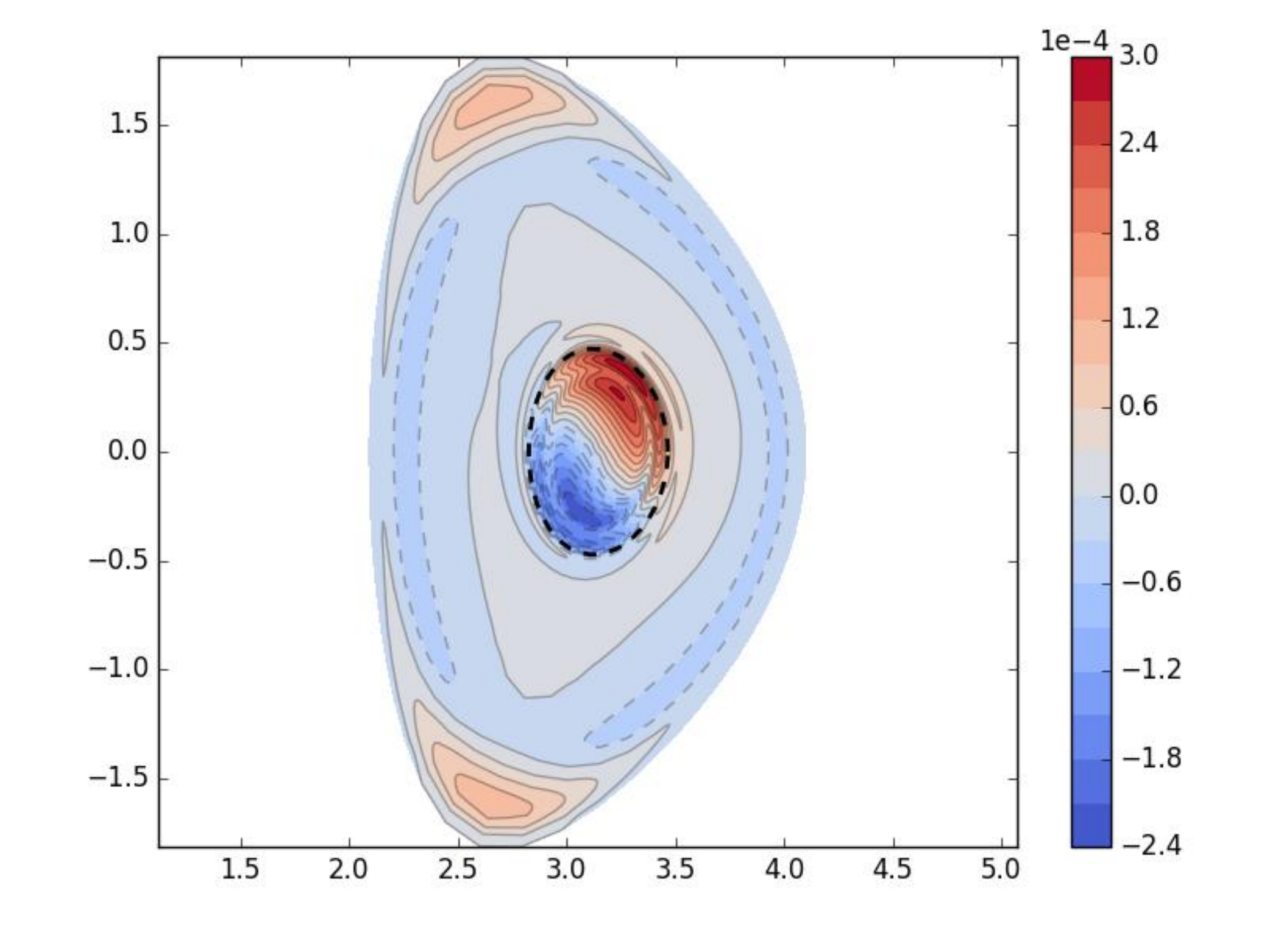}
   \caption{$t/\tau_A = 9676$}
\end{subfigure}
\begin{subfigure}{.24\textwidth} 
   \centering
   \includegraphics[scale=0.22]{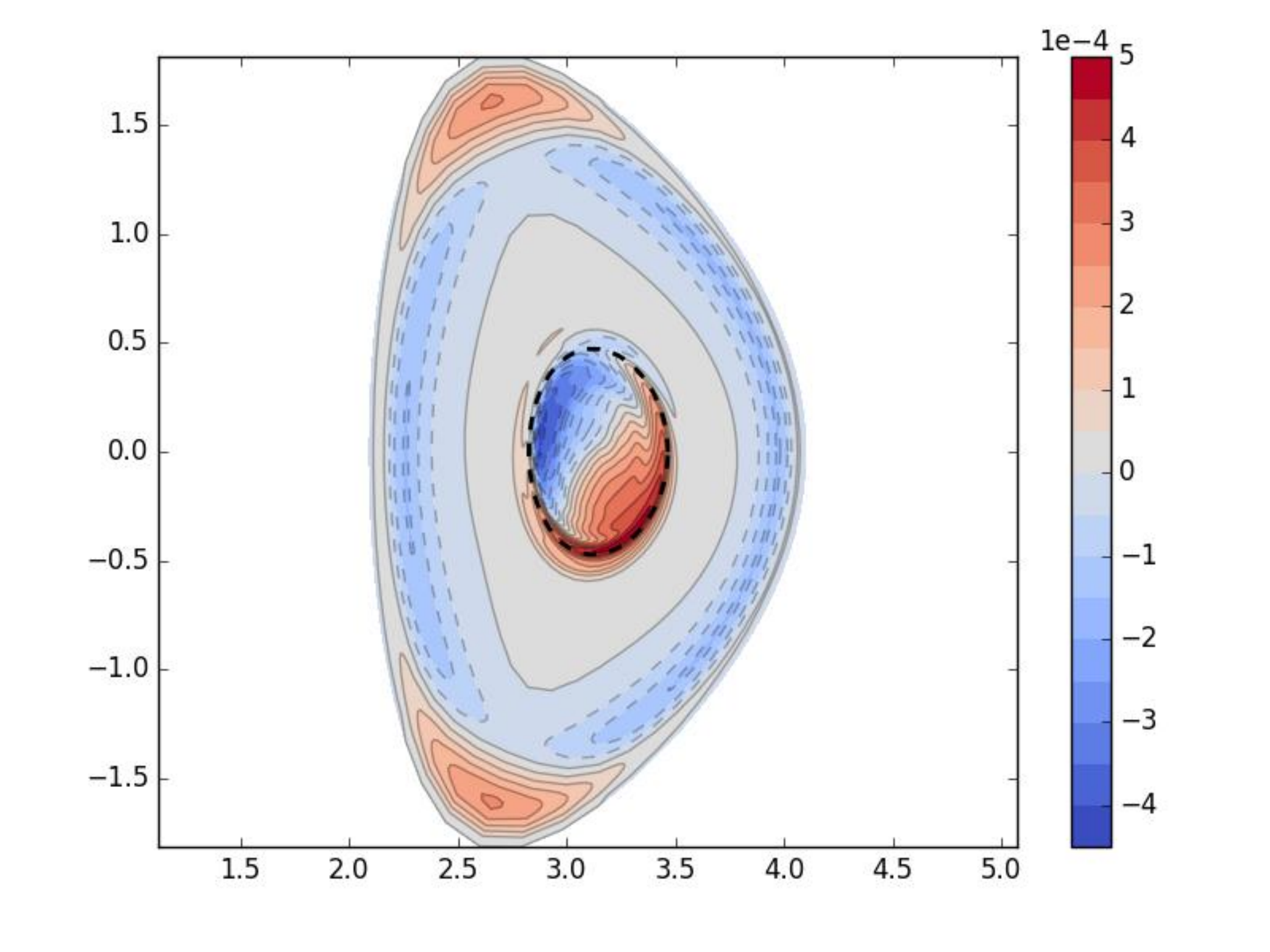}
   \caption{$t/\tau_A = 22058$}
\end{subfigure}
\begin{subfigure}{.24\textwidth} 
   \centering
   \includegraphics[scale=0.22]{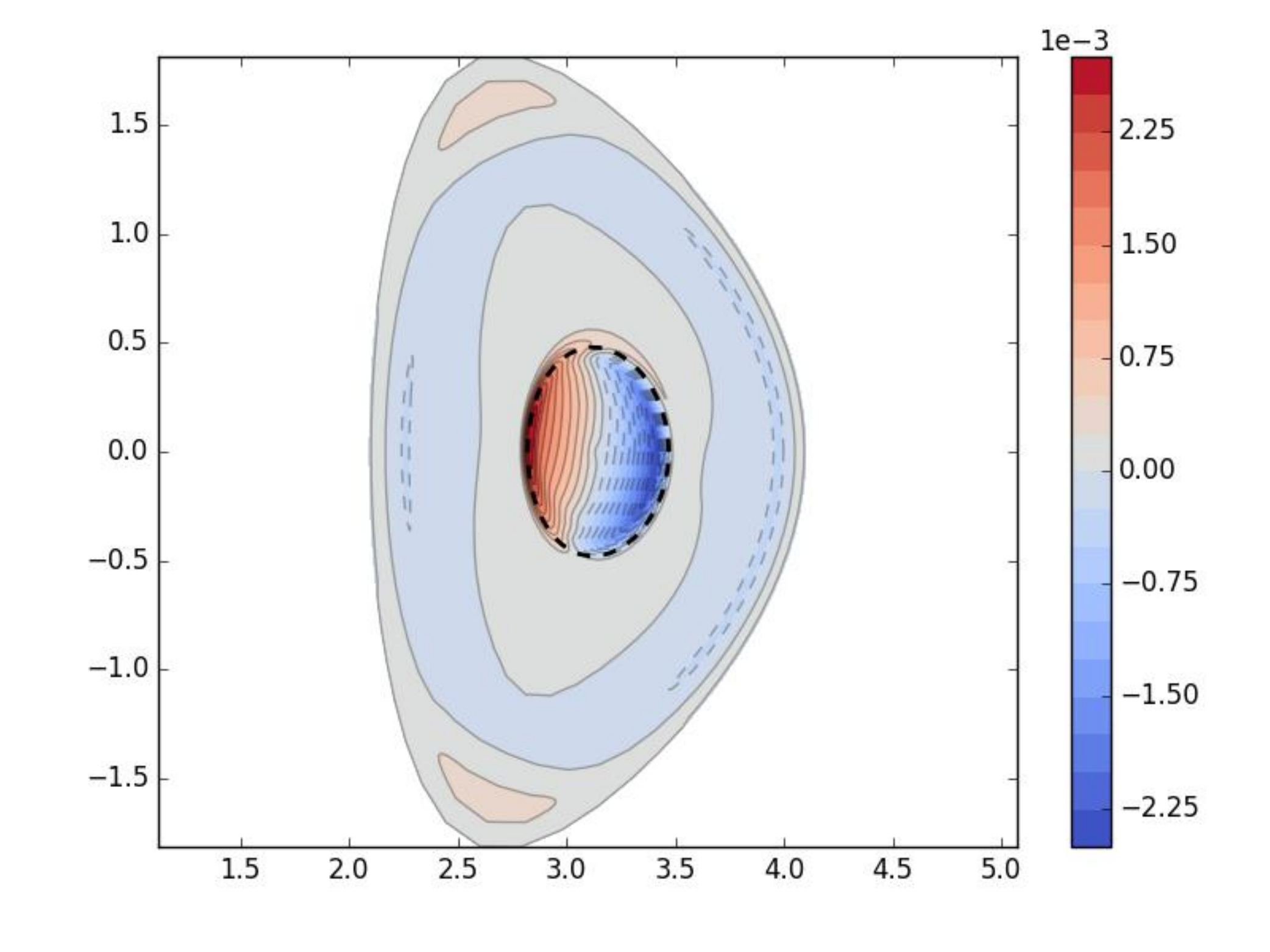}
   \caption{$t/\tau_A = 33161$}
\end{subfigure}
\begin{subfigure}{.24\textwidth} 
   \centering
   \includegraphics[scale=0.31]{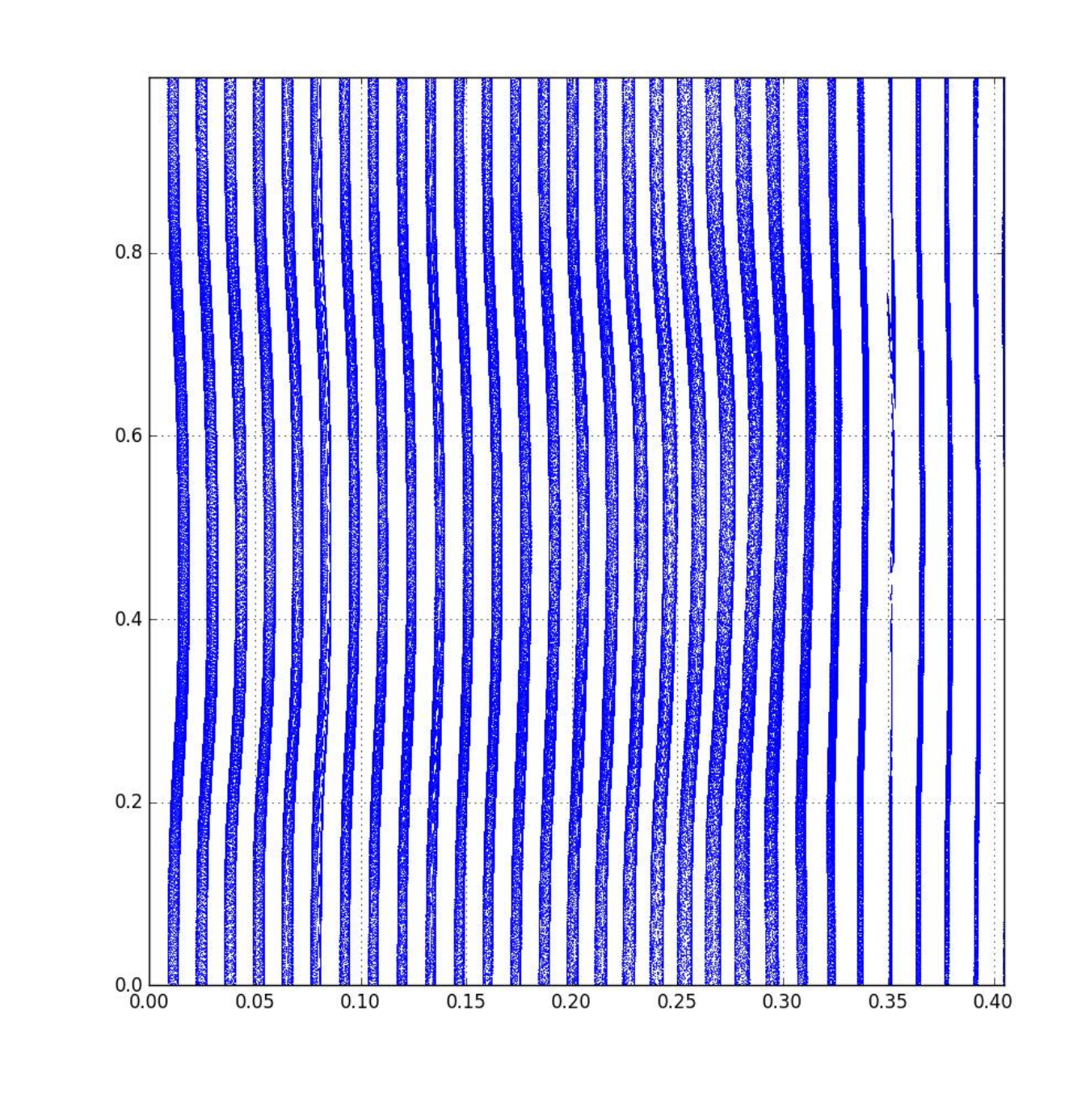}
   \caption{$t/\tau_A = 6281$}
\end{subfigure}
\begin{subfigure}{.24\textwidth} 
   \centering
   \includegraphics[scale=0.31]{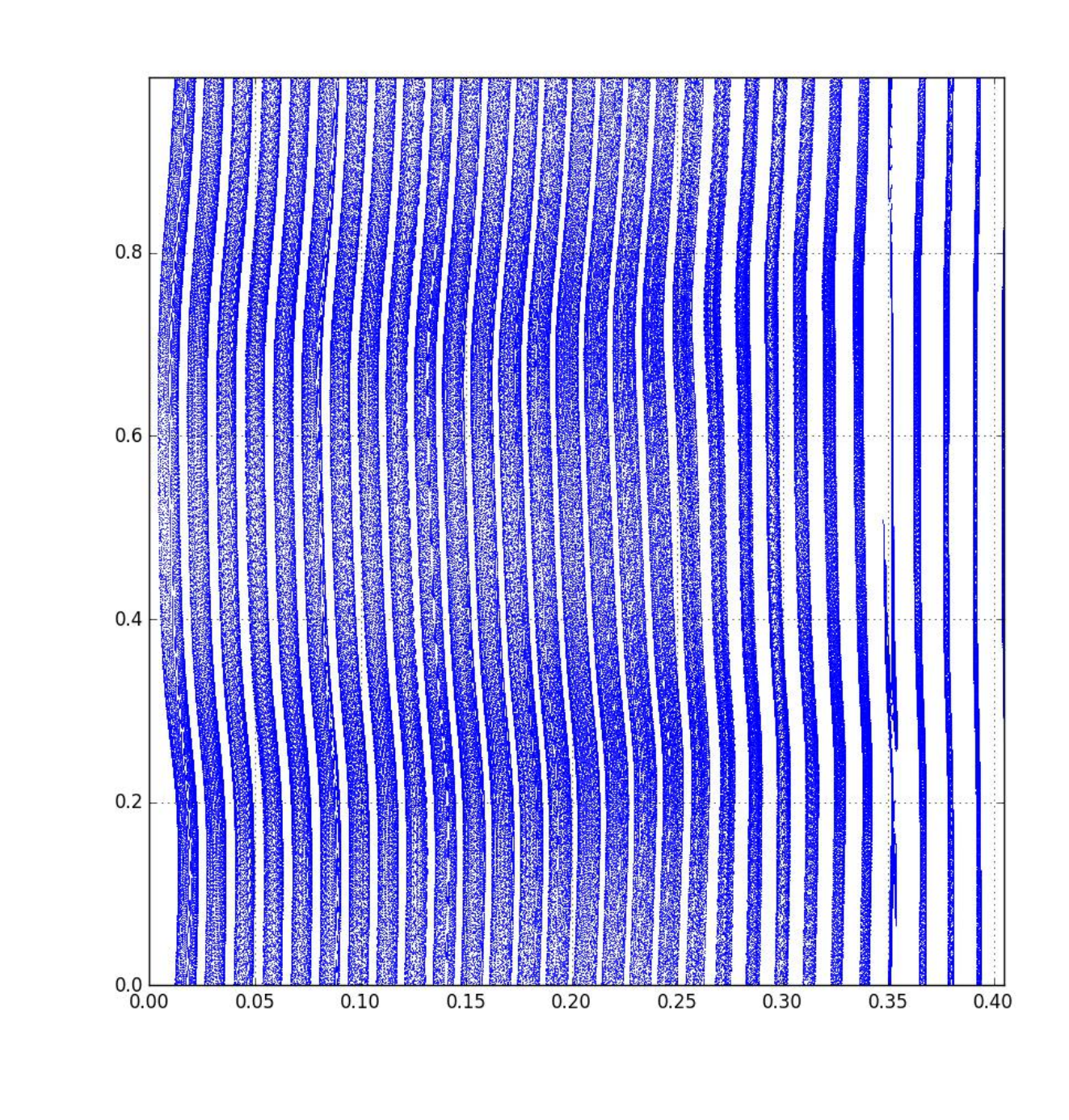}
   \caption{$t/\tau_A = 9676$}
\end{subfigure}
\begin{subfigure}{.24\textwidth} 
   \centering
   \includegraphics[scale=0.31]{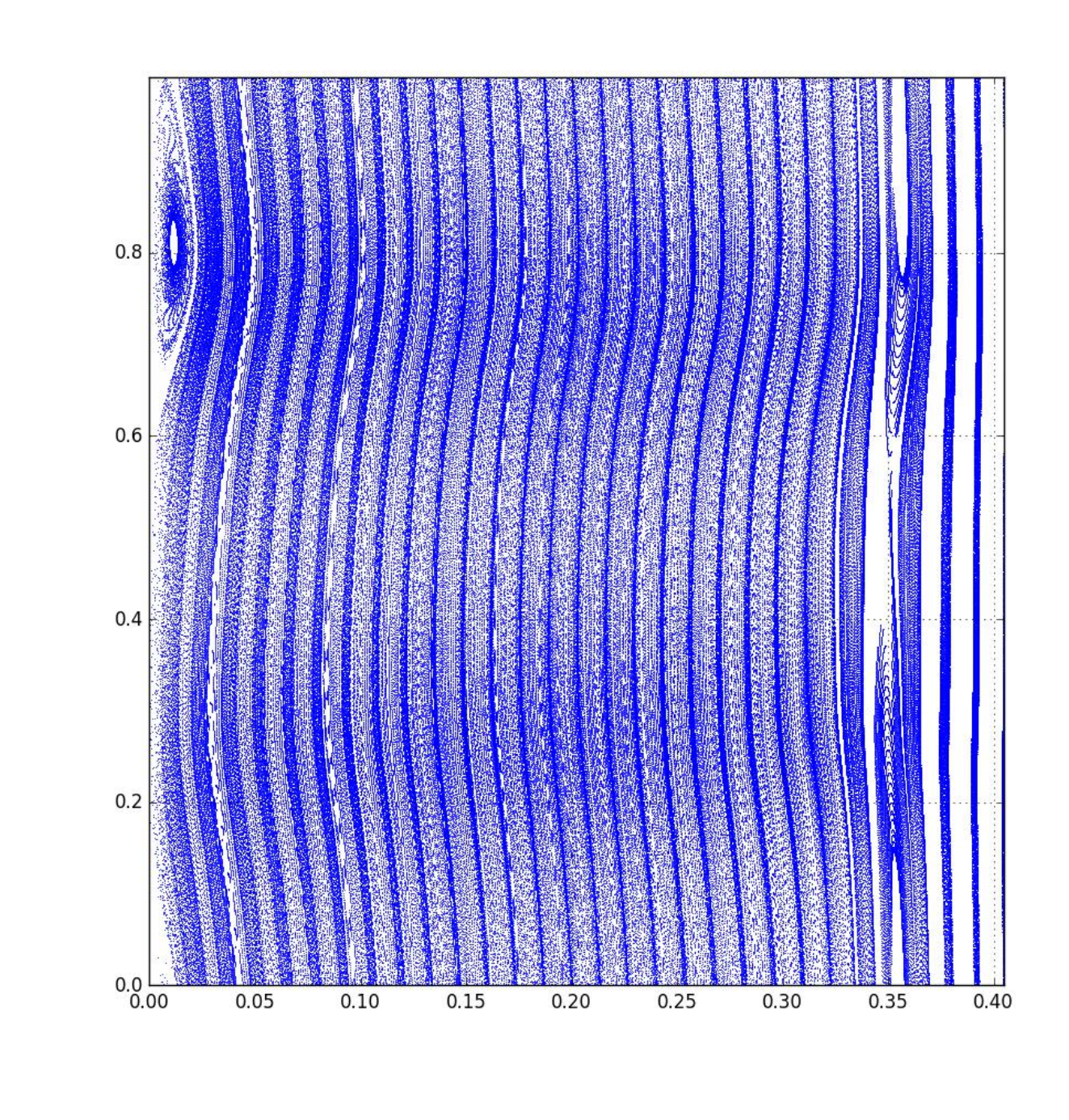}
   \caption{$t/\tau_A = 22058$}
\end{subfigure}
\begin{subfigure}{.24\textwidth} 
   \centering
   \includegraphics[scale=0.31]{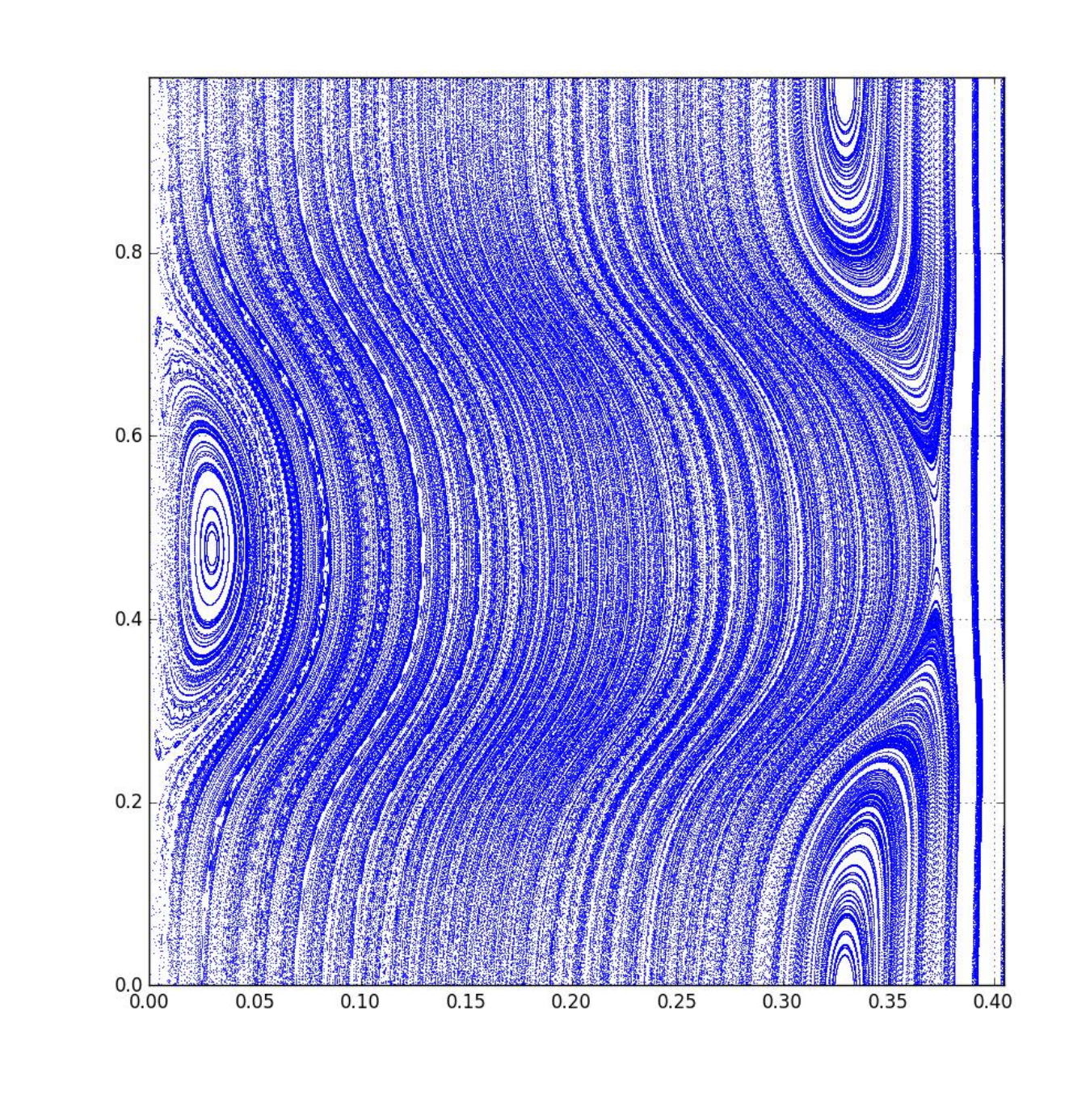}
   \caption{$t/\tau_A = 33161$}
\end{subfigure}   
\caption{Time evolution of the mode amplitude and of the Poincar\'e plots in the weak drive regime. Top figures correspond to the perturbation of the electron temperature in the poloidal plane (R/a,Z/a). Bottom ones correspond to the Poincar\'e plots in polar $(r,\theta)$ coordinates.}
\label{modeWD}
\end{figure}
\\ \\
In contrast to the strong drive scenario, the 1,1 mode frequency in laboratory frame (Figure \ref{omega} (b)) chirps mostly downward in the fishbone phase. It can be noted that a weak additional down-chirping harmonic appears in Figure \ref{omega} (b) for $t\in[1\times10^4,2\times10^4]\tau_A$, with an amplitude slightly lower than the fundamental mode frequency. Furthermore, a sub-dominant up-chirping branch appears in Figure \ref{omega} (b) at the beginning of the fishbone nonlinear phase. The presence of an up-chirping branch in this regime is predicted by near-threshold nonlinear theoretical models \cite{Berk1999}, and have been observed numerically as well with other global hybrid simulations \cite{Shen2017}. \\ The sheared plasma rotation is also non-negligible during the nonlinear fishbone phase as displayed in Figure \ref{omega} (d), even though it is not dominant compared to the 1,1 rotation (Figure \ref{omega} (f)) in the plasma core, as found in the strong drive case. The sheared rotation is mostly carried by the poloidal rotation. 
The evolution of the 0,0 rotation around $t\sim 2\times 10^4\tau_A$ is also linked here to a small increase of the $n=0$ mode kinetic energy in Figure \ref{KE} (b), due to nonlinear energy transfer between the $n=0$ and the $n=1$ modes. In this near-threshold limit, fluid nonlinearities are expected to play a part in mode saturation \cite{Odblom2002}\cite{Chen2016}. \\ \\ In the weak kinetic drive regime, the 1,1 mode structure remains dominant during the whole simulation, as can be seen in Figure \ref{modeWD} (a-d). Moreover, a characteristic feature of the near threshold regime can be observed in Figure \ref{modeWD} (b). When the $n=1$ mode kinetic energy reaches a local minimum near $t\sim9700\tau_A$ in Figure \ref{KE} (b), a second structure forms inside of $q=1$ in Figure \ref{modeWD} (b). It is characteristic of a double step MHD displacement, first observed in \cite{Odblom2002} and \cite{Idouakass2016} during the nonlinear evolution of the near threshold fishbone instability. \\
 At the end of the weak drive simulation, near $t\sim3\times10^4\tau_A$, the fishbone and the internal kink instability co-exist, as can be seen in Figure \ref{modeWD} (h), where a 1,1 magnetic island of significative size is present on $q=1$.
\subsection{Saturation of the fishbone modes}
\subsubsection{Resonances positions in phase space}
\begin{figure}[h!]
\begin{subfigure}{.49\textwidth} 
   \centering
   \includegraphics[scale=0.22]{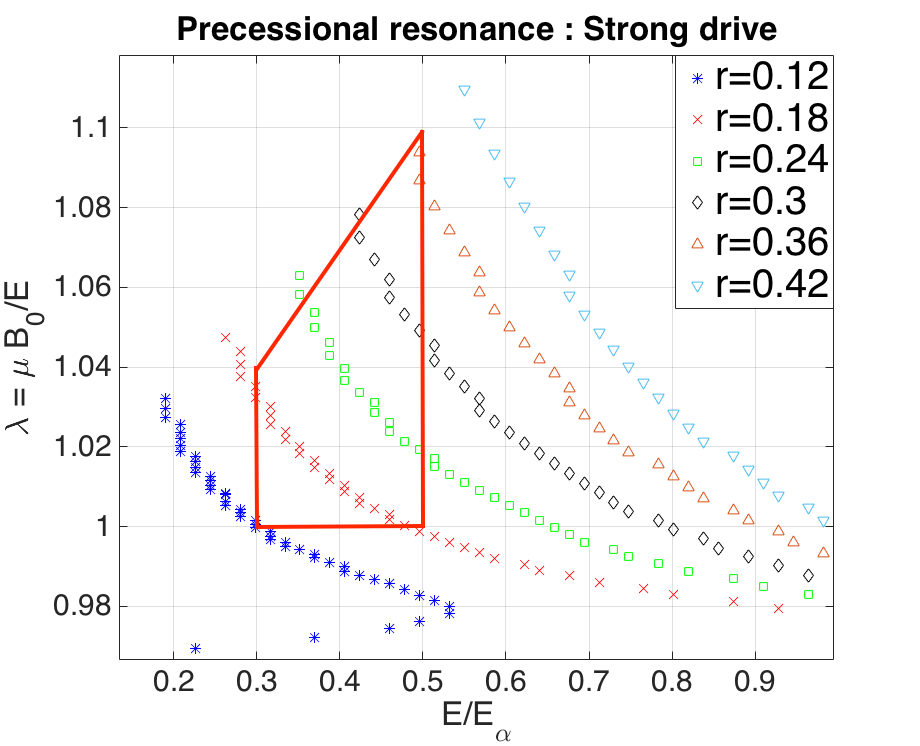}
   \caption{}
\end{subfigure}
\begin{subfigure}{.49\textwidth} 
   \centering
   \includegraphics[scale=0.22]{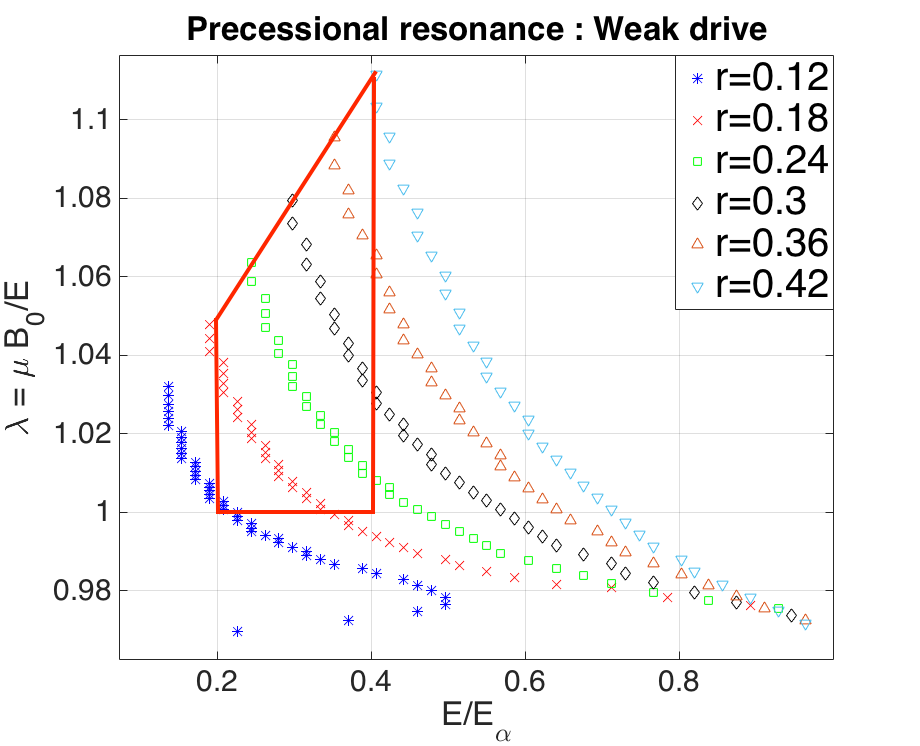}
   \caption{}
\end{subfigure}    
\begin{subfigure}{.49\textwidth} 
   \centering
   \includegraphics[scale=0.22]{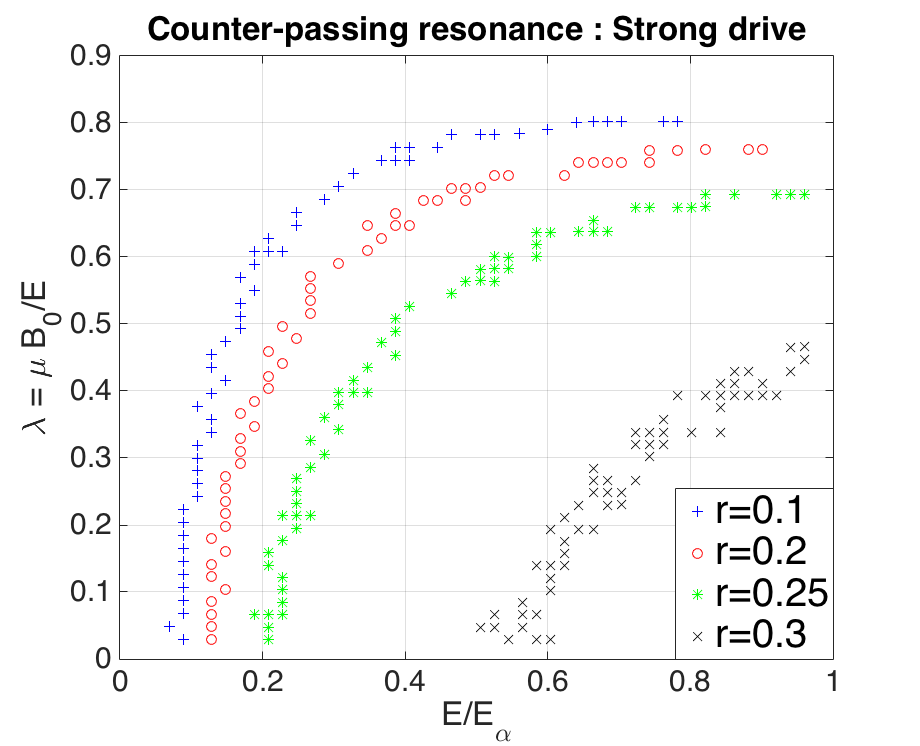}
   \caption{}
\end{subfigure}
\begin{subfigure}{.49\textwidth} 
   \centering
   \includegraphics[scale=0.22]{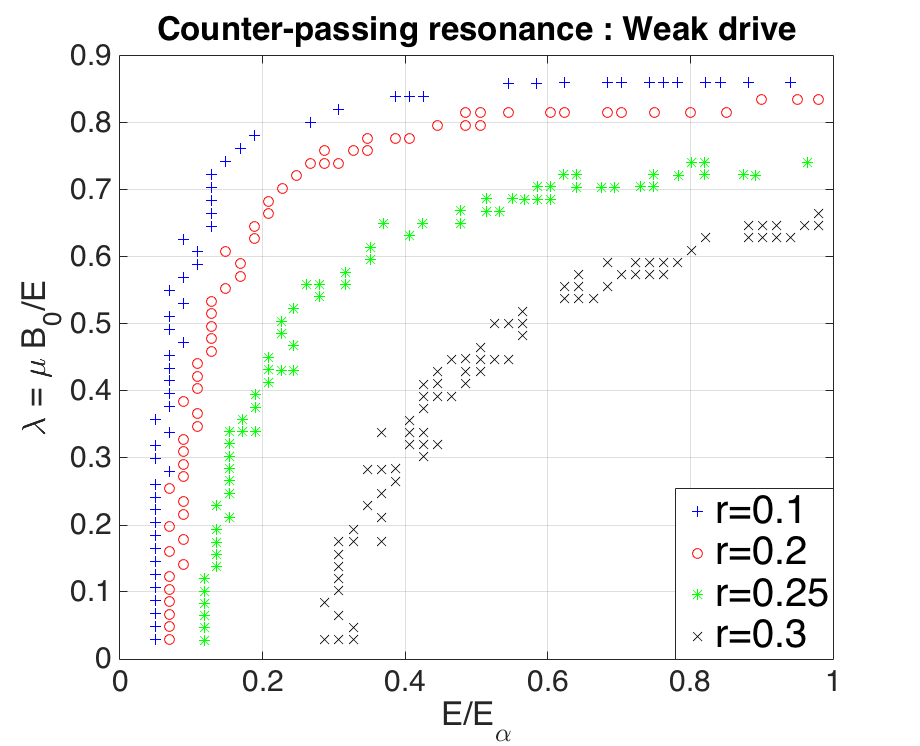}
   \caption{}
\end{subfigure}   
\caption{Resonance curves in the phase space diagram $(E,\lambda)$, taken at different radial position $\bar{r}$. Figures corresponding to the strong drive regime are displayed on the left, and on the right for the weak drive regime. (a,b) Precessional resonance, obtained from the equation $\omega - \omega_d(r,E,\lambda)=0$. The red solid line corresponds to the phase space zone used in Figure 8 to illustrate the resonant particle transport. (c,d) Counter-passing resonance, obtained from the equation $\omega - (1-q[\bar{r}])|\omega_b(r,E,\lambda)|  - \omega_d(r,E,\lambda)= 0.$}
\label{linres}
\end{figure}

Even if fluid nonlinearities contribute to the fishbone mode saturation in both simulations, the main drive for the saturation is found to be the resonant transport of trapped and passing alpha particles. In order to highlight such a process, the linear position of the precessional and passing resonances need to be obtained in phase space. The precessional and passing resonances were identified in \cite{Brochard2020a}\cite{Brochard2018} to be the only relevant resonance conditions for the alpha fishbone instability regarding the general resonance condition 
\begin{equation}\label{rescond}
\omega(r) - [l+\epsilon_bnq(\bar{r})]\omega_b(E,\lambda,\bar{r}) - n\omega_d(E,\lambda,\bar{r}) = 0 
\end{equation}
with $n$ the MHD toroidal harmonic, $l$ the bounce harmonic, $E$ the kinetic particles energy, $\lambda$ their pitch angle, $r = \sqrt{\psi/\psi_{edge}}$ their instantaneous radial position, $\bar{r} = \langle r \rangle$ the radial position of their reference flux surface and $\epsilon_b =0$ for trapped particles and 1 for passing ones. $\omega_b(E,\lambda,r)$ and $\omega_d(E,\lambda,r)$ are respectively the bounce/transit frequency and the precessional frequency of the kinetic particles, defined for a given triplet of invariants $(E,\lambda,\bar{\psi})$ as \cite{Brochard2020a}\cite{Brochard2018}
\begin{equation}\label{charafreq}
\omega_b(E,\lambda,\bar{\psi}) \equiv 2\pi\bigg(\oint_{\mathcal{C}}\frac{dl}{v_{\parallel}}\bigg)^{-1}, \ \ \omega_d(E,\lambda,\bar{\psi}) = \omega_b\oint_{\mathcal{C}}dl\bigg( \textbf{v}_d\cdot\nabla\varphi - q(\bar{\psi})\textbf{v}_d\cdot\nabla\theta + \frac{dq}{d\psi}(\bar{\psi})\hat{\psi}\frac{d\theta}{dt}\bigg)/v_{\parallel}
\end{equation}
with $\bar{\psi}$ the flux surface associated to $\bar{r}$, $\mathcal{C}$ the contour followed by kinetic particles in the poloidal plane, $dl$ a length element of $\mathcal{C}$, $v_{\parallel}(E,\lambda,\bar{\psi})$ the kinetic particles parallel velocity and $\textbf{v}_d(E,\lambda,\bar{\psi})$ their drift velocity. The precessional resonance is associated to trapped particles with $l=0$ such as $\omega = \omega_d$, and the passing resonance to passing particles with $l=-1$ such as $\omega = \omega_d - (1-q)\omega_b$. The considered mode frequency $\omega(r)$ is computed in the plasma frame, taking into account the sheared 0,0 rotation, during the linear phase of both simulations.\\ \\
The solution of the two linear resonance conditions in phase space are obtained from a separate XTOR-K module that advances in time alpha particles on a static electromagnetic field. This field is taken from XTOR-K's initial $n=0$ electromagnetic field, computed on the basis of the Kinetic-MHD equilibrium solution of the Grad-Shafranov equation, obtained from the code CHEASE. The total pressure used in CHEASE is the sum of fluid pressure profiles and the envelop of kinetic pressure profiles. This operation enables accounting for the shaped geometry of the ITER configuration, contrarily to the analytical expressions of Eq. (\ref{charafreq}) derived in \cite{Brochard2020a}\cite{Brochard2018} that assumed a circular geometry. The mode frequency $\omega(r)$ in Eq. (\ref{rescond}) is obtained from the linear phase of the hybrid XTOR-K simulation. \\ \\ For the determination of the resonances in the $(E,\lambda)$ plane, the characteristic frequencies of a set of alpha particles are examined. They are initialized to fill the $(E,\lambda)$ diagram for a given radial position $\bar{r}$. Values for the precessional and bounce/transit frequencies in this diagram are obtained from the time evolution of the toroidal angle $\varphi(t)$ and the poloidal angle $\theta(t)$ of each particle. The bounce/transit frequency $\omega_b$ is computed using a Fourier transform of $\theta(t)$ ($\theta$ is defined modulo $2\pi$). The computation of $\omega_d$ requires more care, since the toroidal motion of passing and trapped particles is different. In the general case, the toroidal angle of kinetic particles verifies $\varphi \propto \Omega_3 t$, where $\Omega_3 = \omega_d + \epsilon_bq(\bar{r})\omega_b$ is the third characteristic frequency in the angle-action formalism (\cite{Brochard_PhD} chapter 2). $\Omega_3$ can be obtained by a linear regression of $\varphi(t)$, and the solutions of the resonance conditions are computed  from $\omega - \Omega_3= 0$ for the precessional resonance, and $\omega + \Omega_2 - \Omega_3 = 0$ for the passing resonance, with $\Omega_2 = \omega_2$ the second characteristic frequency in the angle-action formalism. \\ \\ Once a 2D mapping of $\Omega_2,\Omega_3$ has been obtained in the $(E,\lambda)$ diagram, the solutions of Eq. (\ref{rescond}) are found by interpolation. At a fixed $\lambda$, $E$ is varied in $[0,E_{\alpha}]$ and the couples $(E,\lambda)$ solutions of Eq. (\ref{linres}) are identified when the quantities $\omega - \Omega_3$ and $\omega + \Omega_2 - \Omega_3$ change sign. The operation is repeated for all pitch angles in order to cover the $(E,\lambda)$ diagram.\\ \\
The linear positions of the precessional and passing resonance in the phase space diagram $(E,\lambda)$ at a given radial position $\bar{r}$ are displayed in Figure \ref{linres} for both simulations. Several radial positions $\bar{r}$ have been considered. Regarding the passing resonance, only counter-passing particles were found to have resonance positions in the phase diagram $(E,\lambda)$ for the considered equilibria. \\ \\ The dependencies of the resonant curves on the invariants $(E,\lambda,\bar{r})$ are similar to the ones obtained with the simplified analytical expressions of $\omega_b$ and $\omega_d$ in \cite{Brochard2018}. In the trapped phase space diagram (Figure \ref{linres} (a-b)), since analytically $\omega_d\propto E\lambda/r$, the precessional resonance moves to higher energy and pitch-angle values when larger radial position $\bar{r}$ are considered with constant mode frequency $\omega(r)$. Consequently, when the mode frequency tends to chirp down, at fixed radial position, the resonance is pushed towards lower energy and pitch angles values. \\ \\
In the passing phase space diagram (Figure \ref{linres} (c-d)), the dependency of the resonant frequency $\omega_d + (1-q)|\omega_b|$ over the invariants $(E,\lambda,\bar{r})$ is more complicated, since it is a linear combination between the transit and precessional frequencies. However, it can be noted on Figure \ref{linres} (c-d) that the passing resonance moves to higher energy and lower pitch-angles values when larger $\bar{r}$ are considered. It implies that at fixed radial position, the resonance evolves to lower energy and higher pitch-angle values as the mode frequency chirps down. The dependency of the precessional and passing resonances over the mode frequency $\omega$ in phase space will be useful when analysing the nonlinear phase space dynamics in section 3.2.
\subsubsection{Resonant flattening of the alpha density profiles}
\begin{figure}[h!]
\begin{subfigure}{.24\textwidth} 
   \centering
   \includegraphics[scale=0.15]{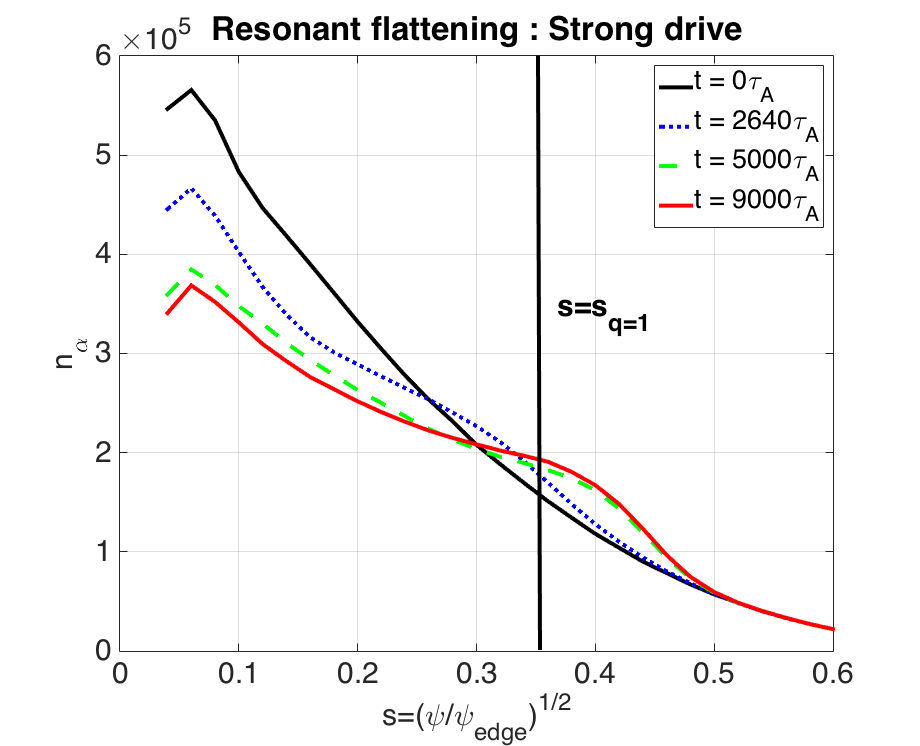}
   \caption{}
\end{subfigure}
\begin{subfigure}{.24\textwidth} 
   \centering
   \includegraphics[scale=0.15]{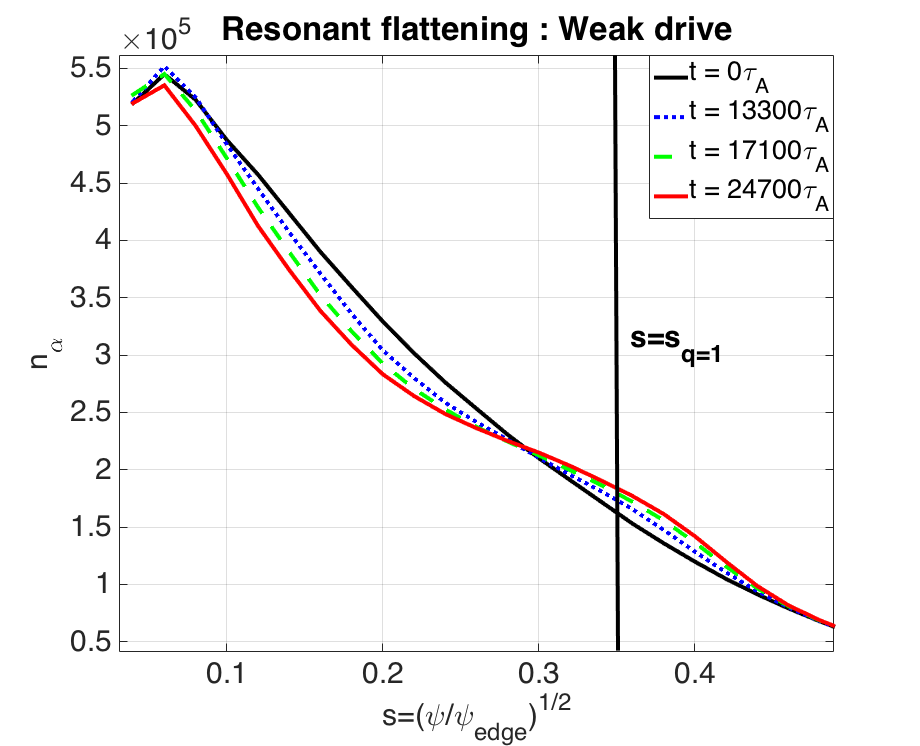}
   \caption{}
\end{subfigure}   
\begin{subfigure}{.24\textwidth} 
   \centering
   \includegraphics[scale=0.15]{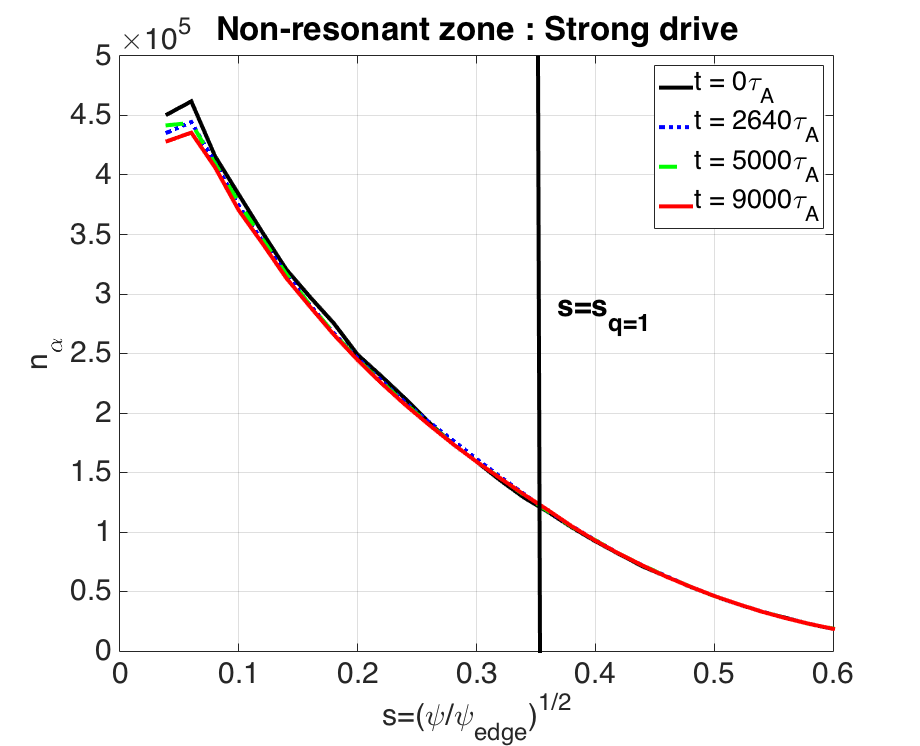}
   \caption{}
\end{subfigure}
\begin{subfigure}{.24\textwidth} 
   \centering
   \includegraphics[scale=0.15]{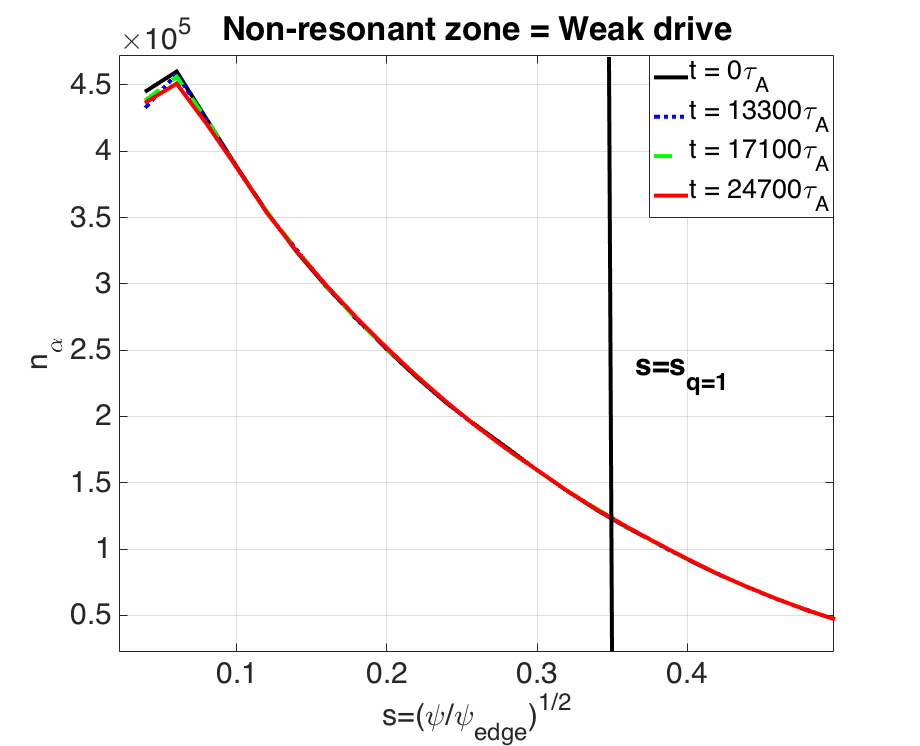}
   \caption{}
\end{subfigure}     
\caption{Resonant flattening of the alpha particles density profile. (a,b) Time evolution of the alpha density profile in the resonant phase space zones illustrated in Figure 7. (c,d) Time evolution of alpha density profile in the non-resonant phase space zones.}
\label{flat}
\end{figure}

In both hybrid simulations, the resonant transport of alpha particles affects mostly trapped alphas. For clarity, the resonant flattening of the alpha density profile is illustrated by considering only trapped particles. Two resonant zones of phase space are highlighted in Figure \ref{linres} (a-b) with the solid red trapezes for the strong and weak drive simulations. The time evolution of the alpha density profile in these resonant zones of phase space is displayed in Figure \ref{flat} (a-b). In both the strong and weak drive simulations, a flattening of the alpha density in the entire $q=1$ volume is observed during the nonlinear fishbone phases. It is noted that a significantly larger fraction of alpha particles is transported out of $q=1$ in the strong drive case than in the weak drive case.  \\ \\
In order to demonstrate that the alpha transport is a mainly resonant process, a non-resonant zone of phase space, $E\in [0,0.2]E_{\alpha}$, $\lambda\in[\lambda_{min}(\bar{r}),0.96]$ has been chosen in Figure (\ref{linres}). $\lambda_{min}(\bar{r}) = B_0/B_{max}(\bar{r})$ denotes the pitch-angle at the trapped/passing boundary for a given reference radial position $\bar{r}$, with $B_{max}(\bar{r})$ the maximum value of the magnetic field on the flux surface $\bar{\psi}$. The time evolution of the alpha density profile in this non-resonant zone of phase space is shown in Figure \ref{flat} (c-d) for both hybrid simulations. As expected, the density profile remains stationary during the whole nonlinear fishbone phase of the strong and weak kinetic limits, illustrating that the alpha transport is mainly resonant.

\section{Phase space characterization of alpha transport during the nonlinear phase} 
In this section, the phase space dynamics of the alpha fishbone is examined for both trapped and passing particles. In a first part, the transport of alpha particles over the entire simulations is analyzed in the $(E,\lambda)$ diagrams, highlighting the resonant nature of the alpha transport for trapped and passing particles. Then, the instantaneous transport $\delta_{t_1}^{t_2}F_{\alpha}$ and wave-particle energy exchange $\int_{t_1}^{t_2}\textbf{J}_{\alpha}\cdot\textbf{E} \ dt$ (Annex A) are analyzed in phase space on small time windows. It enables to study the combined dynamical evolution of the resonances position and of the phase space structures obtained with XTOR-K, which is not possible for the total particle transport since the mode frequencies evolve significantly during the nonlinear fishbone phases.
\subsection{Total alpha transport in phase space}
\begin{figure}[h!]
\begin{subfigure}{.33\textwidth} 
   \centering
   \includegraphics[scale=0.26]{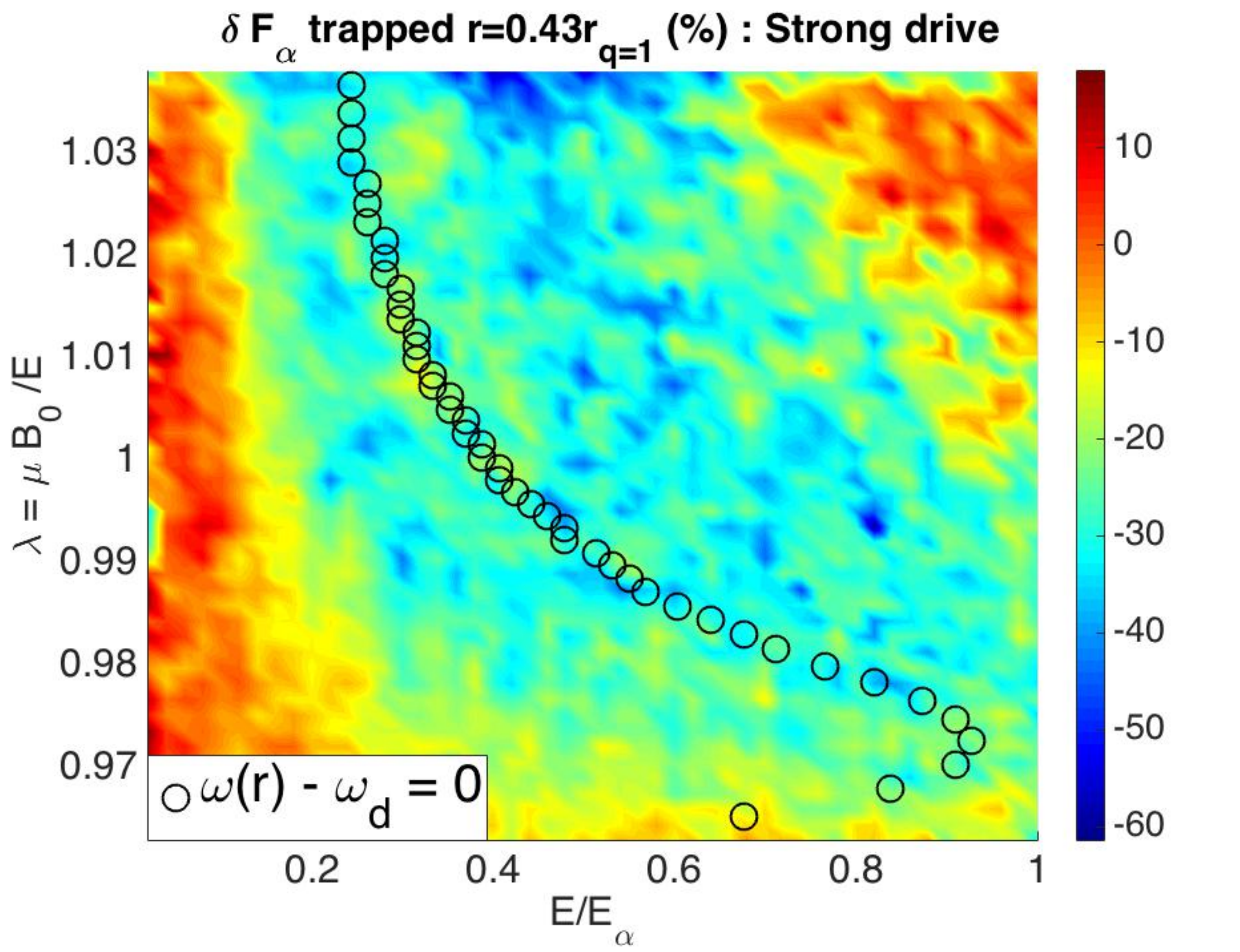}
   \caption{}
\end{subfigure}
\begin{subfigure}{.33\textwidth} 
   \centering
   \includegraphics[scale=0.26]{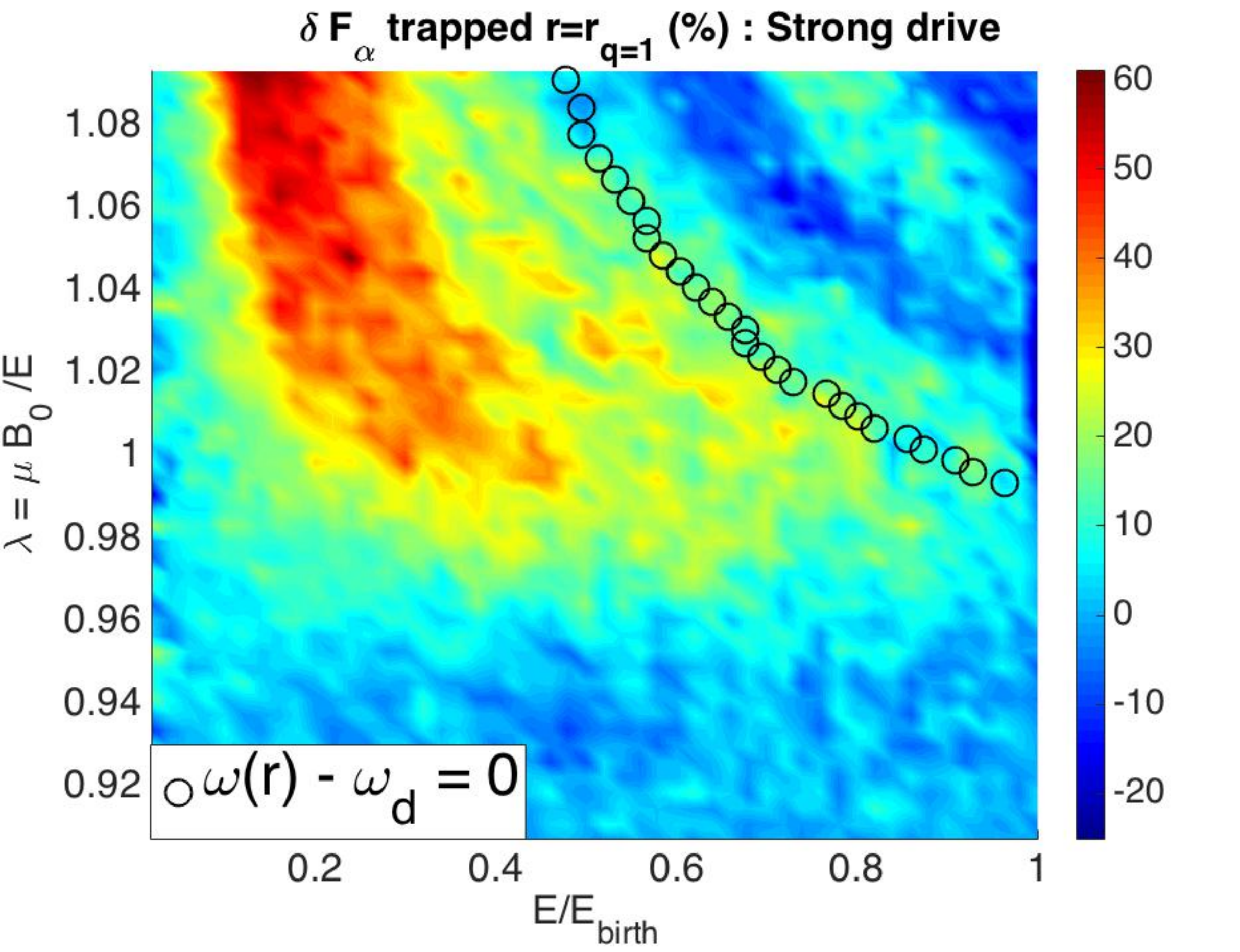}
   \caption{}
\end{subfigure}   
\begin{subfigure}{.33\textwidth} 
   \centering
   \includegraphics[scale=0.26]{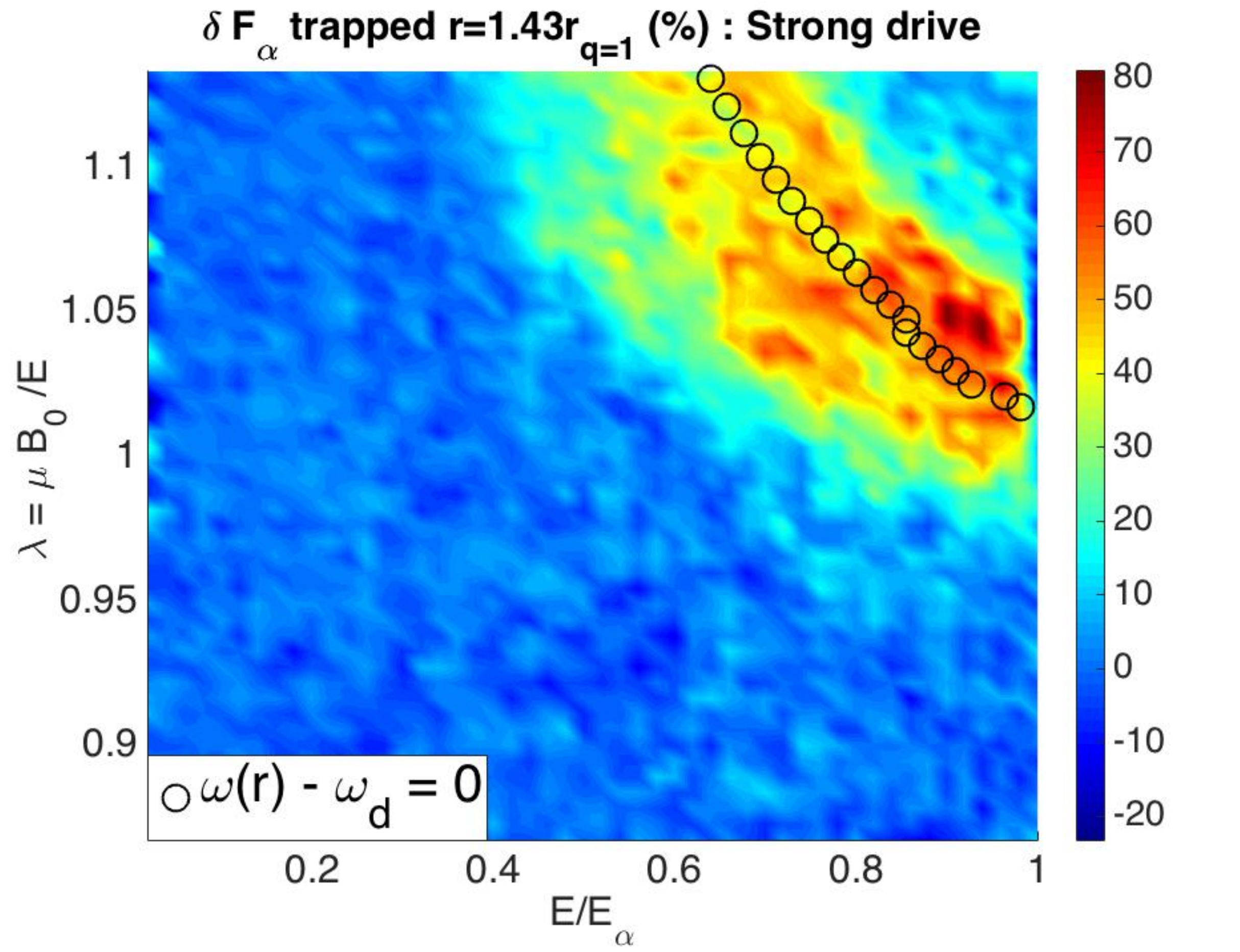}
   \caption{}
\end{subfigure}
\begin{subfigure}{.33\textwidth} 
   \centering
   \includegraphics[scale=0.26]{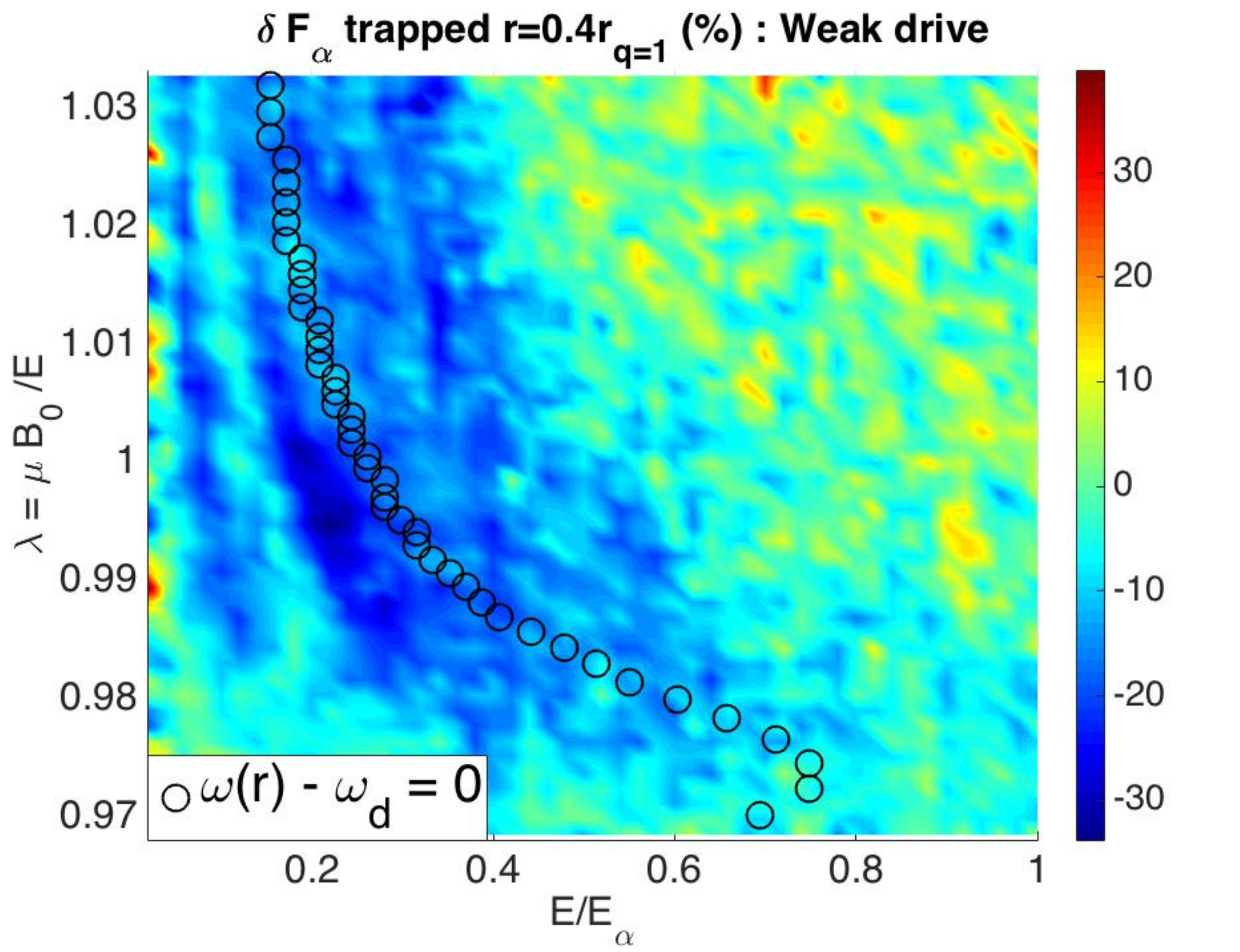}
   \caption{}
\end{subfigure}
\begin{subfigure}{.33\textwidth} 
   \centering
   \includegraphics[scale=0.26]{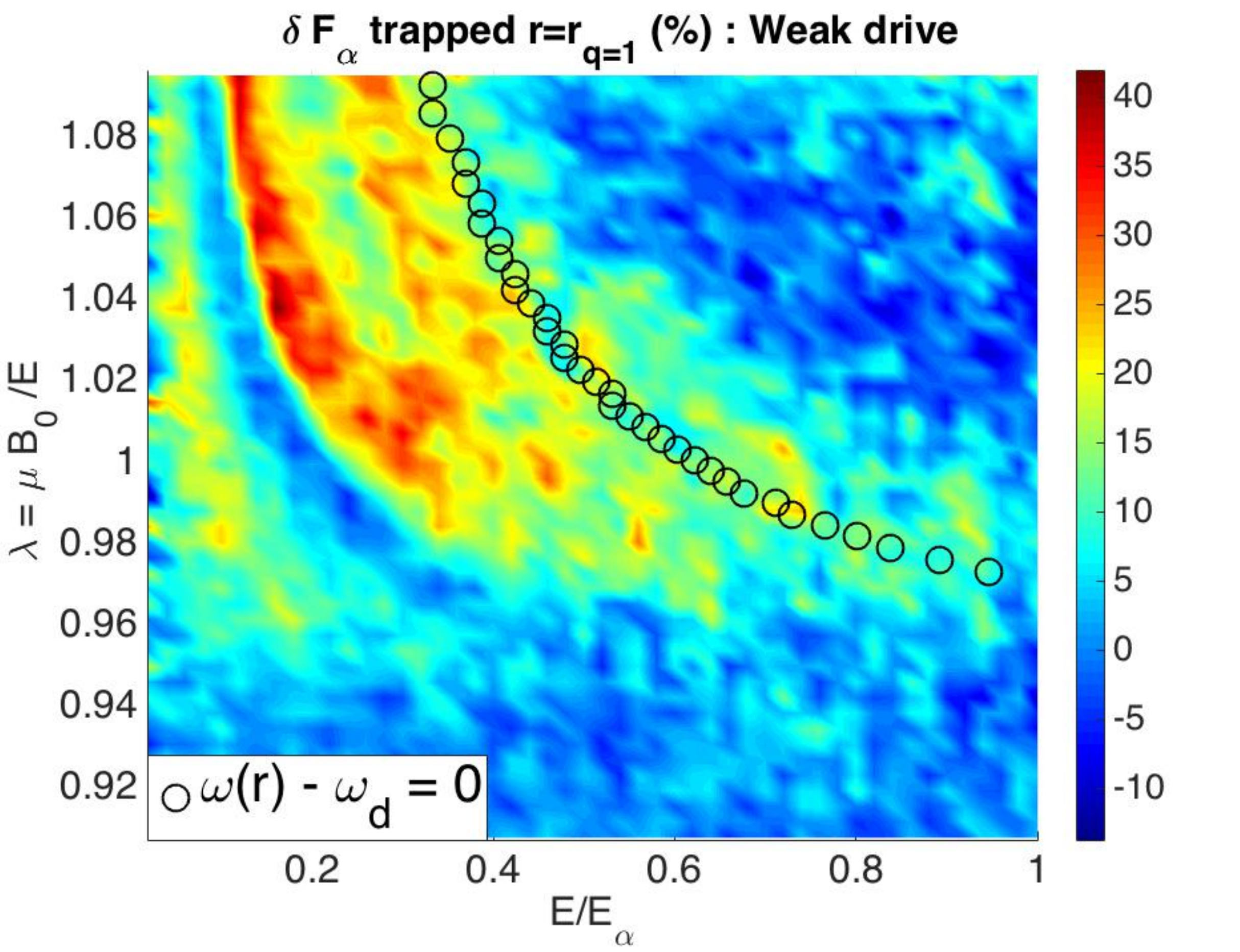}
   \caption{}
\end{subfigure}   
\begin{subfigure}{.33\textwidth} 
   \centering
   \includegraphics[scale=0.26]{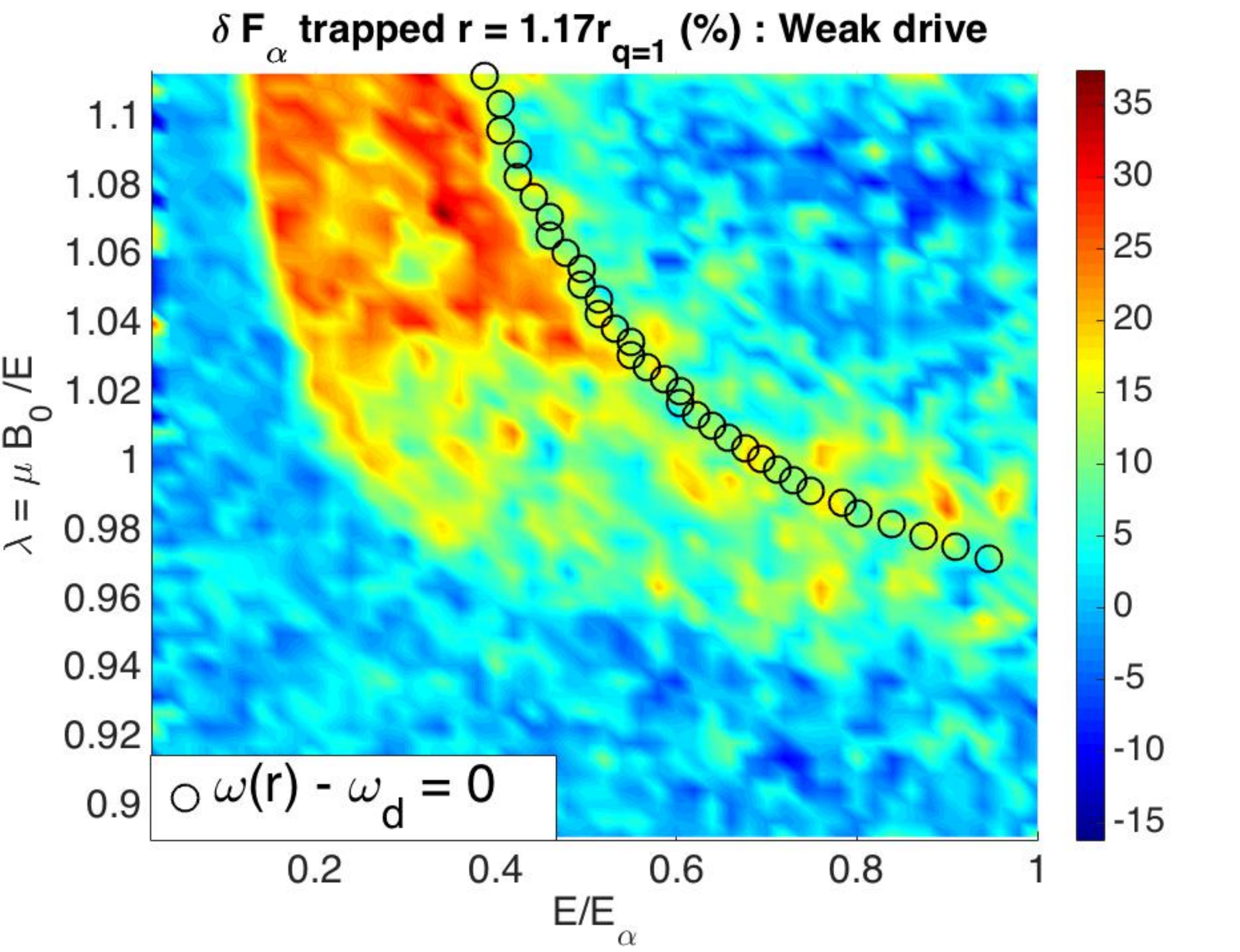}
   \caption{}
\end{subfigure}
\caption{Time evolution of the perturbed alpha distribution function $\delta F_{\alpha}$ on the trapped phase diagram $(E,\lambda)$, at different radial position $\bar{r}$. Top figures correspond to the strong drive regime, bottom ones to the weak drive regime. (a) $r = 0.43 r_{q=1}$ (b)$ r= r_{q=1}$ (c) $r= 1.43 r_{q=1}$ (d) $r=0.4 r_{q=1}$ (e) $r = r_{q=1}$ (f) $r=1.17r_{q=1}$}
\label{totPS_t}
\end{figure}

\begin{figure}[h!]
\begin{subfigure}{.49\textwidth} 
   \centering
   \includegraphics[scale=0.28]{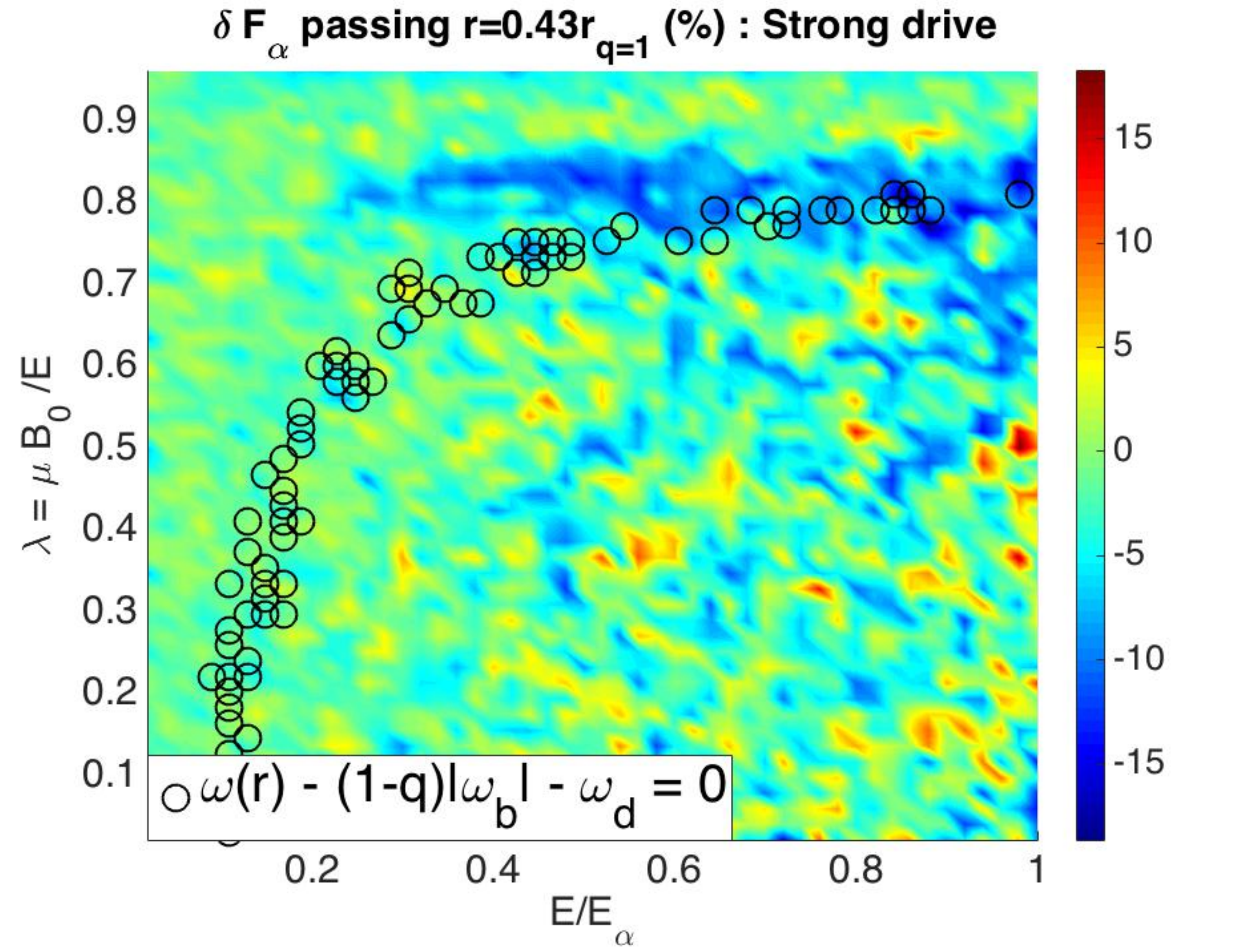}
   \caption{}
\end{subfigure}
\begin{subfigure}{.49\textwidth} 
   \centering
   \includegraphics[scale=0.28]{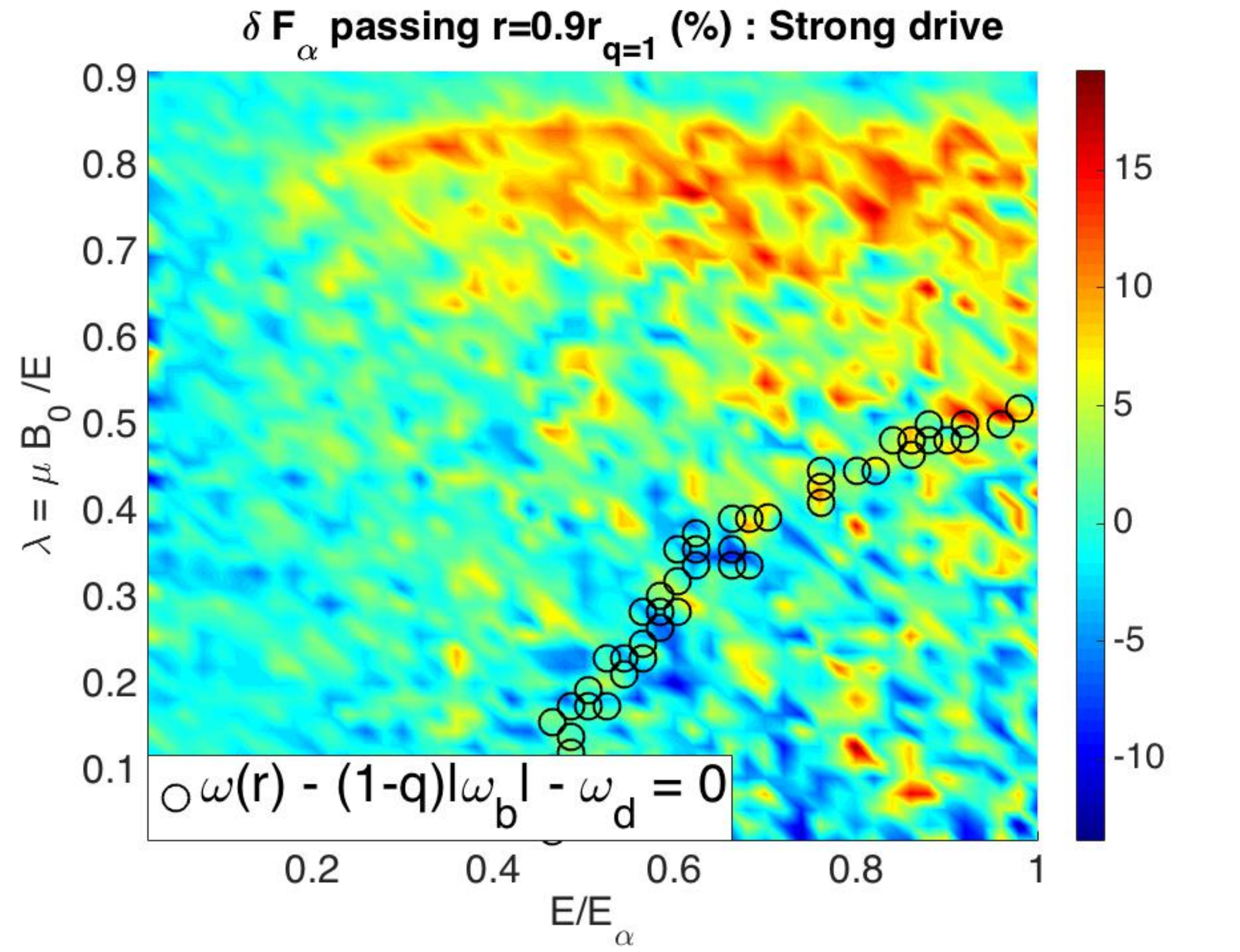}
   \caption{}
\end{subfigure}   
\begin{subfigure}{.49\textwidth} 
   \centering
   \includegraphics[scale=0.28]{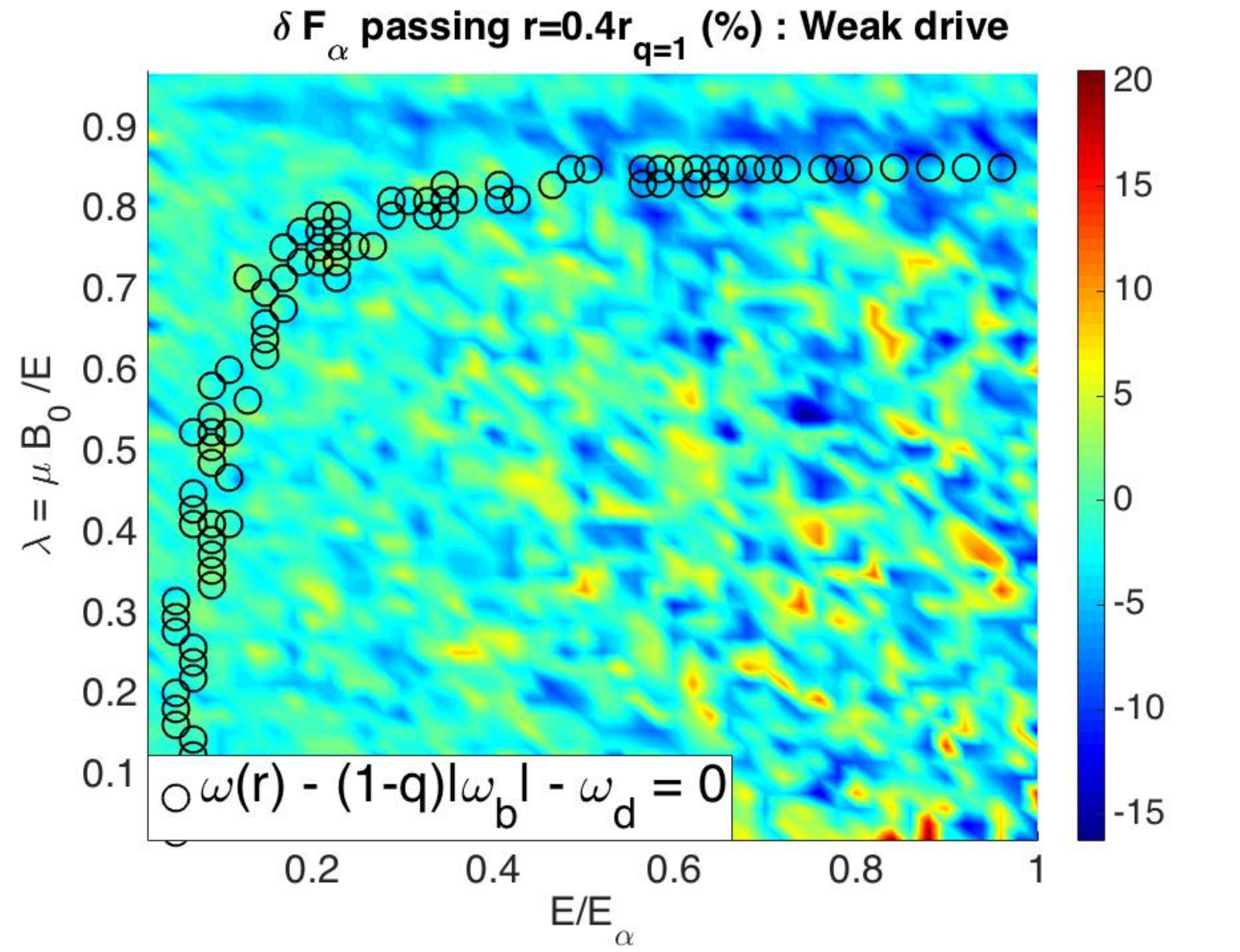}
   \caption{}
\end{subfigure}
\begin{subfigure}{.49\textwidth} 
   \centering
   \includegraphics[scale=0.28]{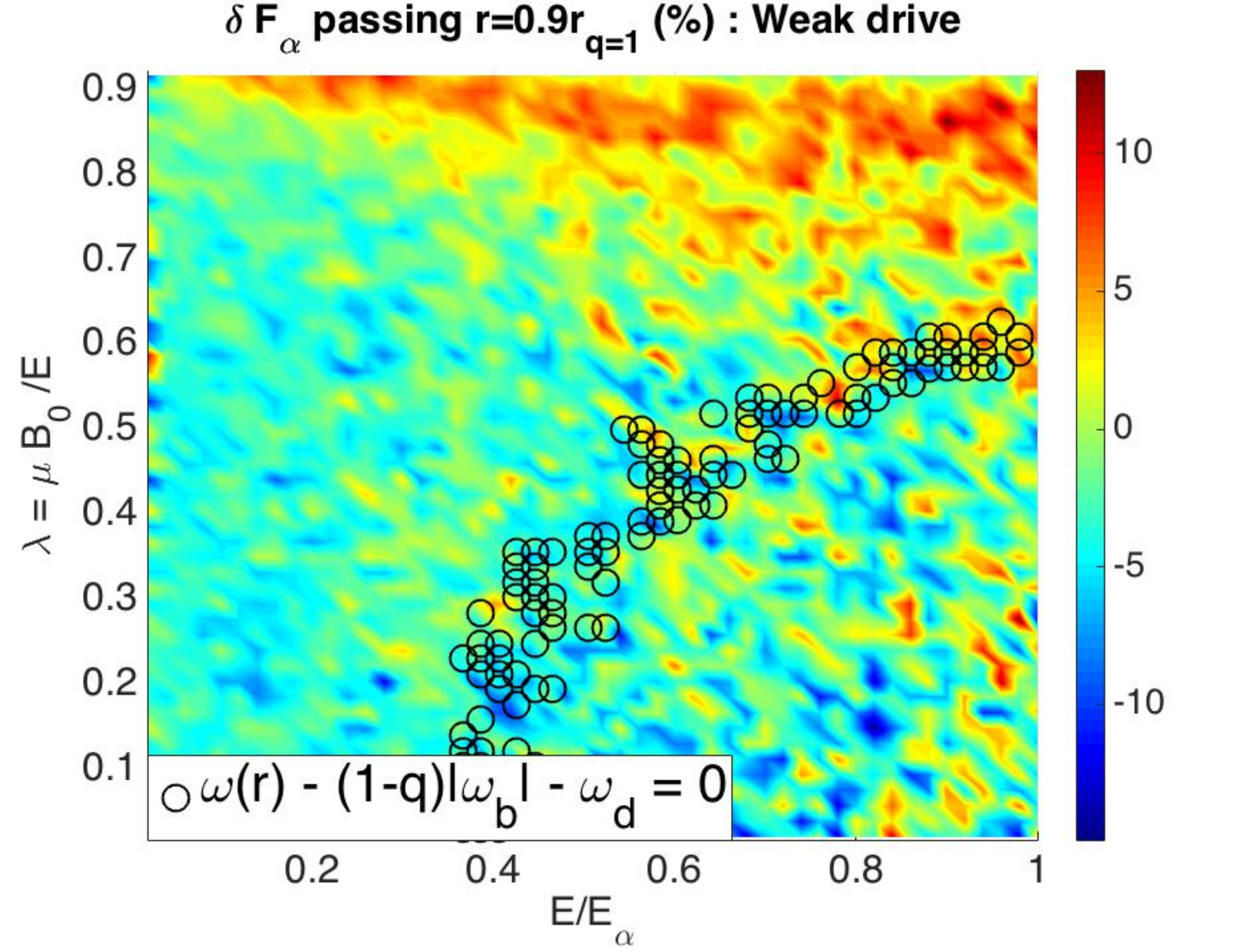}
   \caption{}
\end{subfigure}     
\caption{Time evolution of the perturbed alpha distribution function $\delta F_{\alpha}$ on the passing phase diagram $(E,\lambda)$, at different radial position. Top figures correspond to the strong drive regime, bottom ones to the weak drive regime. (a) $r = 0.43 r_{q=1}$ (b)$ r= 0.9r_{q=1}$ (c) $r = 0.4r_{q=1}$ (d) $r=0.9r_{q=1}$}
\label{totPS_p}
\end{figure}

\begin{figure}[h!]
\begin{subfigure}{.33\textwidth} 
   \centering
   \includegraphics[scale=0.19]{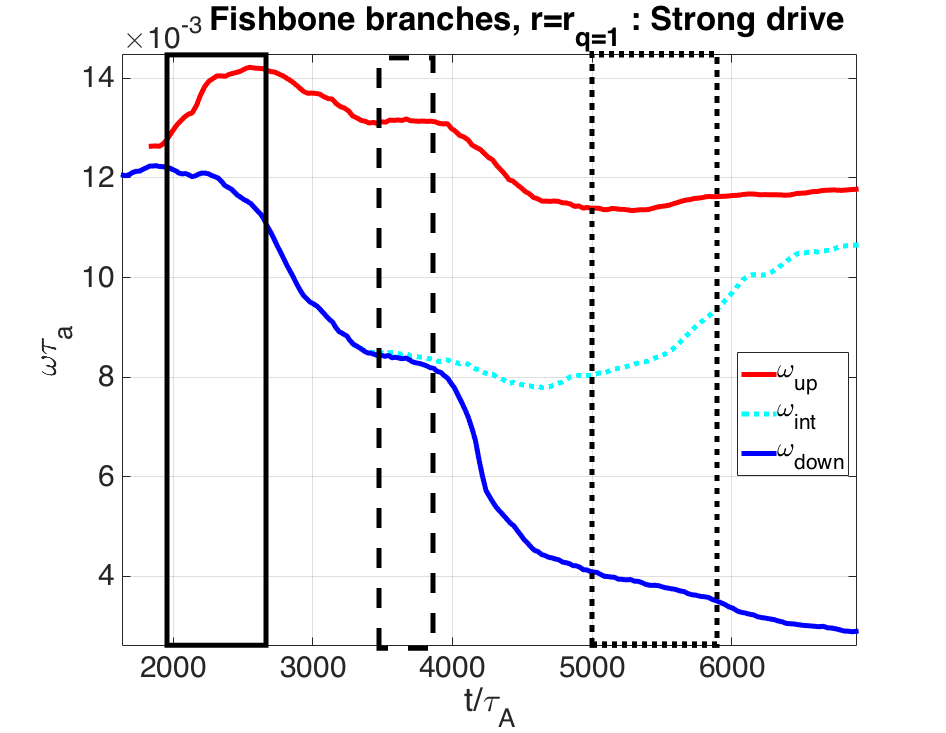}
   \caption{}
\end{subfigure} I
\begin{subfigure}{.33\textwidth} 
   \centering
   \includegraphics[scale=0.19]{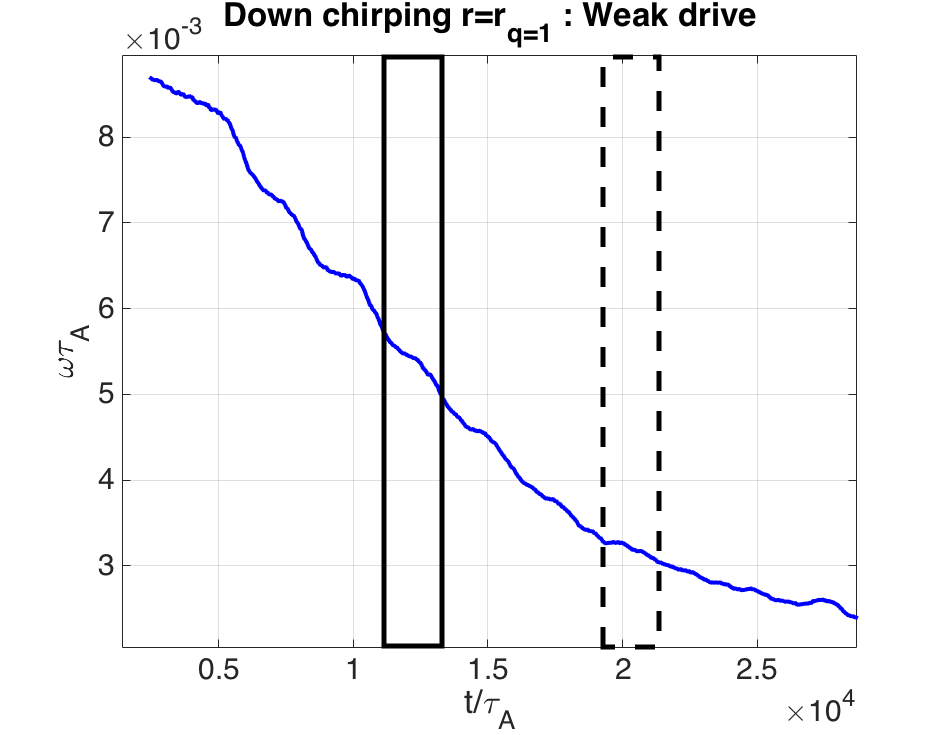}
   \caption{}
\end{subfigure}
\begin{subfigure}{.33\textwidth} 
   \centering
   \includegraphics[scale=0.19]{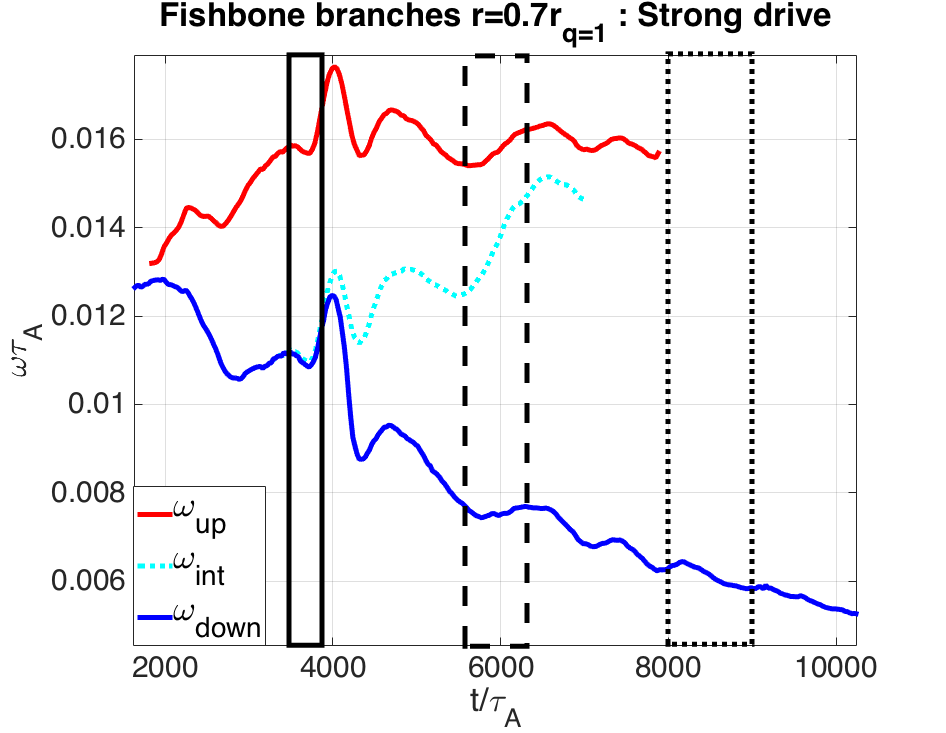}
   \caption{}
\end{subfigure}   
\caption{Considered time windows (black rectangles with solid, dashed and dotted lines) for phase space investigation, highlighting the time evolution of the multiple local 1,1 mode frequencies in plasma frame. (a) Strong drive regime at $r=r_{q=1}$ 
, illustrated in Figure \ref{trappedSD}. (b) Weak drive regime at $r=r_{q=1}$, displayed in Figure \ref{trappedWD}. (c) Strong drive regime at $r=0.7r_{q=1}$, plotted in Figure \ref{passingSD}.}
\label{window}
\end{figure}

The resonant nature of the alpha transport is highlighted here by looking at the perturbed normalized alpha distribution function $\delta F_{\alpha} = [F_{\alpha}(t_{final}) - F_{\alpha}(0)]/F_{\alpha}(0)$ in phase space at the end of each simulations. It is chosen to illustrate the transport by computing $\delta F_{\alpha}$ in the phase space diagram $(E,\lambda)$, at different radial positions spanning from the inner plasma to outside the $q=1$ surface. The trapped and passing parts of phase space are separated. It is expected that alpha particles cannot be transported much further than the $q=1$ surface since beyond this radial position, the mode structure tends to vanish as observed in Figure \ref{modeSD}-\ref{modeWD}, which prevents particles from resonating with the mode. However, as it will be shown in Figure \ref{totPS_t} (c,f), resonant transport beyond $q=1$ does occur. \\ The resonant nature of the transport is underlined by inserting in these diagrams the linear position of either the precessional or passing resonance, before the fishbone mode frequency begins to chirp up and down. \\ \\
In the strong drive simulation, the transport associated to the precessional resonance is displayed in Figure \ref{totPS_t} (a), (b) and (c), where the radial positions used are respectively $r=0.43r_{q=1}$, $r=r_{q=1}$ and $r=1.43r_{q=1}$. The transport due to the passing resonance is illustrated in Figure \ref{totPS_p} (a)-(b), where the considered radial positions are respectively $r=0.43r_{q=1}$ and $r=0.9r_{q=1}$. \\ In the weak drive simulation, the precessional transport is shown in Figure \ref{totPS_t} (d), (e), (f) with respectively $r=0.4r_{q=1}$, $r=r_{q=1}$, $r=1.17r_{q=1}$ and the passing transport is displayed in Figure \ref{totPS_p} (c), (d) with respectively $r=0.4r_{q=1}$, $r=0.9r_{q=1}$.
 \\ \\
The dynamics observed in both simulations for the alpha transport in the trapped and passing phase space diagrams is very similar, even though it appears that more particles are being transported in the strong drive limit, as illustrated in Figure \ref{flat}. A net outward transport is observed in phase space, with a loss of alpha particles inside the $q=1$ surface, and a gain at the $q=1$ surface and further away from it, for both trapped and passing particles. It is observed that the transport of passing particles is weaker than for trapped particles. About 10\% of passing particles are transported in resonant zones of phase space in the strong drive limit, and 5 \% in the weak drive limit. Comparatively, the trapped particle transport is around 40\% in the strong drive regime, and 20\% in the weak drive regime.\\ \\Resonant transport of particles away from the $q=1$ surface can be explained by two factors. First, as observed in Figure \ref{modeSD} and \ref{modeWD}, during the nonlinear fishbone phase, the modes structure expands further away from the $q=1$ surface, enabling particles out of the $q=1$ volume to interact with the mode. Second, even if the reference flux surface $\bar{\psi}$ of energetic particles lies outside the mode structure, they can still resonate with the mode due to their large orbit width \cite{Graves2004}, especially in the case of trapped particles. \\ \\ The structure of the perturbed zones of phase space is similar to those of the precessional and passing linear resonances at the considered radii, revealing the resonant nature of the transport. However, their phase space positions differ.
It can be noticed that the linear resonance does not systematically lies at the center of the perturbed structure (Figure \ref{totPS_t} (a),(b),(e),(f); Figure \ref{totPS_p} (b), (d)) or that the width of the resonant structure is relatively wide around the linear resonance (Figure \ref{totPS_t} (c),(d)). By looking at the total transport in phase space, the nonlinear dynamics between resonant particles and the fishbone modes remains necessarily unclear when looking at the linear resonance positions, since in both simulations, the mode frequency chirps up and/or down notably, due to the nonlinear evolution of both the 1,1 and 0,0 rotations. \\ \\ 
Therefore, in order to investigate this nonlinear dynamics, the phase space resonant structures are examined dynamically, by computing the instantaneous perturbed distribution function $\delta_{t_1}^{t_2}F_{\alpha}$ on specific time windows $t\in[t_1,t_2]$, together with the wave-particle energy exchange $\int_{t_1}^{t_2}\textbf{J}_{\alpha}\cdot\textbf{E} \ dt$ (annex A). The equations describing the nonlinear evolution of resonant particles are presented in annex B. In this annex, Eq. \ref{NL_ev} provides a connection between the time evolutions of the resonant particles kinetic energy $\dot{E}$ and toroidal momentum $\dot{P}_{\varphi}$, such as $\dot{E} = -\dot{P}_{\varphi}/\omega$ when the time evolution of the phase space island width $\partial_t\tilde{h}$ can be neglected. Macroscopically, it implies that in the phase space diagrams, the
quantities $\delta_{t_1}^{t_2}F_{\alpha}$ and $\int_{t_1}^{t_2}\textbf{J}_{\alpha}\cdot\textbf{E} \ dt$ have opposite signs (Eq. (\ref{macro}). This macroscopic link is derived in annex C. This result will be useful when analyzing phase space resonant structures in the next section.
\\ \\ The phase space dynamics of trapped particles is examined in the strong and weak drive regimes. Passing particles phase space dynamics is only studied in the strong drive regime, since the overall transport of passing particles in the other regime is relatively low, which prevents from obtaining a good resolution for $\delta_{t_1}^{t_2} F_{\alpha}$ and $\int_{t_1}^{t_2}\textbf{J}_{\alpha}\cdot\textbf{E} \ dt$ in the $(E,\lambda)$ diagrams. The time windows used to explore the phase space dynamics are displayed in Figure \ref{window}, together with the time evolution of the main co-existing fishbone mode frequencies from Figure \ref{omega}. The time windows used to study the precessional resonance are displayed respectively in Figure \ref{window} (a) and (b) for the strong and the weak drive, and the ones for the passing resonance in Figure \ref{window} (c).
\subsection{Phase space dynamics of resonant trapped particles}
The phase space dynamics of the precessional resonance is studied in a radial layer centered at $r=r_{q=1}$, since according to Figure \ref{totPS_t} (b),(e), this zone of phase space features incoming transport of particles coming from inner radial layers, and outgoing transport of particles leaving the $r_{q=1}$ layer. For the same reason, the layer $r=0.7r_{q=1}$ is used for the passing phase space diagrams. The instantaneous particle transport $\delta_{t_1}^{t_2}{F}_{\alpha} = [F_{\alpha}(t_2)-F_{\alpha}(t_1)]/F_{\alpha}(0)$ and the wave-particle energy exchange $\int_{t_1}^{t_2}\textbf{J}_{\alpha}\cdot\textbf{E}  \ dt$ are computed in the $(E,\lambda)$ diagrams, in the time windows displayed in Figure \ref{window}, to examine the time evolution of the phase space resonant structures. On all phase space diagrams, the positions of the instantaneous resonances are plotted, taking into account the mean value of the available fishbone frequencies inside the time windows displayed in Figure \ref{window}. These phase space positions are computed using the $n=m=0$ components of the electromagnetic field. By performing this operation, it is assumed that the amplitude of the perturbed electromagnetic field in the nonlinear phases is negligible regarding the equilibrium field. In both simulations, this assumption is well respected since in the strong drive limit, $|\delta B_{\perp}|/B_0 \sim 2\times 10^{-3}$ and in the weak drive limit, $|\delta B_{\perp}|/B_0 = 1\times 10^{-4}$. 
\subsubsection{Strong drive regime}
\begin{figure}[h!]
\begin{subfigure}{.49\textwidth} 
   \centering
   \includegraphics[scale=0.22]{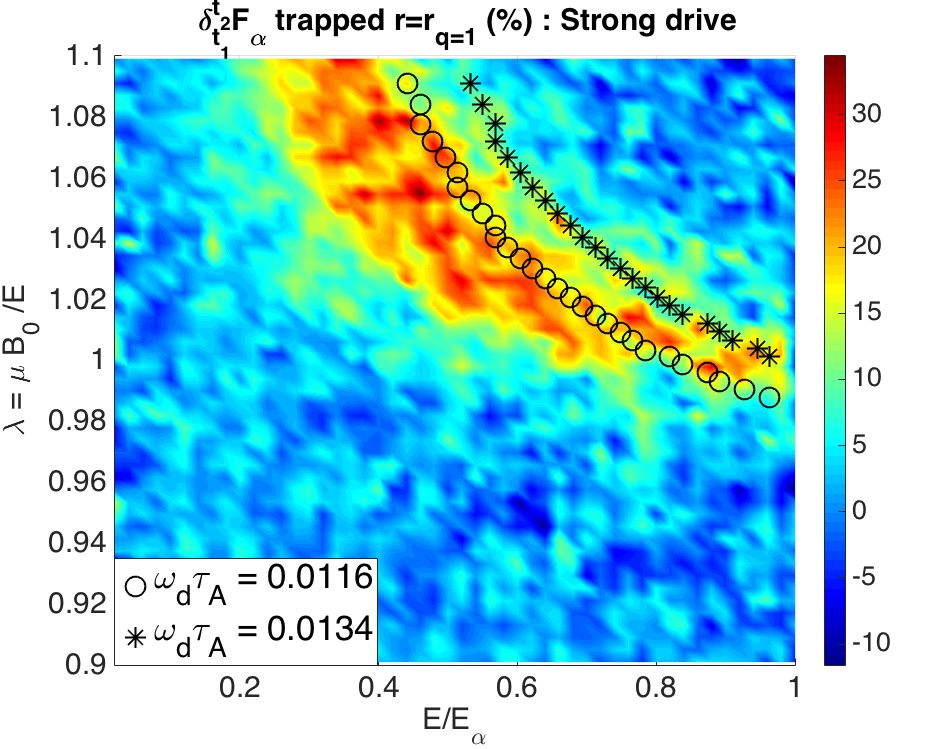}
   \caption{$t\in [1900,2700] \tau_A$}
\end{subfigure}
\begin{subfigure}{.49\textwidth} 
   \centering
   \includegraphics[scale=0.22]{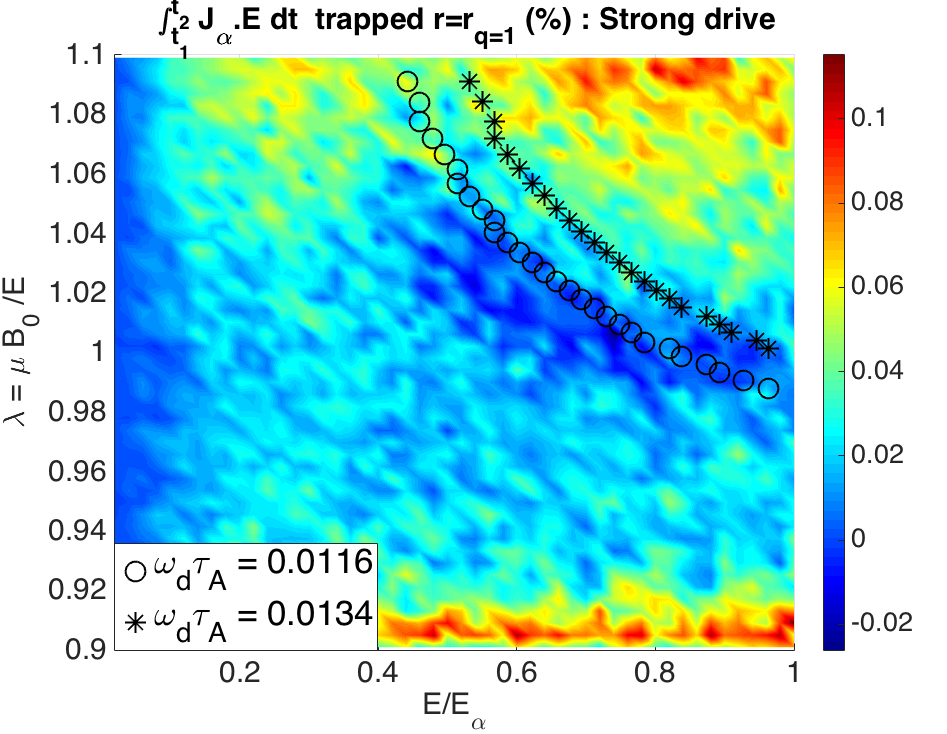}
   \caption{$t\in [1900,2700] \tau_A$}
\end{subfigure}   
\begin{subfigure}{.49\textwidth} 
   \centering
   \includegraphics[scale=0.22]{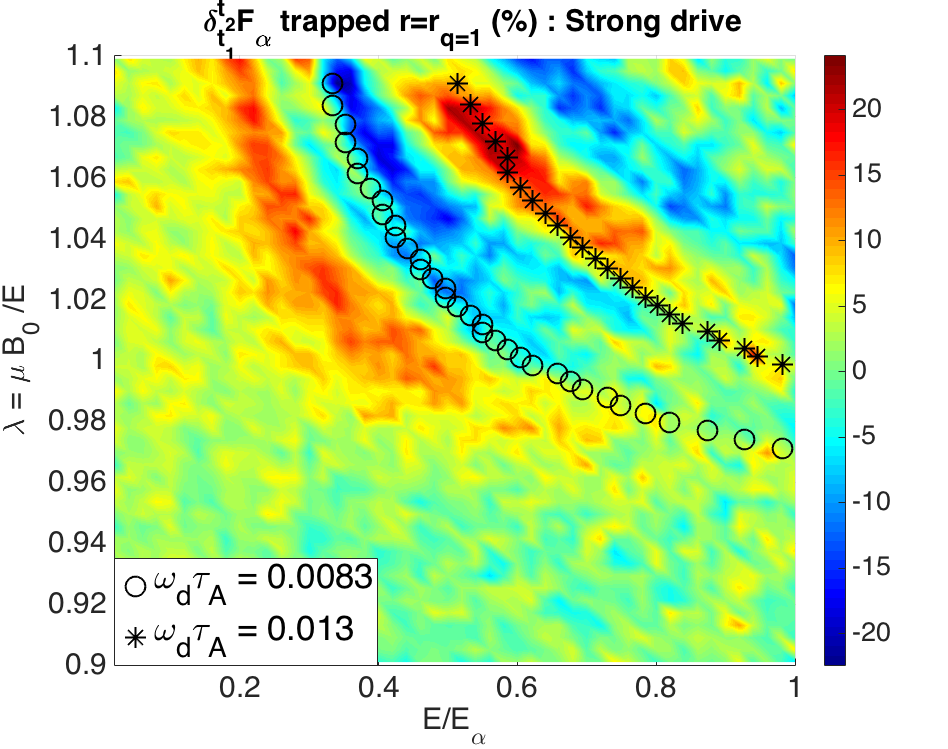}
   \caption{$t\in [3500,3900] \tau_A$}
\end{subfigure}
\begin{subfigure}{.49\textwidth} 
   \centering
   \includegraphics[scale=0.22]{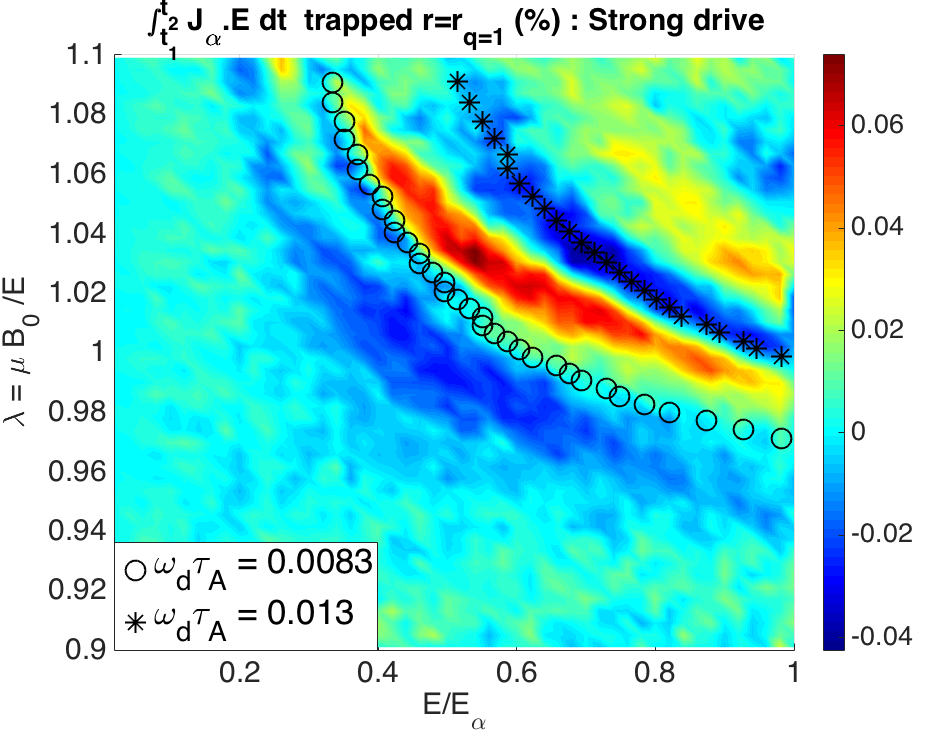}
   \caption{$t\in [3500,3900] \tau_A$}
\end{subfigure}     
\begin{subfigure}{.49\textwidth} 
   \centering
   \includegraphics[scale=0.22]{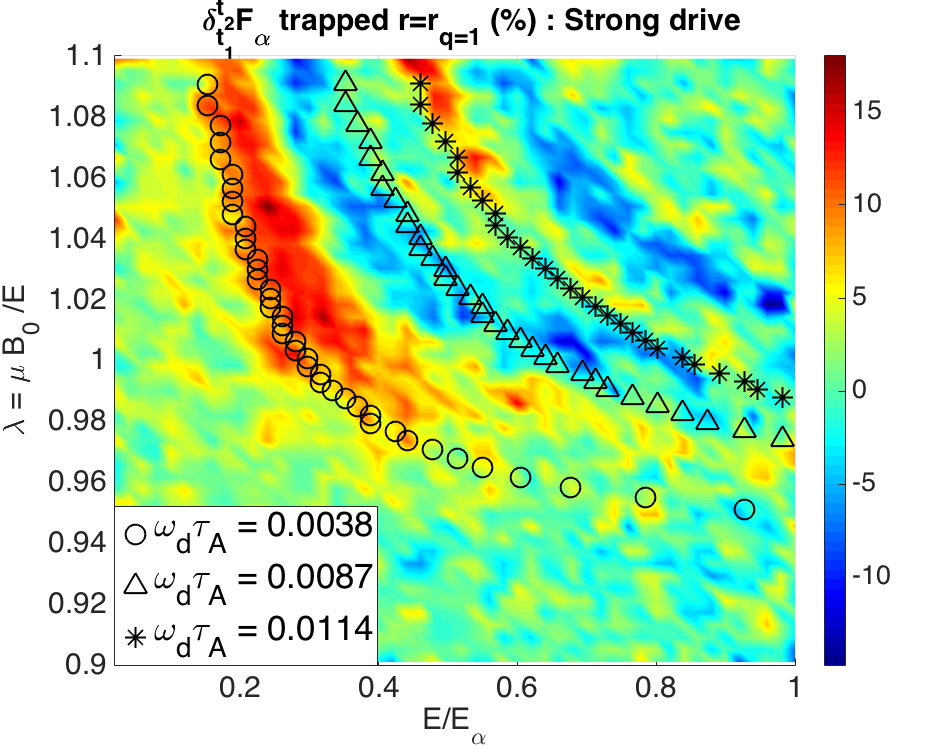}
   \caption{$t\in [5000,5900] \tau_A$}
\end{subfigure}
\begin{subfigure}{.49\textwidth} 
   \centering
   \includegraphics[scale=0.22]{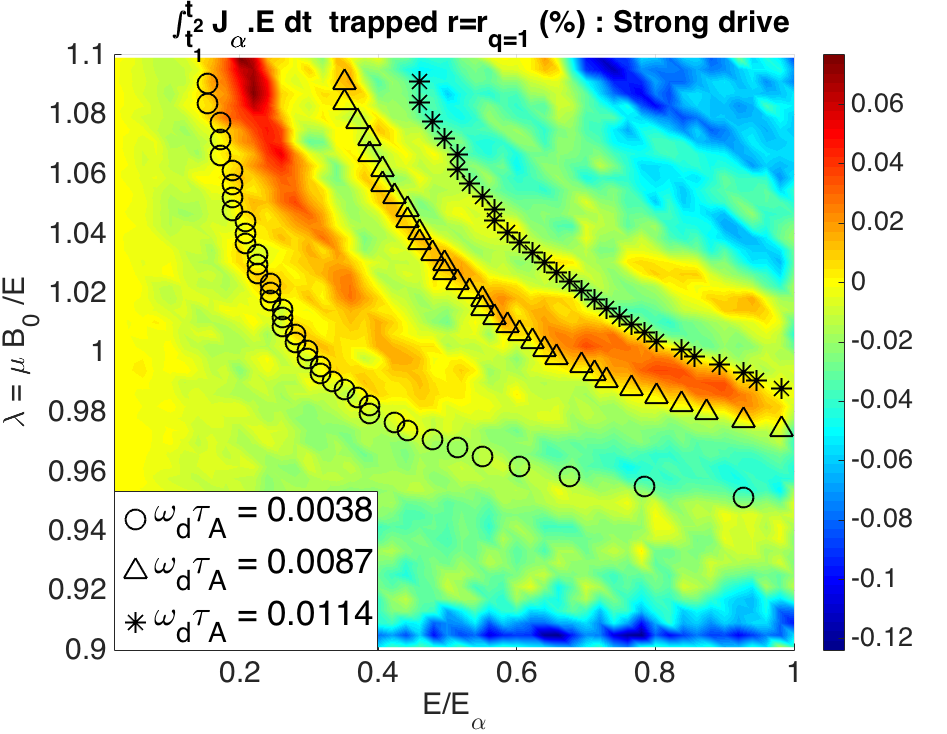}
   \caption{$t\in [5000,5900] \tau_A$}
\end{subfigure}     
\caption{Time evolution of the instantaneous perturbed alpha distribution function $\delta_{t_1}^{t_2} F_{\alpha}$ (left figures) and energy exchange $\int_{t_1}^{t_2}\textbf{J}_{\alpha}\cdot\textbf{E} \ dt$ (right figures) on the trapped $(E,\lambda)$ diagram at $r=r_{q=1}$, in the strong drive regime. The time frames $t\in[t_1,t_2]$ used here correspond to : (a,b) Figure \ref{window} (a) solid rectangle. (c,d) Figure  \ref{window} (a) dashed rectangle. (e,f) Figure  \ref{window} (a) dotted rectangle. The black circles and black stars correspond respectively to the solution of $\omega-\omega_d(r,E,\lambda)$ on the trapped $(E,\lambda)$ diagram for the down and up chirping branches. Black triangles on subfigures (e,f) correspond to the resonance position when considering the second up-chirping branch. } 
\label{trappedSD}
\end{figure}

The phase dynamics of the precessional resonance in the strong drive regime is illustrated in Figure \ref{trappedSD}. The early nonlinear fishbone phase is represented in Figure \ref{trappedSD} (a-b), at the beginning of the mode chirping and particle transport. An incoming transport of alphas can be seen in Figure \ref{trappedSD} (a), at slightly lower energy and pitch angle values than the resonance positions of the up and down chirping branches. This transport corresponds to a flux of particles coming from inner radial positions just below the $q=1$ surface. The phase space localization of this transport is coherent with the radial dependency of the precessional resonance, illustrated in Figure \ref{linres} (a-b). Together with this incoming transport, a similar structure can be observed at the same location in phase space in Figure \ref{trappedSD} (b), where the wave-particle energy exchange $\int_{t_1}^{t_2} \textbf{J}_{\alpha}\cdot\textbf{E} \ dt$ is negative. It implies that resonant particles are transported to larger radial positions while giving energy to the fishbone mode. For trapped particles, the signs of $\dot{P}_{\varphi}$ and $\dot{\bar{r}}$ are the same since $P_{\varphi} \equiv mRv_{\varphi} - Ze\psi = -Ze\bar{\psi}$, and that $\bar{\psi} \propto -\bar{r}^2$ is the convention taken in XTOR-K. Therefore, the observed transport is consistent with the nonlinear evolution described in Eq. \ref{NL_ev} annex B. The time evolution of the precessional island's width in phase space is then negligible at that time. 
\\ \\ Later in the nonlinear phase in Figure \ref{trappedSD} (c-d), when the separation between the up and down chirping branches becomes significant, two pairs of hole and clump structures appear in phase space. These structures are centered around the instantaneous resonant positions of both branches, and the respective signs of $\delta_{t_1}^{t_2} F_{\alpha}$ and $\int_{t_1}^{t_2} \textbf{J}_{\alpha}\cdot\textbf{E} \ dt$ in phase space are still in agreement with Eq. (\ref{NL_ev}) annex B. The emergence of such structures was initially proposed by Berk and Breizman \cite{Berk1999}, and later observed in nonlinear hybrid simulations \cite{Shen2017}. It is however noted that hole and clump structures in \cite{Berk1999} are a feature of Energetic Particle Modes near marginal stability, which is not the case here since the strong drive limit is investigated. In the present study, the formation of holes and clumps is explained by the structure of the precessional resonance in phase space (Figure \ref{linres} (a-b)). Around each resonance, a zone of incoming transport (particles arriving in the $q=1$ radial layer) can be observed at lower energy and pitch values, and a outgoing one (particles leaving the $q=1$ radial layer) at higher energies and pitch angles. Such a process is consistent with the radial dependency of the precessional resonance depicted in Figure \ref{linres} (a-b). The precessional resonance tends to move at higher energy and pitch angles values when the radial position is increased, explaining the incoming and outgoing particle fluxes on Figure \ref{trappedSD} (c-d). \\ \\
It is also noted that in contrast to the Berk-Breizman hole and clump theory, and recent numerical results \cite{Shen2017}, the hole and clump do not separate to follow respectively the down and up chirping branches. Instead, as it can be observed by comparing the resonant structures position in phase space at different times in Figure \ref{trappedSD}, each pair of hole and clump are following their respective resonance position as the different mode frequencies evolve. \\ \\
In the late nonlinear fishbone phase, as shown in Figure \ref{omega} (a), the down chirping branch is dissociated into two new branches, that chirp up and down as well around $t\sim 5000\tau_A$. This leads to the emergence of a third precessional resonance, as illustrated in Figure \ref{trappedSD} (e-f). Three pairs of hole and clump structures can be observed on the trapped phase space diagrams, each one corresponding to a resonance position. However, in that phase, the sign of the $\delta_{t_1}^{t_2} F_{\alpha}$ and $\int_{t_1}^{t_2} \textbf{J}_{\alpha}\cdot\textbf{E} \ dt$ structures do not strictly follow Eq. (\ref{NL_ev}) anymore. It indicates that in the late nonlinear fishbone phase, the island width of the precessional resonance in phase space evolves significantly.
\subsubsection{Weak drive regime}
\begin{figure}[h!]
\begin{subfigure}{.49\textwidth} 
   \centering
   \includegraphics[scale=0.22]{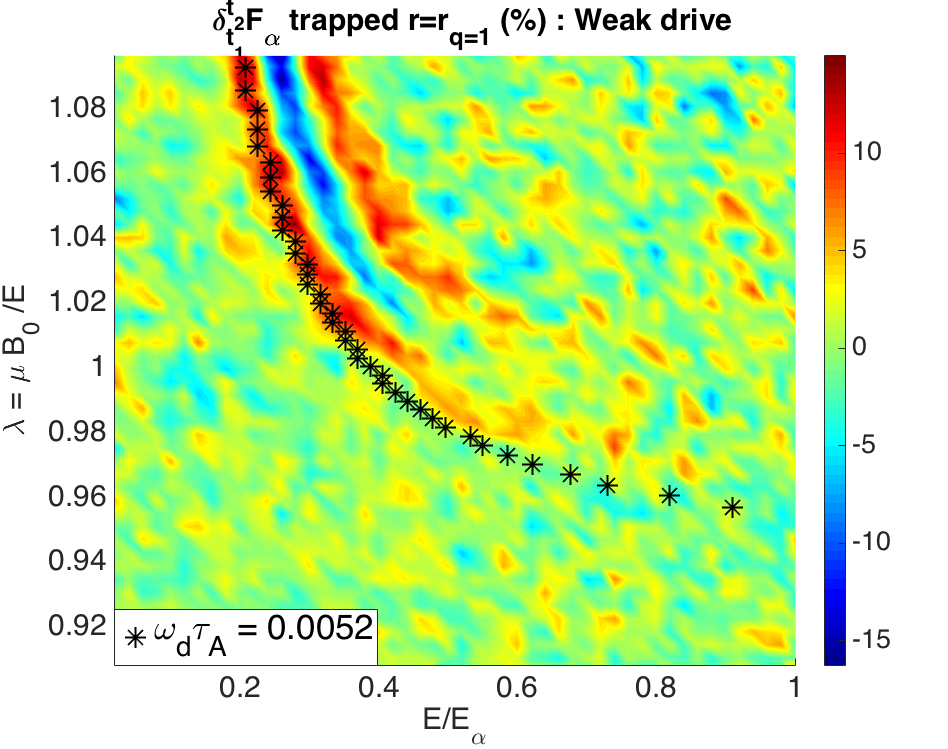}
   \caption{$t\in [11000,13000] \tau_A$}
\end{subfigure}
\begin{subfigure}{.49\textwidth} 
   \centering
   \includegraphics[scale=0.22]{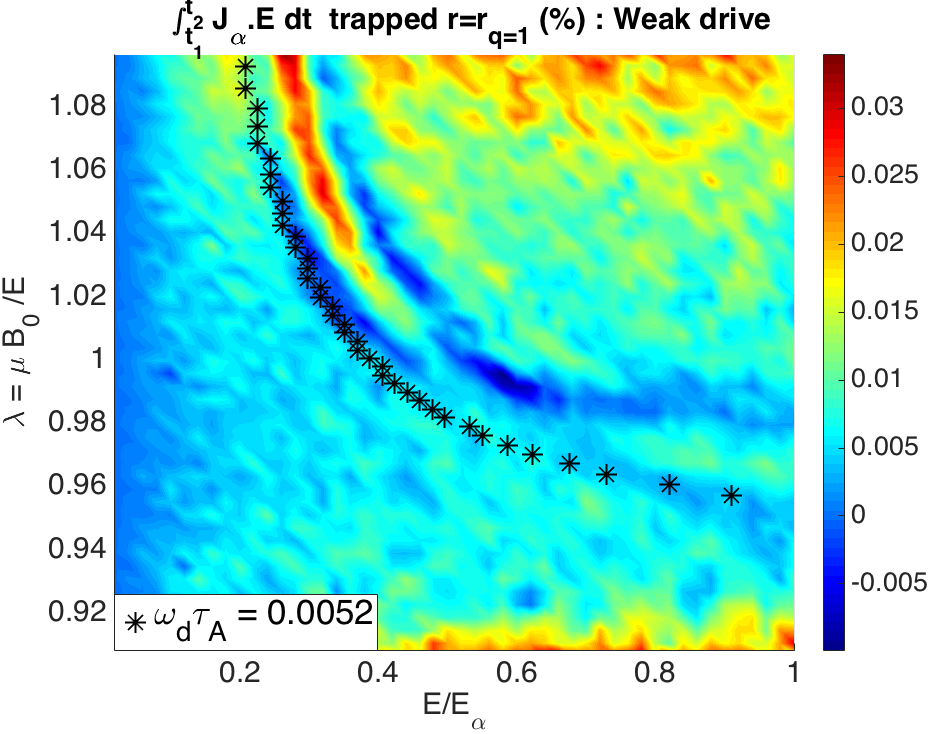}
   \caption{$t\in [11000,13000] \tau_A$}
\end{subfigure}   
\begin{subfigure}{.49\textwidth} 
   \centering
   \includegraphics[scale=0.22]{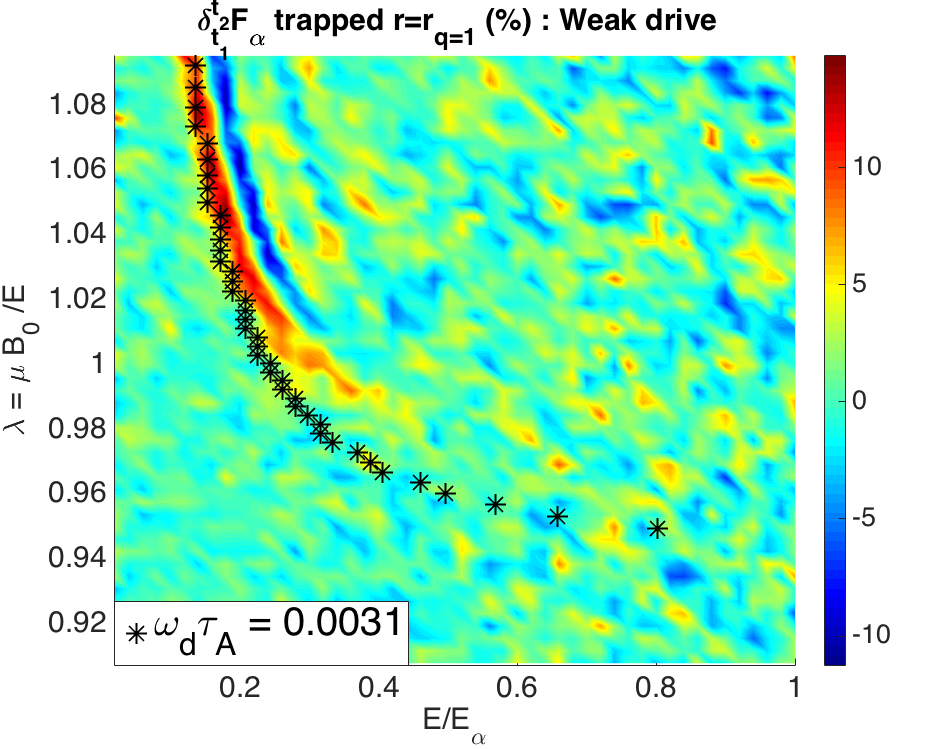}
   \caption{$t\in [19000,21500] \tau_A$}
\end{subfigure}
\begin{subfigure}{.49\textwidth} 
   \centering
   \includegraphics[scale=0.22]{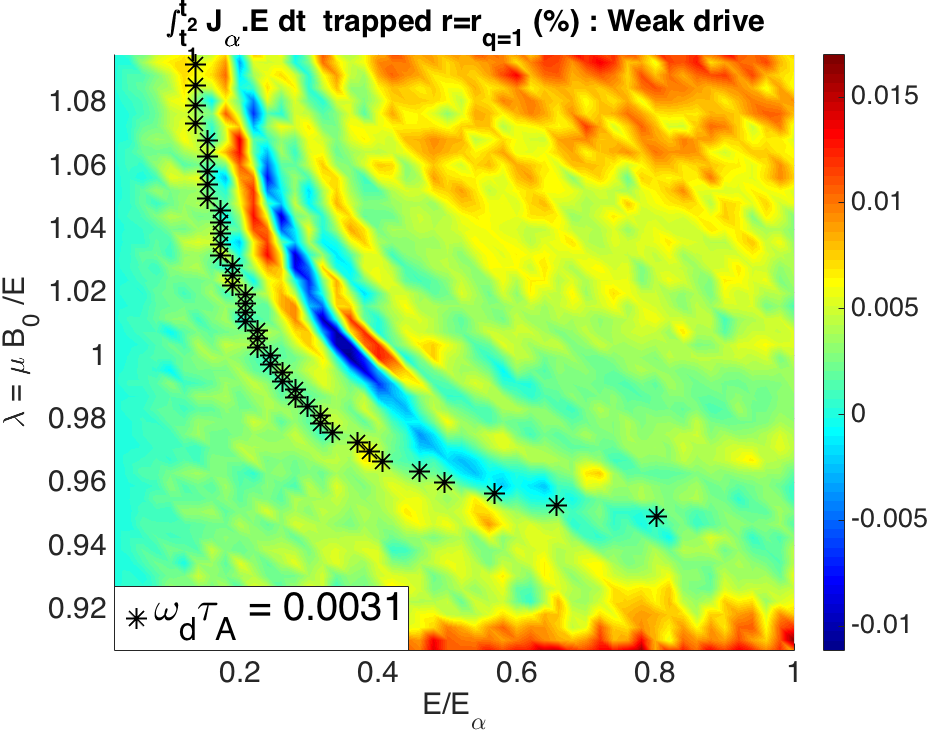}
   \caption{$t\in [19000,21500] \tau_A$}
\end{subfigure}     
\caption{Time evolution of the instantaneous perturbed alpha distribution function $\delta_{t_1}^{t_2} F_{\alpha}$ (left figures) and energy exchange $\int_{t_1}^{t_2}\textbf{J}_{\alpha}\cdot\textbf{E} \ dt$ (right figures) on the trapped $(E,\lambda)$ diagram at $r=r_{q=1}$, in the weak drive regime. The time frames $t\in[t_1,t_2]$ used here correspond to (a,b) Figure  \ref{window} (c) solid rectangle. (c,d) Figure  \ref{window} (c) dashed rectangle. (e,f) Figure  \ref{window} (c) dotted rectangle. The black stars correspond to the solution of $\omega-\omega_d(r,E,\lambda)$ on the trapped $(E,\lambda)$ diagram for the down chirping branch} 
\label{trappedWD}
\end{figure}

The phase space dynamics of the precessional resonance in the weak drive regime is fairly similar to the strong drive case. At the beginning of the nonlinear fishbone phase, formation of coherent structures in phase space have not been observed, due to the relatively slow dynamics of the resonant transport in this regime. However, at the middle of the fishbone phase (Figure \ref{trappedWD} (a-b)), a hole and clump structure respecting Eq. (\ref{NL_ev}) appears, centered around the resonance position. A weak additional hole and clump structure can be observed at just higher energies and pitch angles. It corresponds to the additional down-chirping branch illustrated in Figure \ref{omega} (b), which has a slightly higher amplitude than the main mode frequency. \\ \\

As the mode frequency chirps down in the late fishbone phase, the hole and clump follows the precessional resonance (Figure \ref{trappedWD} (a-b)), while still respecting equation \ref{NL_ev}. The absence of a dominant up-chirping branch in Figure \ref{omega} (b) is confirmed by the fact that no corresponding hole and clump structures can be noticed in Figure \ref{trappedWD}. The following differences between the strong and weak drive regime can be noted : the width of the resonant structures is smaller in the weak drive case, as well as the proportion of particles transported around the resonance positions. The width of the phase space island does not evolve significantly in the weak drive regime contrarily to the other one. Such differences are consistent with those observed in the previous section, namely a slower dynamics with weaker amplitude when looking at the mode kinetic energy (Figure \ref{KE}) and amplitude of the modes structure in Figure \ref{modeSD}-\ref{modeWD}.
\subsection{Phase space dynamics of resonant passing particles}
\begin{figure}[h!]
\begin{subfigure}{.49\textwidth} 
   \centering
   \includegraphics[scale=0.3]{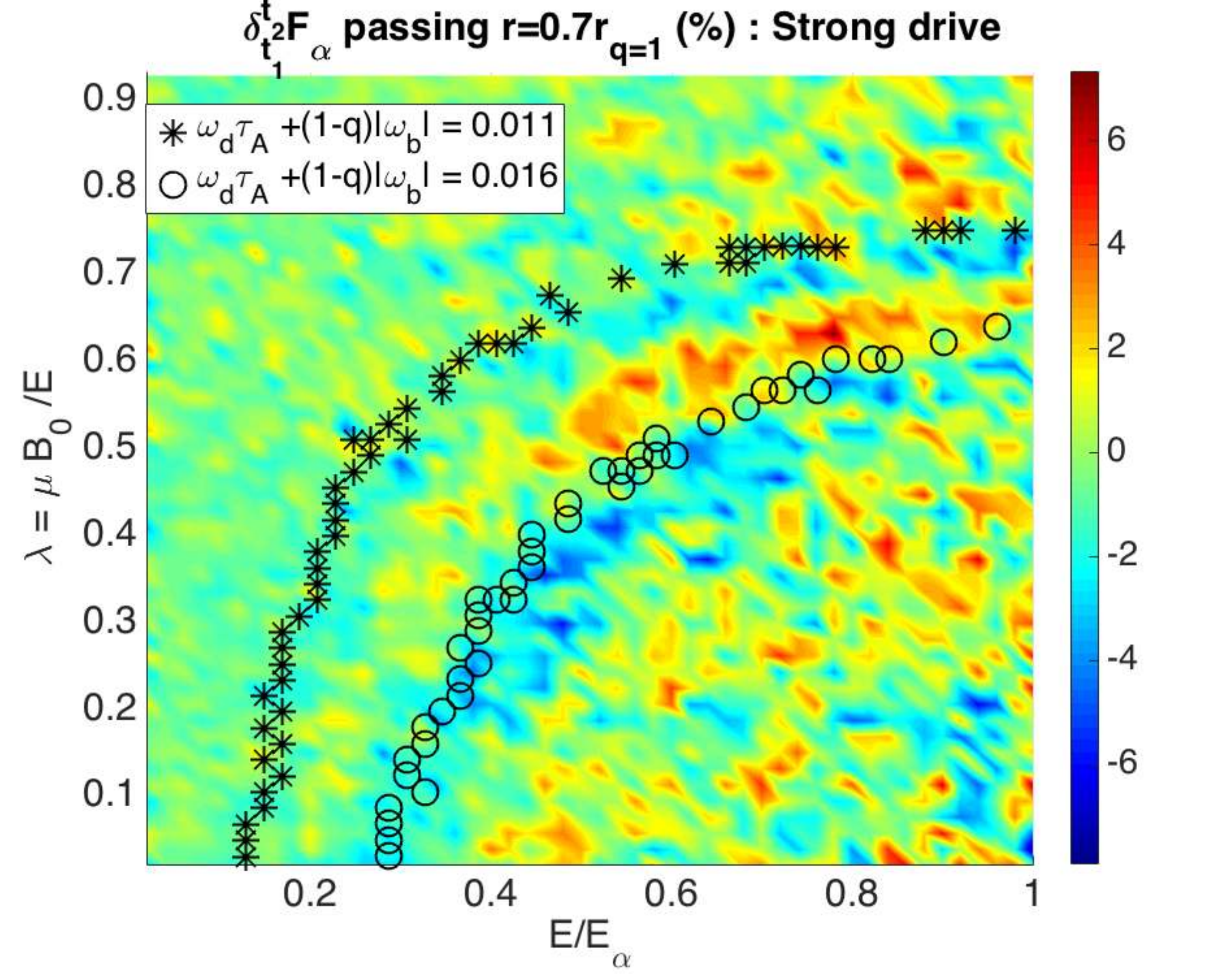}
   \caption{$t\in [3500,3900] \tau_A$}
\end{subfigure}
\begin{subfigure}{.49\textwidth} 
   \centering
   \includegraphics[scale=0.3]{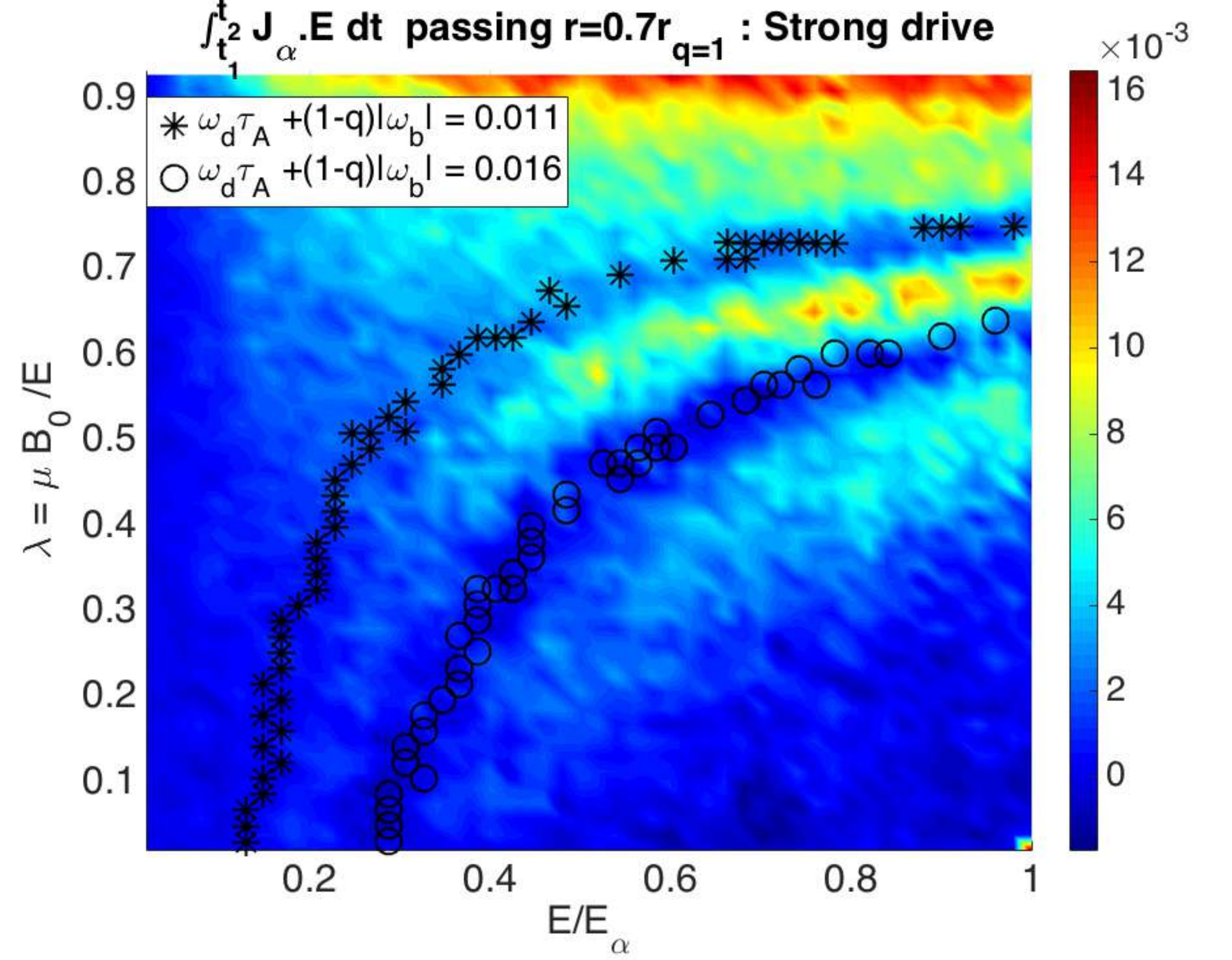}
   \caption{$t\in [3500,3900] \tau_A$}
\end{subfigure}   
\begin{subfigure}{.49\textwidth} 
   \centering
   \includegraphics[scale=0.3]{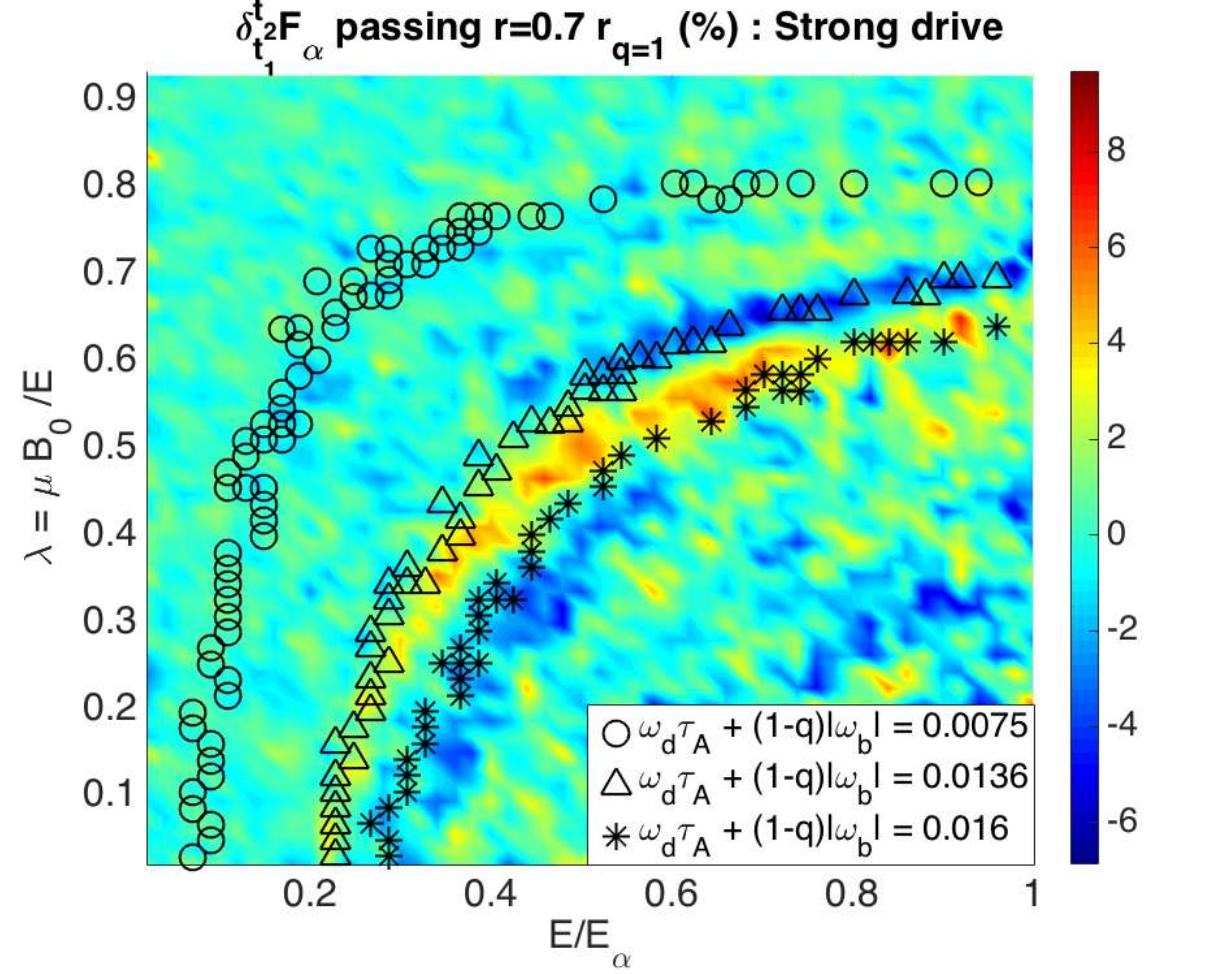}
   \caption{$t\in [5700,6300] \tau_A$}
\end{subfigure}
\begin{subfigure}{.49\textwidth} 
   \centering
   \includegraphics[scale=0.3]{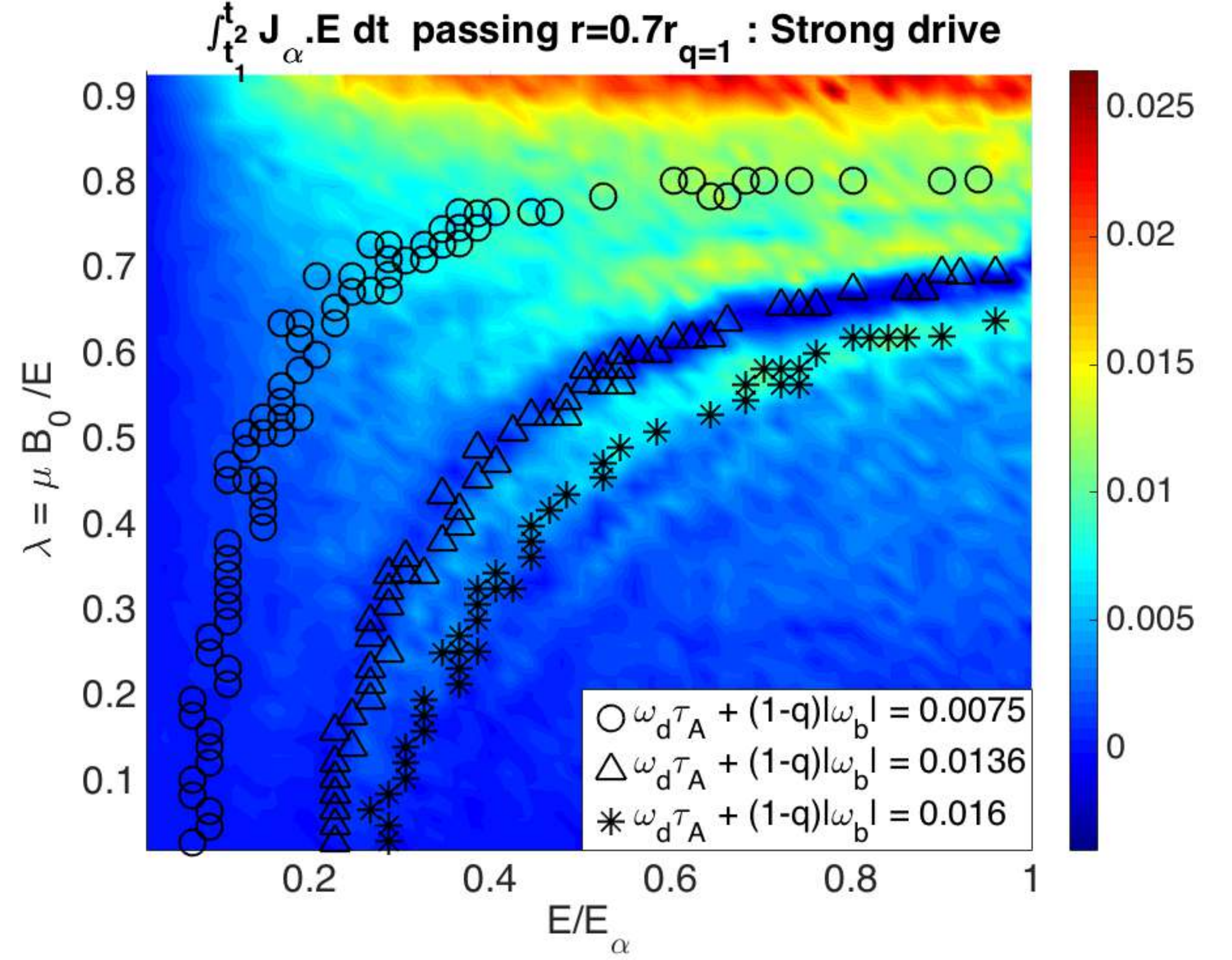}
   \caption{$t\in [5700,6300] \tau_A$}
\end{subfigure}     
\begin{subfigure}{.49\textwidth} 
   \centering
   \includegraphics[scale=0.3]{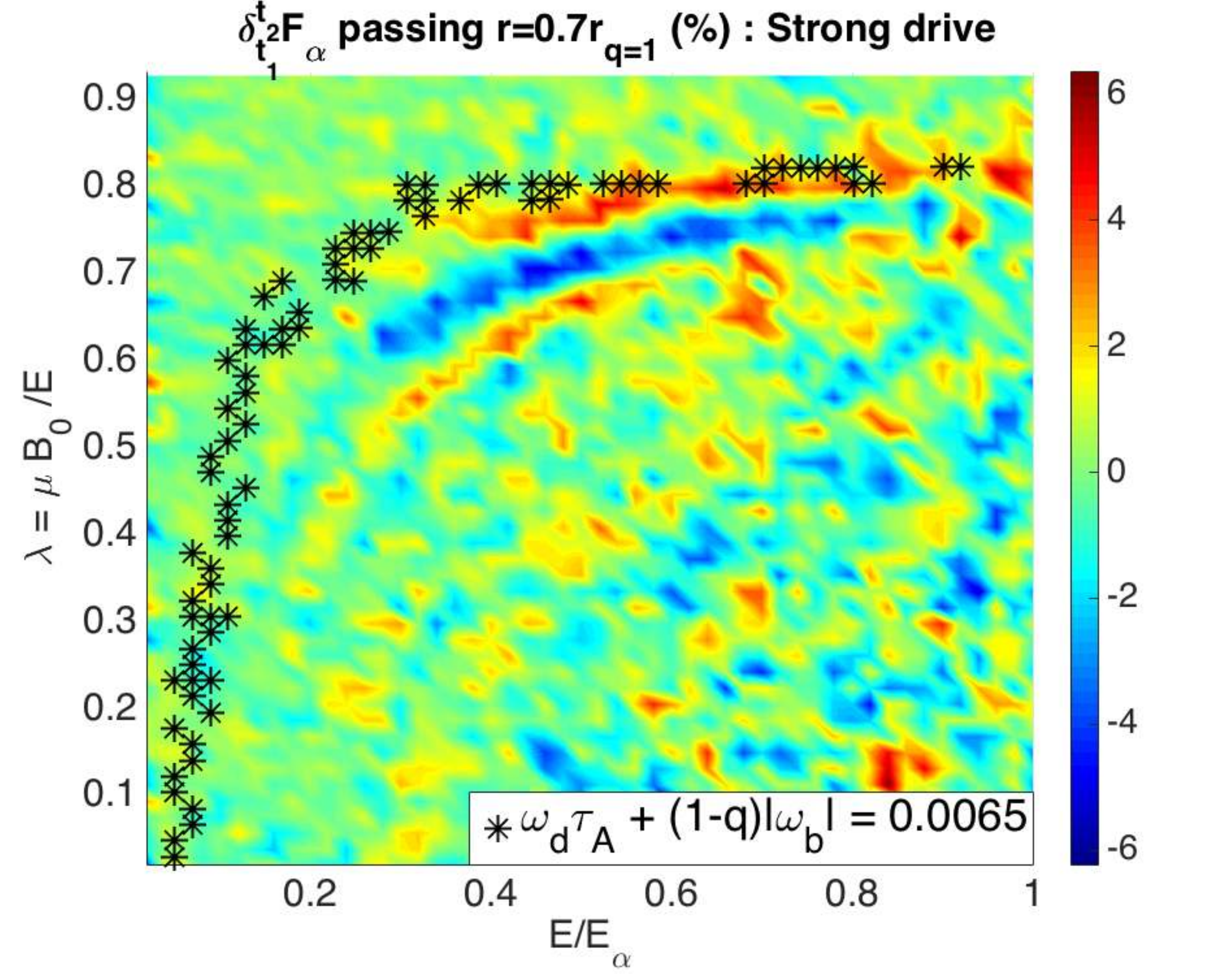}
   \caption{$t\in [8000,9000] \tau_A$}
\end{subfigure}
\begin{subfigure}{.49\textwidth} 
   \centering
   \includegraphics[scale=0.3]{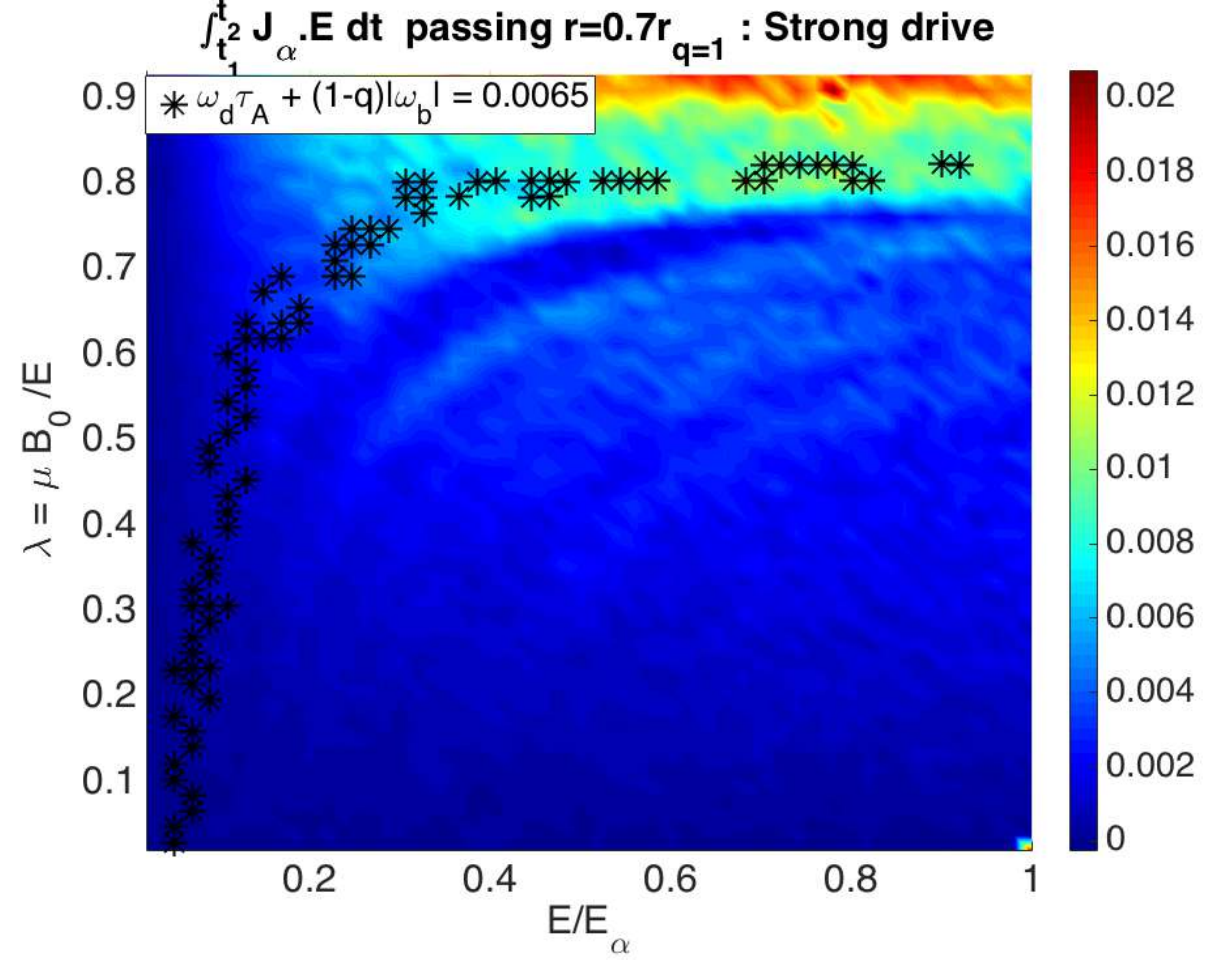}
   \caption{$t\in [8000,9000] \tau_A$}
\end{subfigure}     
\caption{Time evolution of the instantaneous perturbed alpha distribution function $\delta_{t_1}^{t_2} F_{\alpha}$ (left figures) and energy exchange $\int_{t_1}^{t_2}\textbf{J}_{\alpha}\cdot\textbf{E} \ dt$ (right figures) on the passing $(E,\lambda)$ diagram at $0.71r=r_{q=1}$, in the strong drive regime. The time frames $t\in[t_1,t_2]$ used here correspond to (a,b) Figure  \ref{window} (b) solid rectangle. (c,d) Figure  \ref{window} (b) dashed rectangle. (e,f) Figure  \ref{window} (b) dotted rectangle. The black circles and black stars correspond respectively to the solution of $\omega-(1-q)|\omega_b|(r,E,\lambda)-\omega_d(r,E,\lambda)$ on the passing $(E,\lambda)$ diagram  for the down and up chirping branches} 
\label{passingSD}
\end{figure}
The phase space dynamics of the passing resonance in the strong drive regime is illustrated in Figure \ref{passingSD}. At the beginning of the nonlinear fishbone phase, no coherent structures are observed. It is only near $t\sim4000\tau_A$ (Figure \ref{passingSD} (a-b)), when the mode frequencies in plasma frame evolve significantly due to the 0,0 rotation (Figure \ref{omega} (c),(e)), that two pairs of hole and clump appear, here again centered around their respective passing resonance position. The formation of holes and clumps is also here due to the phase space structure of the passing resonance, whose radial dependency is highlighted in Figure \ref{linres} (c-d). \\ \\
As the mode frequencies evolve, these structures stay attached to the passing resonances (Figure \ref{passingSD} (c-d)), until one dominant mode frequency persists at the very end of the fishbone phase, associated to only one pair of hole and clump (Figure \ref{passingSD} (e-f)). The phase space evolution of the passing resonances is consistent with a down-chirping of the mode frequency, as explained in section 2.3.1. \\ \\
However, contrarily to the dynamics of the precessional resonance, the $\delta_{t_1}^{t_2} F_{\alpha}$ and $\int_{t_1}^{t_2} \textbf{J}_{\alpha}\cdot\textbf{E} \ dt$ structures have the same signs in the passing phase space diagrams, thus not respecting equation \ref{NL_ev}. Even if for passing particles, the toroidal momentum $P_{\varphi}$ is not directly proportional to $\bar{\psi}$, locally in the phase space $(E,\lambda)$ diagram (at constant $v_{\parallel}$), the quantity $\dot{P}_{\varphi}$ and $\dot{\bar{r}}$ should have the same sign. Therefore, it implies that during the entire time evolution of the passing resonance, the temporal variation of its island width $\partial_t \tilde{h}$ is non-negligible and should be taken into account in Eq. (\ref{NL_ev}).
\section{Coupling mechanism between mode chirping and particle transport}
Now that the detailed dynamics of the alpha fishbone has been studied in both physical space and phase space, the individual nonlinear behavior of representative resonant particles is studied in this section, in order to understand how they couple with the fishbone mode. For simplicity, only the dynamics of individual trapped particles is analyzed, since the resonant transport is dominated by the trapped fraction of the alpha particles distribution function. From the individual nonlinear behavior observed in both regimes, together with the fluid and kinetic dynamics observed in section 2 and 3, a partial mechanism is proposed, explaining the nonlinear coupling between particle transport and dominant mode down-chirping.
\subsection{Dynamics of an individual resonant trapped particle in the strong drive regime}
When performing a XTOR-K nonlinear simulation, the entire 3D electromagnetic field is stored at each fluid time step. It enables, once the simulation has been achieved, to look at the time evolution of arbitrarily initialized test particles by taking into account the time evolution of the electromagnetic field. This procedure is very useful since it allows to post-process the time evolution of particles in any zones of phase space, especially those which would not have been thought interesting prior to the hybrid simulation. In nonlinear phases, the evolution of the resonance positions is hardly predictable, due to the time evolution of both the mode frequency (0,0 and 1,1) and of the particles invariants of motion (see \cite{Zonca2015} appendix A-2).
\begin{figure}[h!]
\begin{subfigure}{.24\textwidth} 
   \centering
   \includegraphics[scale=0.15]{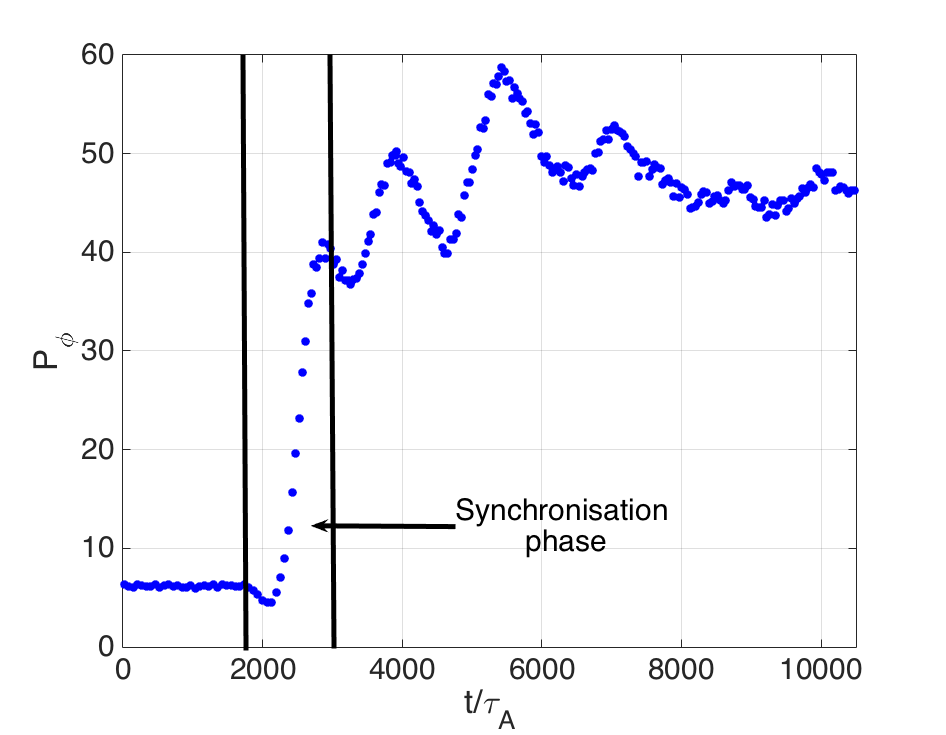}
   \caption{}
\end{subfigure}
\begin{subfigure}{.24\textwidth} 
   \centering
   \includegraphics[scale=0.15]{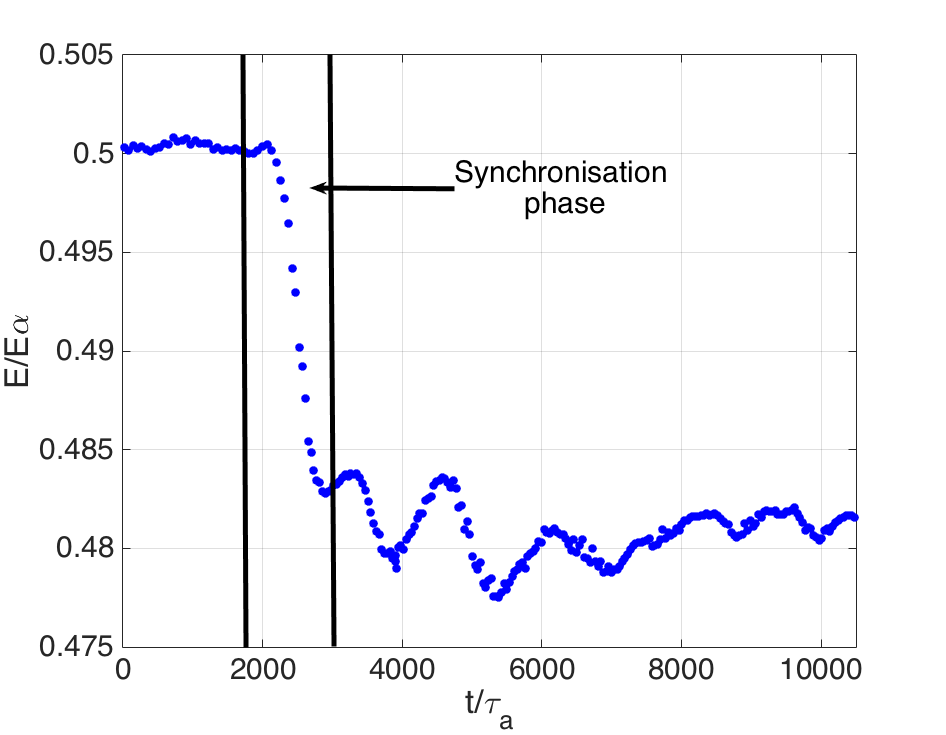}
   \caption{}
\end{subfigure}   
\begin{subfigure}{.24\textwidth} 
   \centering
   \includegraphics[scale=0.15]{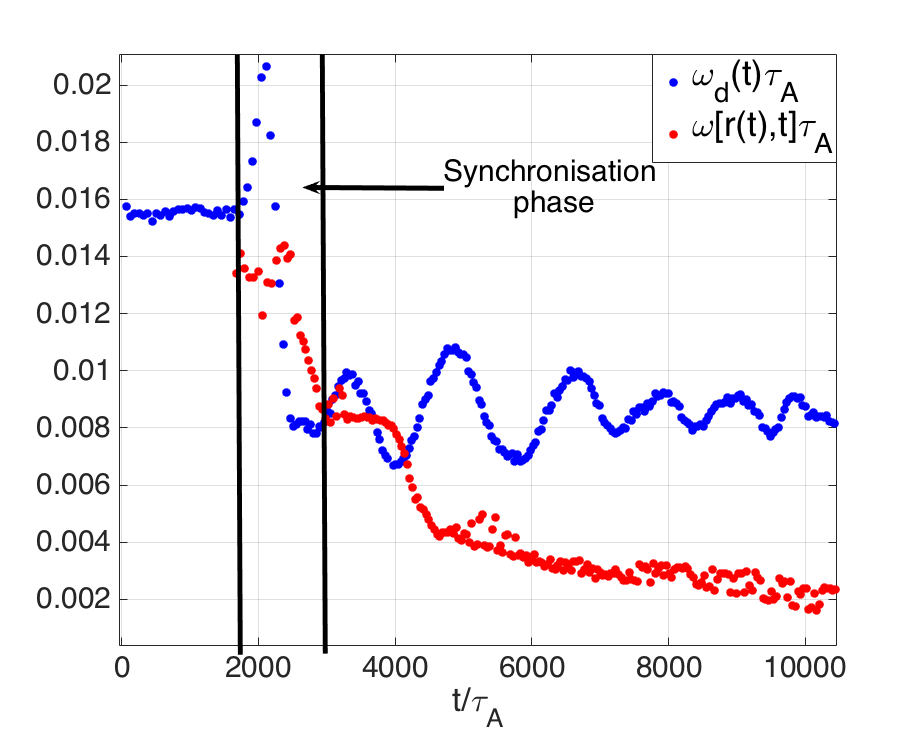}
   \caption{}
\end{subfigure}
\begin{subfigure}{.24\textwidth} 
   \centering
   \includegraphics[scale=0.15]{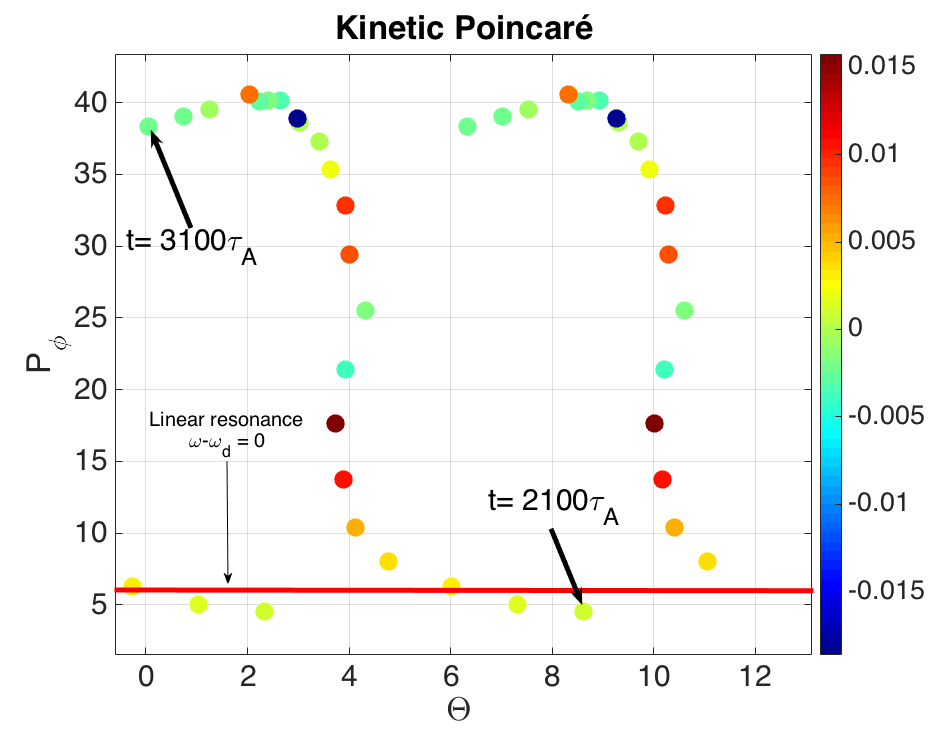}
   \caption{}
\end{subfigure}   
\caption{Time evolution of a typical resonant trapped particle in the strong drive regime. (a) Time evolution of $P_{\varphi}$. (b) Time evolution of the particle kinetic energy (c) Time evolution of the particle precessional frequency and of the local mode frequency in the plasma frame. (d) Kinetic Poincar\'e. The phase $\Phi$ is computed modulo $2\pi$, the Poincar\'e is duplicated to $\Phi \in [2\pi,4\pi]$ to enhance clarity. Colors correspond to the particle instantaneous power exchange $w_i = q_{\alpha}\textbf{v}_{\alpha,i}\cdot\textbf{E}$ with the mode.}
\label{indSD}
\end{figure}
 \\ \\
Using this stored set of fields, in the strong drive regime, an alpha particle is initialized with $E=0.5E_{\alpha},\lambda=1,\bar{r}=0.15$, so that its initial precessional frequency $\omega_d\tau_A = 1.6\times10^{-2}$ is just above the local linear mode frequency $\omega(\bar{r})\tau_A = 1.4\times 10^{-2}$. Its time evolution is illustrated in Figure \ref{indSD}. Along the particle trajectory, its invariants of motion $P_{\varphi}$ (Figure \ref{indSD} (a)) and $E$ (Figure \ref{indSD} (b)) start evolving significantly near $t\sim 2200\tau_A$. It is noted that the variations of these quantities are opposed, as described in equation \ref{NL_ev}. The alpha particle yields energy to the mode as it is transported to higher radial positions. This transport occurs during a phase identified as a synchronisation phase between the mode frequency $\omega$ and the precessional frequency $\omega_d$ in Figure \ref{indSD} (c). For $t\in[1900,3000]\tau_A$, phase-locking arises between the resonant particle and the mode. Such a process has also been identified in \cite{Wang2016}\cite{Vlad2013} to be the underlying mechanism for coupled mode chirping and particle transport. \\ In Figure \ref{indSD} (d), a Kinetic-Poincar\'e plot of the individual particle has been performed in the diagram $(\Theta,P_{\varphi})$, during the synchronisation phase $t\in[2100,3100]\tau_A$. Similarly to \cite{Briguglio2014}, the phase $\Theta$ is computed once per bounce period when $\theta = 0$ at the largest major radius position. It enables computing the phase as $\Theta = \varphi - \int_0^tdt'\omega(t')$, since for trapped particles with $\theta = 0$, $\alpha_3 = \varphi$ in the angle-action formalism \cite{Brochard_PhD}\cite{Brochard2018}. The color bar used in Figure \ref{indSD} (d) corresponds to the instantaneaous wave-particle power exchange $w_i(\textbf{r}_{i,\alpha},\textbf{v}_{i,\alpha}) = q_{\alpha} \textbf{v}_{i,\alpha}\cdot\textbf{E}(\textbf{r}_{i,\alpha})$. In the $(\Theta,P_{\varphi})$ diagram, it is noted that the particle is transported at relatively constant $\Theta$ while giving energy to the mode, which implies that the particle synchronises with the mode by minimizing resonance detuning since $\dot{\Theta} \approx 0$ in the synchronization phase, as observed in \cite{Vlad2013}. The trajectory of the resonant particle on the $(\Theta,P_{\varphi})$ diagram is replicated on the range $2\pi<\Theta<4\pi$ to enhance the readability of the Kinetic Poincar\'e plot. \\ Such a wave-particle synchronisation is consistent with the parametric dependencies of the crossed field and precessional frequencies. Since $\omega_d\propto 1/r$ and $\omega_E\propto 1/r$, according to Eq. (\ref{rescond}), kinetic particles can only stay in resonance when the mode chirps dominantly $(\dot{\omega} < 0)$ down by being transported radially outward $(\dot{r} > 0)$.
\\ \\
It can also be noted in Figure \ref{indSD} that the resonant particle remains shortly synchronized with the mode. A different behavior is observed in Ref. \cite{Wang2016}, Figure 13, where some resonant particles with precessional frequency just above the linear mode frequency remain trapped in the mode until the end of the simulation. Such a difference was expected since the strong drive regime is considered here, and the fishbone instability considered in \cite{Wang2016} is in the weak drive regime. The nonlinear transport time $\tau_{NL}$ is similar to the bounce time in the phase space island $\tau_B$ in the weak drive limit, as presented in section 2.1. The resonant particle therefore exits the phase space island before performing a single bounce, as depicted in Figure \ref{indSD} (d).
\subsection{Dynamics of an individual resonant trapped particle in the weak drive regime}
\begin{figure}[h!]
\begin{subfigure}{.24\textwidth} 
   \centering
   \includegraphics[scale=0.15]{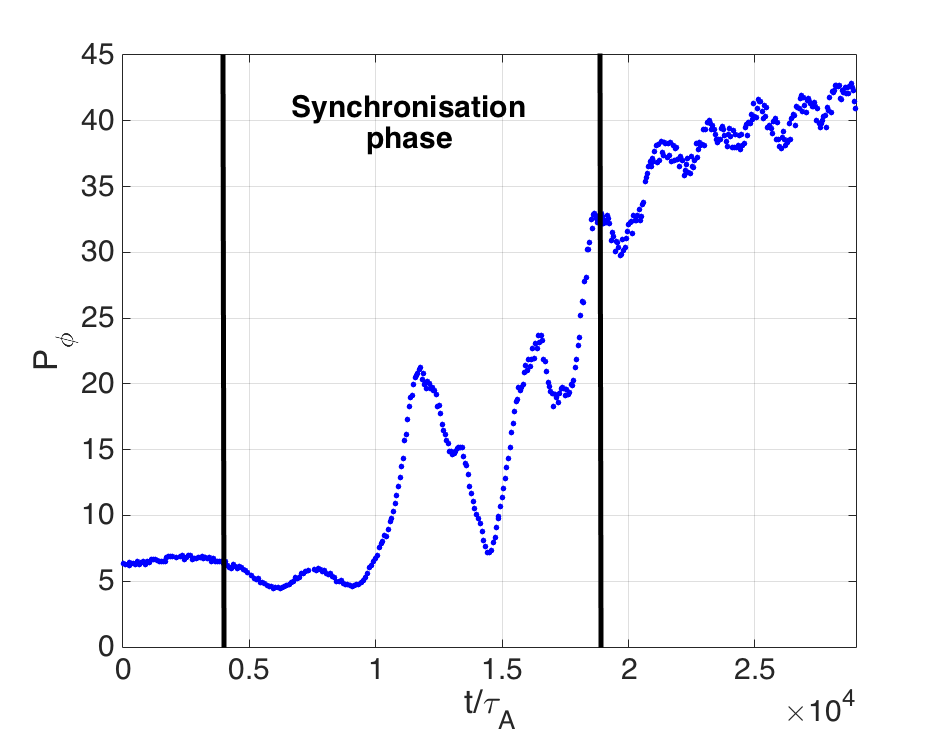}
   \caption{}
\end{subfigure}
\begin{subfigure}{.24\textwidth} 
   \centering
   \includegraphics[scale=0.15]{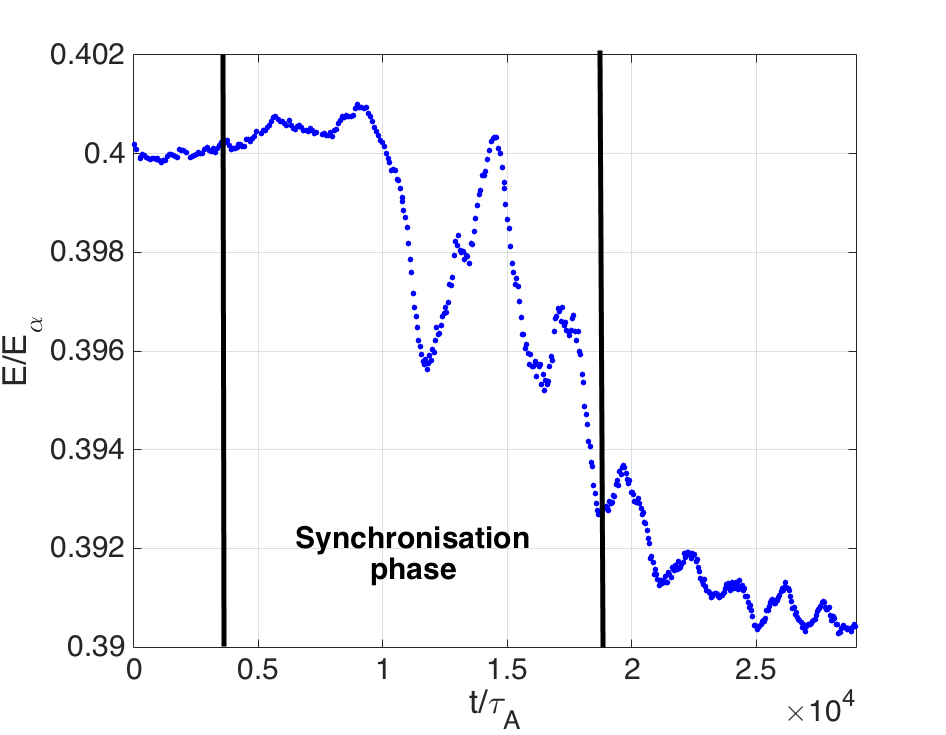}
   \caption{}
\end{subfigure}   
\begin{subfigure}{.24\textwidth} 
   \centering
   \includegraphics[scale=0.15]{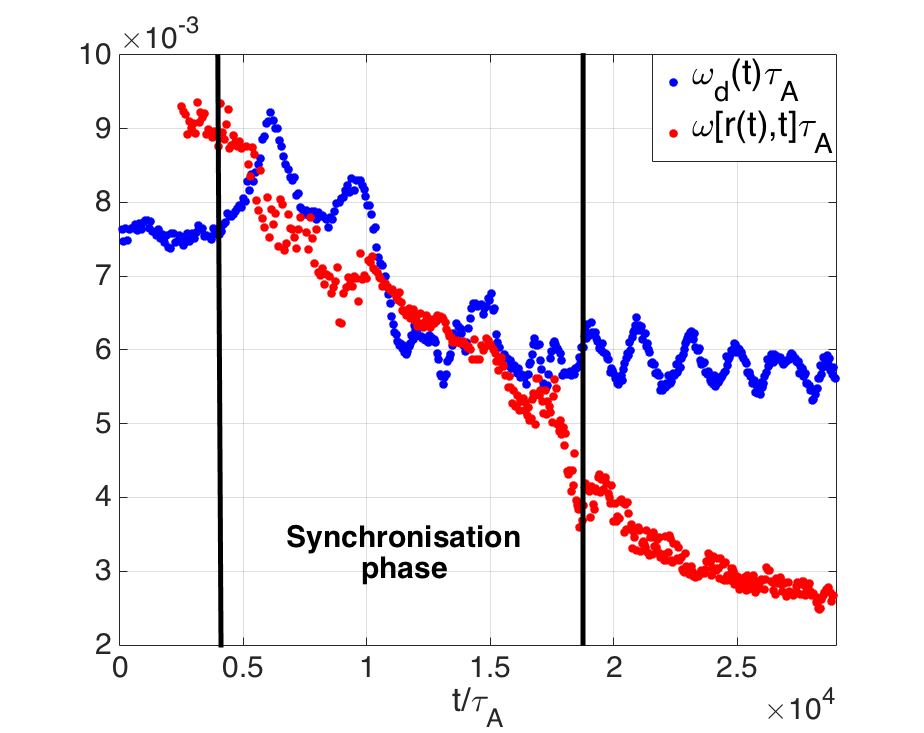}
   \caption{}
\end{subfigure}
\begin{subfigure}{.24\textwidth} 
   \centering
   \includegraphics[scale=0.15]{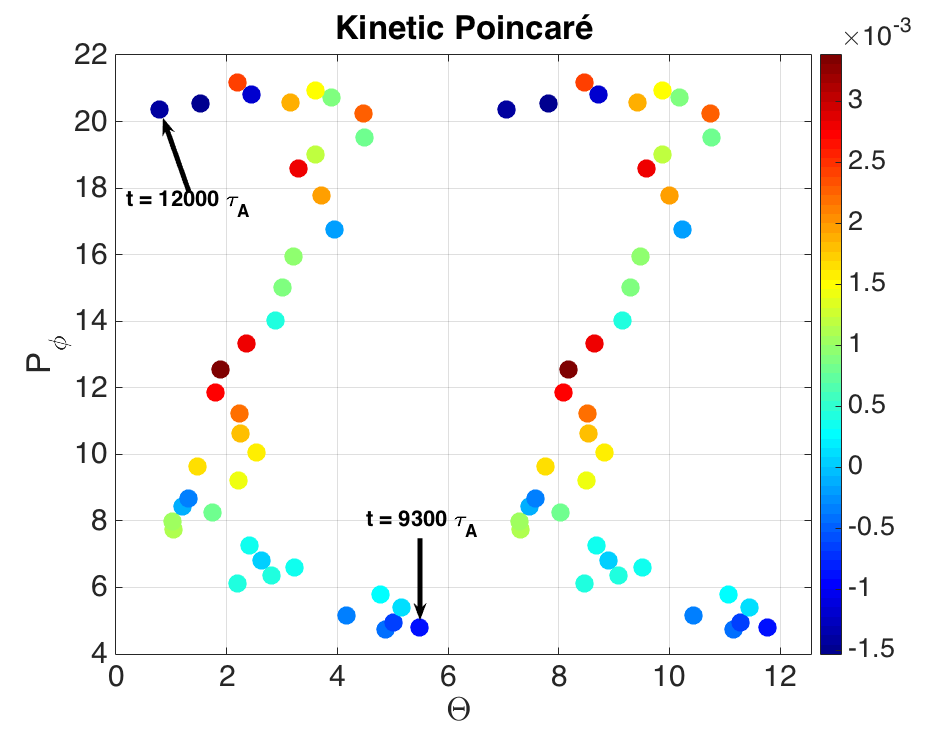}
   \caption{}
\end{subfigure}   
\caption{Time evolution of a typical resonant trapped particle in the strong drive regime. (a) Time evolution of $P_{\varphi}$. (b) Time evolution of the particle kinetic energy (c) Time evolution of the particle precessional frequency and of the local mode frequency in the plasma frame. (d) Kinetic Poincar\'e. The phase $\Phi$ is computed modulo $2\pi$, the Poincar\'e is duplicated to $\Phi \in [2\pi,4\pi]$ to enhance clarity. Colors correspond to the particle instantaneous power exchange $w_i = q_{\alpha}\textbf{v}_{\alpha,i}\cdot\textbf{E}$ with the mode.}
\label{indWD}
\end{figure}
Using the electromagnetic field of the weak drive simulation, a trapped alpha particle is initialized with $E = 0.4E_{\alpha}$, $\bar{r} = 0.15$ and $\lambda=1$ so that its initial precessional frequency $\omega_d\tau_A = 7.5\times10^{-3}$ is just below the local linear mode frequency at $\omega(\bar{r})\tau_A = 9\times10^{-3}$. Its evolution along the electromagnetic field self-consistently advanced with XTOR-K is illustrated in Figure \ref{indWD}. A synchronisation phase is also identified in Figure \ref{indWD} for $t\in[4000,19000]\tau_A$. This phase is still associated to the outward transport of the resonant particle, during which it yields on average its energy to the mode. As depicted in Figure \ref{indWD}, phase-locking also occurs, along with a minimization of the resonance detuning since $\dot{\Theta} \approx 0$ in that phase. \\ \\
Some features of the individual nonlinear behavior in the weak drive regime are however different compared to the strong one. First, the synchronization phase lasts ten times longer than in the strong drive limit, which was expected due to the slower dynamics observed in that regime. Moreover, the particles does not exit the mode before performing a single bounce in the phase space island. As can be seen in Figure \ref{indWD}, two bounces are performed before the particle is lost. It can however be noted that the number of bounces remains relatively low, since in the weak drive regime, it is expected that $\tau_B \ll \tau_{NL}$. In Ref. \cite{Wang2016} Figure 12, some resonant trapped particles just below the resonance condition tune with the mode for at least seven bounces. It can then be concluded that the simulation is not deeply in the weak kinetic drive regime, even if the condition $|\gamma-\gamma_L|/\gamma_L \ll 1$ is met. It implies that the interval in $\beta_{\alpha}$ over which the weak drive regime exists is very narrow, since the fishbone beta threshold lies at $\beta_{\alpha}/\beta_{tot} = 6\%$, and the present simulation used $\beta_{\alpha}/\beta_{tot} = 7\%$.
\subsection{Partial nonlinear mechanism of the alpha fishbone}
A partial mechanism that explains the coupling between the mode dominant down chirping and the resonant transport is now proposed, based on the dynamics observed in both simulations. First, at the end of the linear phase, as it was observed with individual trapped particles, the precessional frequency of resonant particles synchronises with the mode frequency, allowing the particles to be trapped in a phase space island. While being trapped, the particles give irreversibly their energy to the mode, which induces their outward transport out of the mode, as predicted by Eq. ($\ref{NL_ev}$). The reason why particles give away on average their energy is however not clear. The presence of a resonant island in phase space does not necessarily lead to a preferential transfer of energy from the particles trapped in the mode \cite{ONeil}. This open problem can be investigated by performing Kinetic Poincar\'e plots with a large number of test particles, initialized such that they fill a desired zone of the $(\Theta,P_{\varphi})$ diagram to reveal the dynamical evolution of the phase space island, as done in \cite{Vlad2013}\cite{Idouakass2016}. This is left for future studies with XTOR-K. \\ \\
The resonant transport of particles induces a flattening of the alpha density profile in phase space inside the $q=1$ volume, as observed in Figure \ref{flat},\ref{totPS_t} and \ref{totPS_p}. According to the fishbone linear model developed in \cite{Brochard2020a}\cite{Brochard2018}, the linear mode frequency $\omega$ is proportional to the imaginary part of the kinetic particles contribution to the fishbone dispersion relation (equation (16) in \cite{Brochard2020a}). Since the kinetic contribution itself verifies $\lambda_K \propto \int dr \nabla n_{\alpha}(r)$, the flattening of the alpha particle density profile lowers the amplitude of $Im[\lambda_K]$, which in turns decreases the mode frequency. This leads to the dominant down-chirping of the fishbone mode observed in both simulations in Figure \ref{omega}. \\ \\
Due to the dominant mode down chirping, the precessional and passing resonance positions explore new zones of phase space as observed in Figure \ref{trappedSD},\ref{trappedWD} and \ref{passingSD}, allowing initially non-resonant particles to be trapped in phase space islands and to be transported. This leads to further increase the mode down-chirping and resonant transport, until the mode frequency approaches zero, which prevents new zones of phase space to interact with the mode.
\section{Implication of the alpha transport on alpha power}
From the quantification of the global transport of alpha particles, conclusions can be drawn regarding the alpha heating power lost on ITER during a single burst of alpha fishbone. In this section, the separate global transport of trapped and passing particles is first examined. Then, the overall transport of alpha particles in each nonlinear regime is quantified.
\subsection{Overall transport of trapped and passing particles}
\begin{figure}[h!]
\begin{subfigure}{.49\textwidth} 
   \centering
   \includegraphics[scale=0.3]{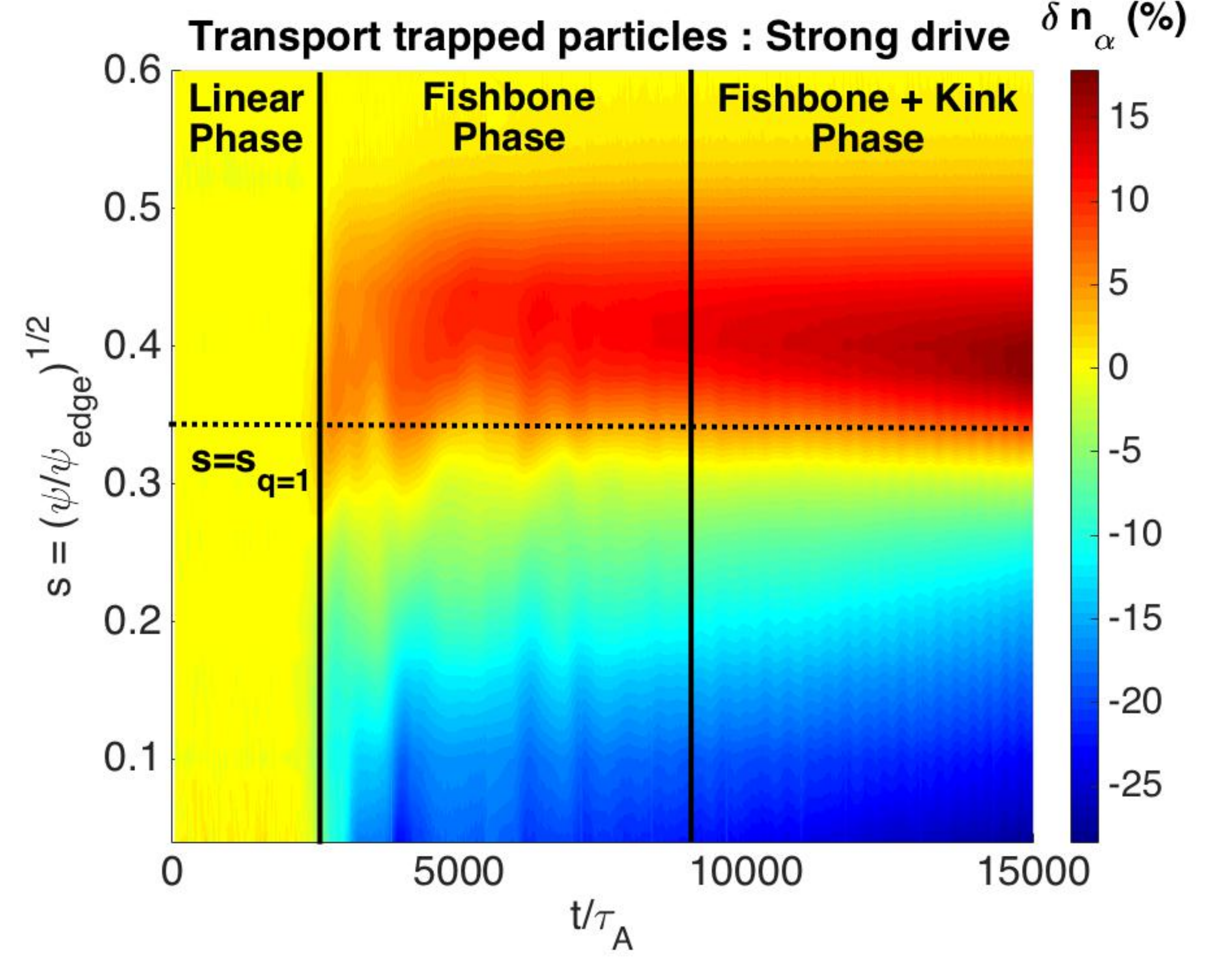}
   \caption{}
\end{subfigure}
\begin{subfigure}{.49\textwidth} 
   \centering
   \includegraphics[scale=0.3]{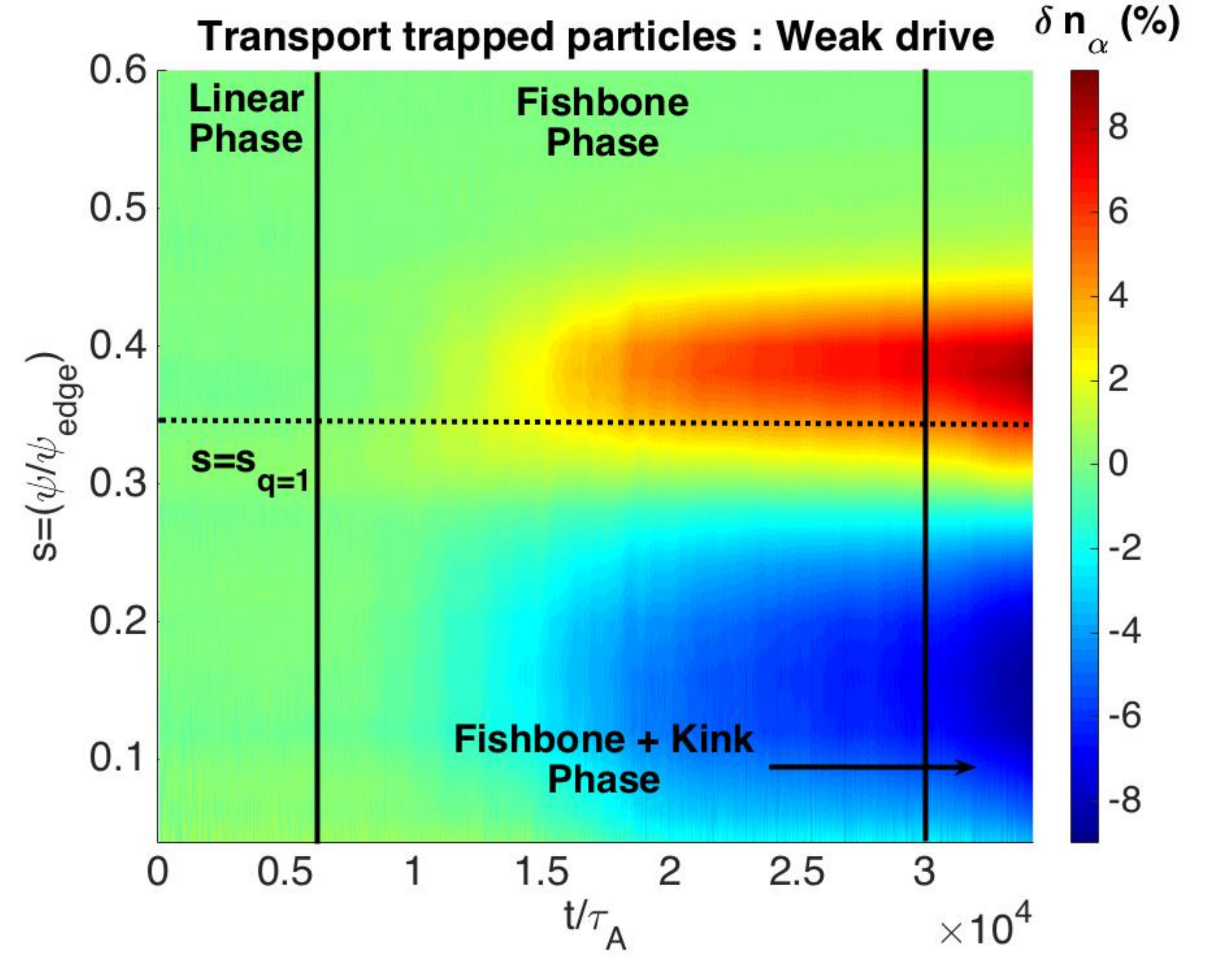}
   \caption{}
\end{subfigure}   
\begin{subfigure}{.49\textwidth} 
   \centering
   \includegraphics[scale=0.3]{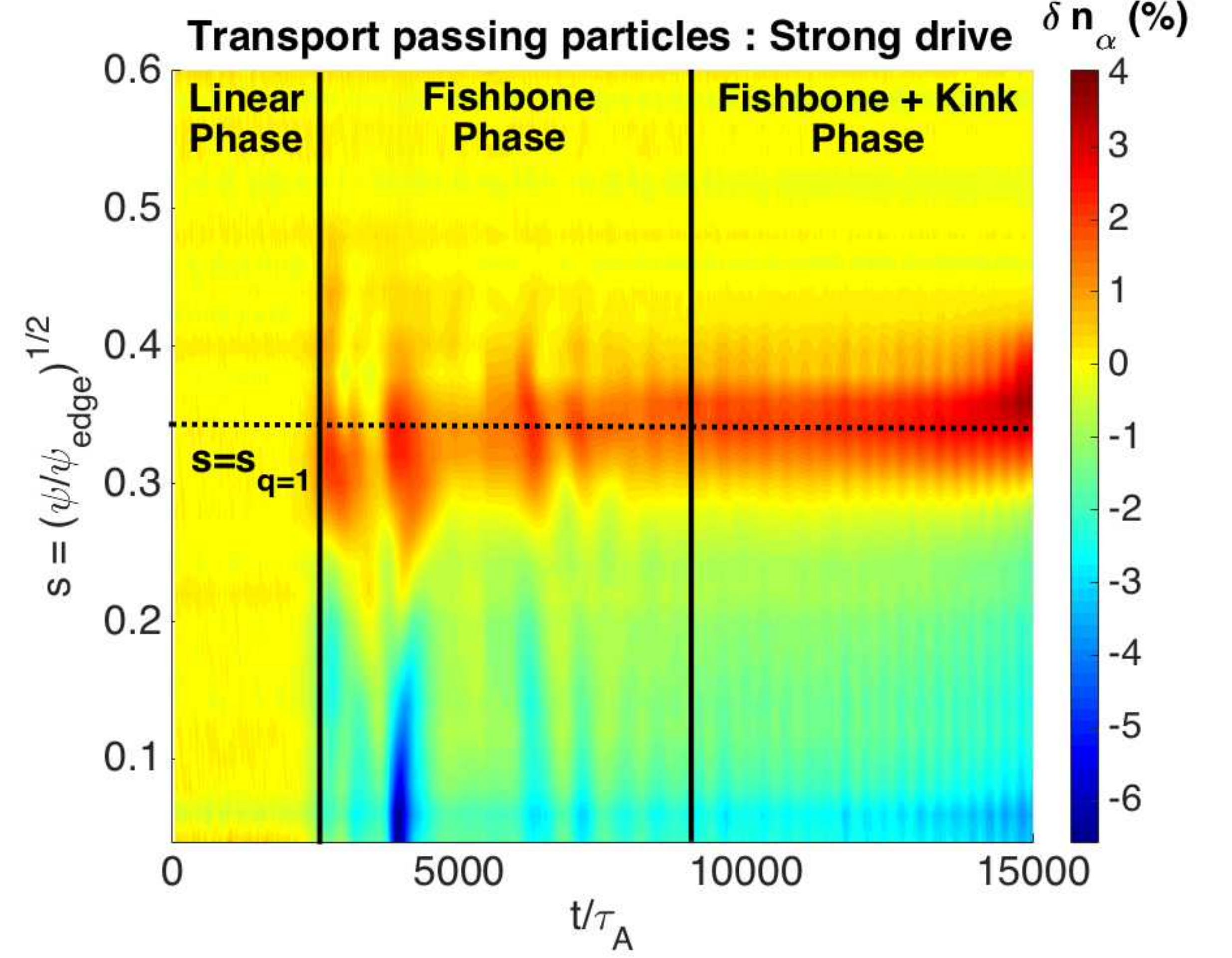}
   \caption{}
\end{subfigure}
\begin{subfigure}{.49\textwidth} 
   \centering
   \includegraphics[scale=0.3]{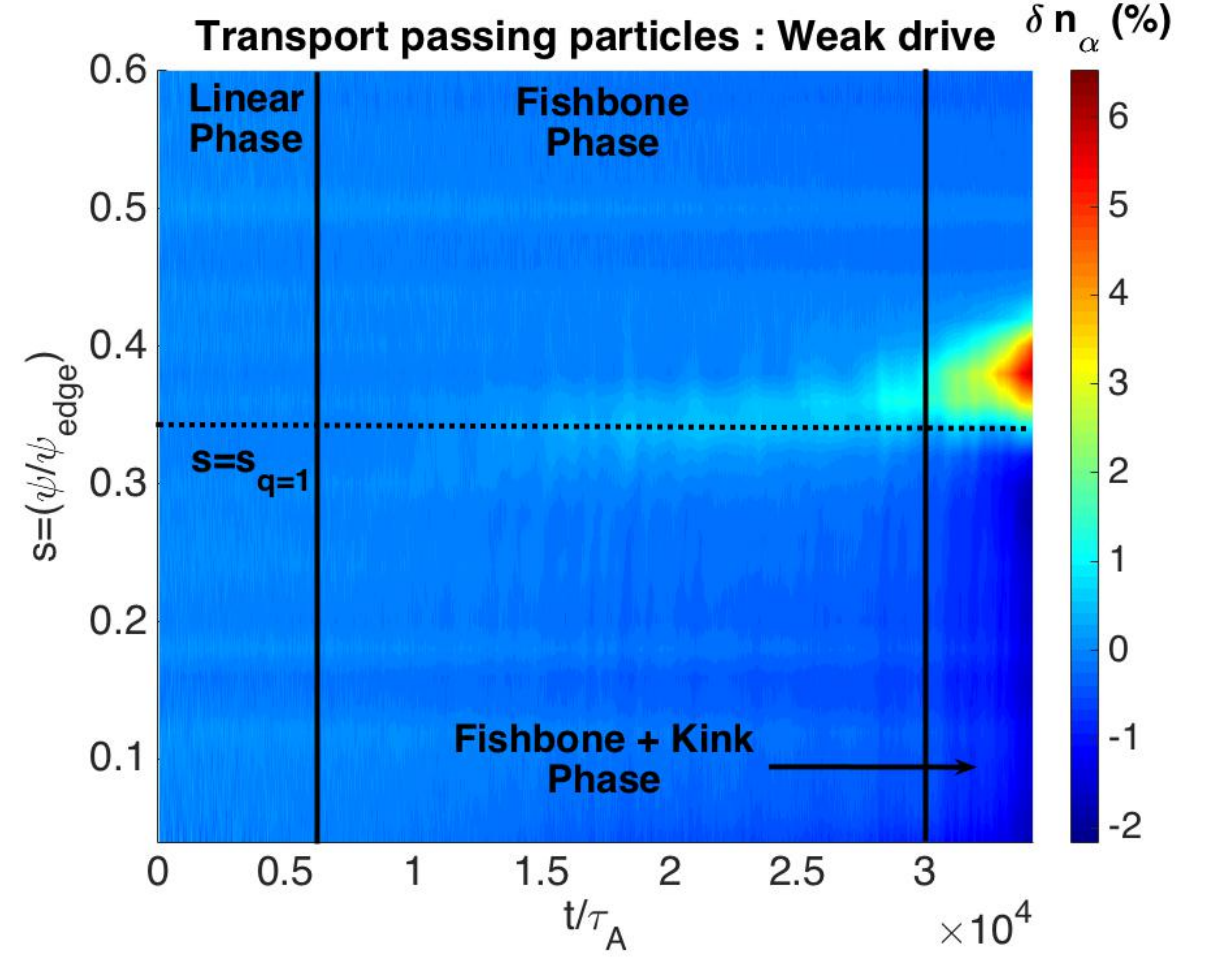}
   \caption{}
\end{subfigure}   
\caption{Time evolution of the perturbed alpha density profile $\delta n_{\alpha}$ for trapped particles (top figures) and passing particles (bottom figures). Left figures correspond to the strong drive regime, and right ones to the weak drive regime.}
\label{totalPS}
\end{figure}

In Figure \ref{totalPS}, the perturbed alpha density $\delta n_{\alpha}(r,t) =[n_{\alpha}(r,t)-n_{\alpha}(r,0)]/n_{\alpha}(r,0)$ in both nonlinear simulations is represented on the diagram $(t/\tau_A,s)$, where $s=\sqrt{\psi/\psi_{edge}}$ is the normalized radial position. The global population of trapped and passing particles are treated separately in both regimes : Figure \ref{totalPS} (a) and (b) correspond respectively to the trapped distribution in the strong and weak drive limits. Figure \ref{totalPS} (c) and (d) are associated respectively to the passing distribution in the strong and weak drive regime. The diagrams have been divided into the three temporal phases (linear, nonlinear fishbone, nonlinear fishbone / linear internal kink) identified in Section 2 for both simulations. The position of the $q=1$ surface is also marked. In all nonlinear simulations, the position $r_{q=1}$ evolves very weakly. 
\\ \\
In the strong kinetic drive regime (Figure \ref{totalPS} (a) (c)), particles start to be transported right after the end of the linear phase. During the nonlinear fishbone phase, a loss of about 10 \% of trapped particles is noticed inside $q=1$, and gain of 10\% outside of it. Trapped particles are transported up to $r=0.55$, whereas $r_{q=1} = 0.35$. As explained in Sections 2 and 3, such a transport far away from the $q=1$ surface is due to both the extension of the fishbone mode structure beyond $q=1$, and the large orbit width of trapped particles. Still in the strong drive regime, only 2 \% of passing particles are transported outside $q=1$. A net gain of $\sim$2\% is observed around the $q=1$ layer. Passing particles orbits have a small radial extension, which explains why they are not transported as far as trapped particles. A stronger transport of trapped particles compared to passing particles is consistent with the phase space dynamics detailed in Section 2.3.2. During the mixed fishbone/kink phase, the transport of alphas is enhanced, with 15\% of trapped and 3\% of passing particles transported. This additional transport is not resonant, and corresponds to an increase of the MHD displacement $\boldsymbol{\xi}$ that pushes the entire core plasma towards the $q=1$ surface as the linear internal kink mode grows. \\ \\
In the weak kinetic drive regime, the resonant transport only begins at the middle of the nonlinear fishbone phase, towards $t\sim 1.1\times10^4\tau_A$. In Figure \ref{KE} (b), it corresponds to the time when the $n=1$ mode reaches its maximum amplitude. The overall amount of transported trapped and passing particles is lower in this regime, as it was expected from section 2.3.2. Around 6\% of trapped particles are transported from inside $q=1$ towards $r\sim 0.4$ in the fishbone phase. It is noted that the maximal radial extent of transported trapped particles is also lower in that regime. Around 1\% of passing particles are transported in the fishbone phase. In the fishbone/kink phase, a slightly larger amount of particles is transported, as observed in the strong drive regime, for the same reasons. 
\subsection{Global transport of alphas and loss of alpha heating power}
\begin{figure}[h!]
\begin{subfigure}{.49\textwidth} 
   \centering
   \includegraphics[scale=0.29]{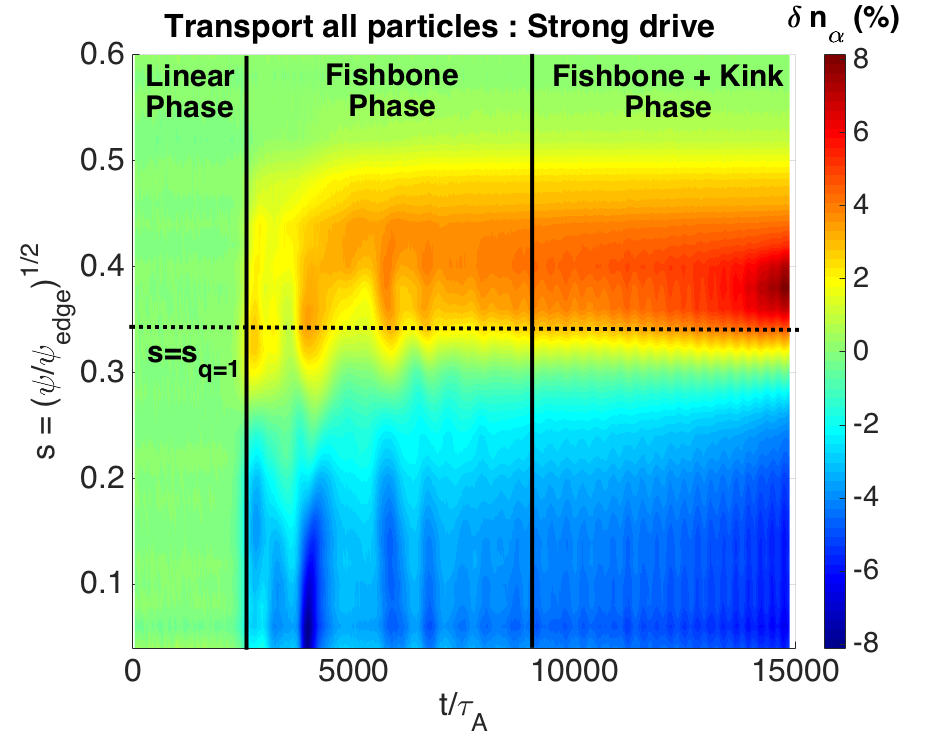}
   \caption{}
\end{subfigure}
\begin{subfigure}{.49\textwidth} 
   \centering
   \includegraphics[scale=0.29]{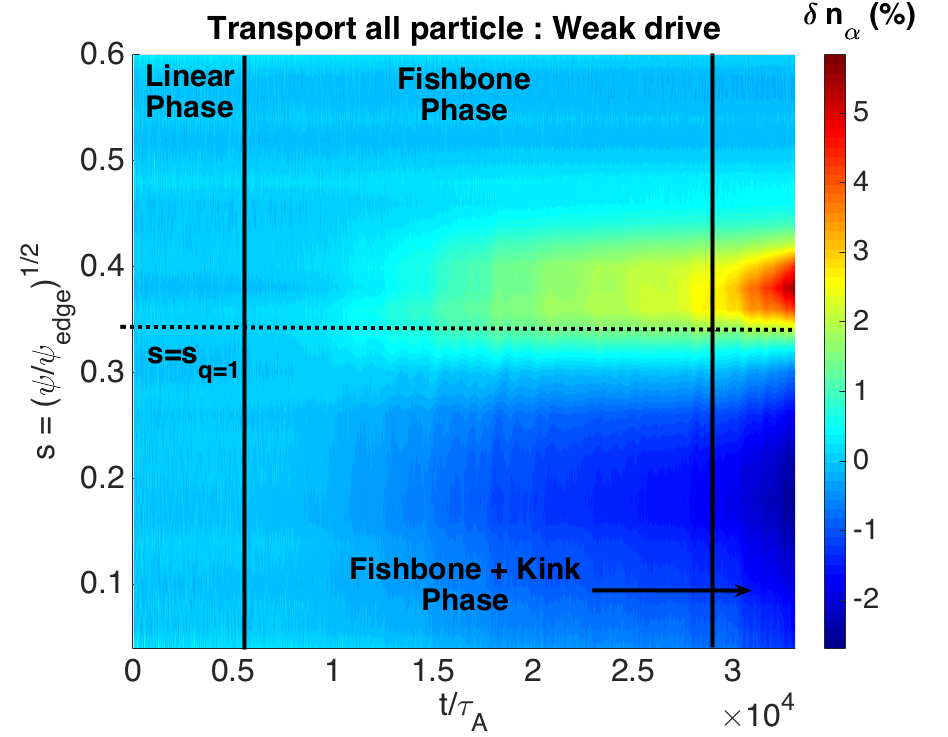}
   \caption{}
\end{subfigure}    
\caption{Time evolution of the perturbed alpha density profile $\delta n_{\alpha}$. (a) Strong drive. (b) Weak drive.}
\label{totall}
\end{figure}
The perturbed density of the entire alpha distribution function is displayed in the $(t/\tau_A,s)$ diagram in Figure \ref{totall} (a) for the strong drive regime, and in Figure \ref{totall} (b) for the weak one. The temporal dynamics of the perturbed alpha density on these diagrams is identical to the one observed for the fractions of trapped and passing distributions in Figure \ref{totalPS}. It is noted that the sudden burst of transport appearing in Figure \ref{totall} (a) near $t\sim3000 \tau_A$, $t\sim4000\tau_A$ and $t\sim6000\tau_A$ is correlated with sharp variations of the 0,0 plasma rotation observed in Figure \ref{omega} (c). \\ \\
Since the fraction of trapped particles is proportional to $\sqrt{r/R}$, it is expected that the overall alpha transport at the plasma core is weaker than the transport of the trapped fraction of alphas displayed in Figure \ref{totalPS}. In fact, in Figure \ref{totall} (a) in the fishbone phase, about 5\% of alphas inside $q=1$ are transported outside this surface, up to $r\sim 0.5$. It is noted that in the fishbone phase, the time window during which alphas are transported is $t\sim[2500,6000]\tau_A$, which lasts $\tau_{trans} \sim 3.5\times10^3 \tau_A$.  In the weak drive regime, $\sim 2\%$ of alphas are transported out of the $q=1$ volume, up to $r\sim 0.45$. The time window on which alphas are transported in the fishbone phase is $t\in[1.6\times10^4,3\times10^4]\tau_A$, with $\tau_{trans} = 1.4\times10^{4} \tau_A$.\\ \\
The partial thermalization time $\tau_{th,part}$ can be evaluated in both simulations by measuring the width $\delta_E$ in the energy direction of the hole and clump structure observed. It is given by $\tau_{th,part} = (\delta_E/\Delta_E) \tau_{th}$, where $\Delta_E = E_{\alpha}-E_c \sim 3$ MeV at the magnetic axis, and $\tau_{th}=4.6\times 10^{5}\tau_A$ the total thermalization time it takes for a fusion born alpha particle to reach the critical energy $E_c\sim 0.5$ MeV at the core plasma. In the strong drive regime, $\delta_E$ can be evaluated from Figure \ref{trappedSD} (c,d) as $\delta_E \sim 0.3 E_{\alpha}$, which leads to $\tau_{th,part} = 1.6\times 10^5 \tau_A \gg \tau_{trans}$. It was indeed valid to neglect collisions in the strong drive regime as introduced in section 2.1. In the weak drive regime, $\delta_E$ can be estimated from Figure \ref{trappedWD} (a,b) as $\delta_E \sim 0.1 E_{\alpha}$. It leads to $\tau_{th,part} = 5.3\times 10^{4}\tau_A > \tau_{trans} = 1.4\times 10^{4}\tau_A$. Therefore, neglecting the collisions in the weak drive regime remains a reasonable approximation.
\\ \\
The alpha heating power can be obtained from $P_{\alpha}(r,t) = p_{\alpha}(r,t)/\tau_{th}(r,t)$, with $p_{\alpha}$ the alpha pressure profile. As detailed in \cite{Brochard2020a} equation (14-15), the pressure profile of the isotropic slowing-down population of alphas is given by
\begin{equation}
p_{\alpha}(r,t) = n_{\alpha}(r,t) E_{\alpha} \frac{2I_{v2}(r,t)}{I_{v1}(r,t)}
\end{equation}
with $2I_{v2}/I_{v1}$ the equivalent alpha temperature profile when considering Maxwellians distributions with $T_{\alpha,0} = E_{\alpha}$, derived in \cite{Brochard2020a} equation (15). The thermalization time $\tau_{th}$ dependencies over $(r,t)$ are given by $\tau_{th}(r,t) \propto T_e^{3/2}(r,t)/n_e(r,t)$, with $T_e$ and $n_e$ the electron temperature and density profile. The normalized perturbed alpha power $\delta P_{\alpha} = [P_{\alpha}(r,t)-P_{\alpha}(r,0)]/P_{\alpha}(r,0) $ becomes
\begin{equation}
\delta P_{\alpha}(r,t) = \delta\bigg[n_{\alpha}(r,t) \times \frac{I_{v2}(r,t)n_e(r,t)}{I_{v1}(r,t)T_e^{3/2}(r,t)}\bigg]
\end{equation}
In the strong drive regime, the quantity $\delta[I_{v2}n_e/(I_{v1}T_e^{3/2})]$ varies only of $\sim 0.5\%$ inside the $q=1$ volume during the nonlinear phase, and of $\sim 0.2\%$ in the weak drive regime. Therefore, the loss of alpha power is mostly dominated by the loss of alpha particles, $\delta P_{\alpha} = \delta n_{\alpha}$. In the ITER 15 MA baseline scenario, a single burst of alpha fishbone activity then induces a decrease of 5\% of alpha power inside the $q=1$ volume in the strong kinetic drive regime, and a loss of 2\% in the weak kinetic drive regime.
\section{Conclusion}
In the present study, the nonlinear dynamics of the alpha fishbone instability in the ITER 15 MA baseline scenario has been thoroughly examined in two limit cases, the strong and weak drive regimes. Since the fishbone mode is an hybrid Kinetic-MHD mode, its dynamics in both physical space and phase space were analyzed in order to understand the nonlinear behavior of the instability. In physical space, it was noted in both simulations that the fishbone mode frequency in laboratory frame chirps up and down, accordingly with earlier theoretical and numerical nonlinear results \cite{Berk1999}\cite{Odblom2002}\cite{Idouakass2016}\cite{Fu2006}\cite{Wang2016}\cite{Shen2017}. The effects of the sheared plasma rotation \cite{Graves2000}\cite{Graves2003} were identified for the first time to have a significant impact on the fishbone mode frequency in plasma frame, inducing an important change in the wave-particle resonance condition. After the main burst of fishbone activity, a phase where a fishbone and an internal kink mode co-exist is identified at the end of each simulations. It is observed that the fishbone dynamics in the strong drive regime is faster than in the weak one, with a saturation level of the $n=1$ mode one order of magnitude above the weak drive regime. \\ \\
In phase space, it is observed in both regimes that the main saturation mechanism of the alpha fishbone is the resonant transport of alpha particles, dominated by trapped particles, as observed in \cite{Fu2006}. Hole and clump structures appear in the vicinity of each local phase space resonant positions, and stay attached to the resonance as the mode frequencies evolve. This feature is original since in theoretical models \cite{Berk1999} and others nonlinear simulations \cite{Shen2017}, the hole and clump were observed to follow respectively the resonance positions associated with down and up chirping. Moreover, these structures arise in both nonlinear regimes, whereas hole and clump were theoretically predicted near marginal stability, in the weak drive regime. \\ \\
The underlying mechanism that induces a coupling between mode chirping and particle transport of trapped particles is observed to be a synchronisation between the particles precessional frequency and the down-chirping mode frequency. It minimizes the resonance detuning with $\dot{\Theta}\approx 0$, as first observed in \cite{Vlad2013}. On this basis, a partial mechanism is proposed to explain the nonlinear dynamics of the alpha fishbone mode. Near resonance particles are getting trapped in phase space islands, while giving out on average their energy to the mode. It leads to an outward transport of resonant particles, which induces a flattening of the alpha density profiles. Since in linear theory, the mode frequency scales as $\omega \propto \int dr \ \nabla n_{\alpha}$, the flattening tends to decrease $\omega$. It enables the resonance position to explore new zones of phase space, trapping more particles in phase space islands, leading to enhanced mode down chirping and resonant transport until the fishbone mode frequency vanishes. \\ \\ 
A quantification of the global transport of alphas on ITER has been given in both nonlinear regimes. The time scale $\tau_{trans}$ over which alphas are transported is significantly lower than the partial thermalization time $\tau_{th,trans}$ of alphas onto the bulk electrons. It justifies the approximation made when neglecting collisions in XTOR-K. The alpha heating power $\delta P_{\alpha}$ lost on ITER during one alpha fishbone burst is found to be directly equal to the loss of alpha particles $\delta n_{\alpha}$. In the strong kinetic drive regime, $5\%$ of the alpha heating power is lost inside the $q=1$ volume, and $2\%$ in the weak kinetic drive regime. \\ \\
If it is valid to neglect collisions during one fishbone burst, the simulation of multiple fishbone bursts however requires to take into account collisions. Experimentally \cite{Nave1991}, several fishbone bursts are usually observed before the internal kink becomes dominant and leads to a sawtooth crash. Over several bursts, a competition will exist between the characteristic times $\tau_{trans}$ and $\tau_{th,trans}$. In order to study the impact of the fishbone instability on a complete sawtooth cycle, the implementation of collisions in hybrid nonlinear codes is necessary. Steps in that direction were recently made in XTOR-K \cite{Orain2020}, with a model based on binary collisions \cite{Bobylev2000}. It will enable XTOR-K to simulate a full sawtooth cycle with kinetic alpha particles in the future.
\appendix
\section{Wave-particle power exchange in XTOR-K}
Let us note the kinetic energy of alpha particles as $\mathcal{E}_{\alpha}$. Its time evolution is, summing over all kinetic particles $i$
\begin{equation}
\partial_t\mathcal{E}_{\alpha} = m_{\alpha}\sum_{i}^N\textbf{v}_{i,\alpha}\cdot(\partial_t\textbf{v}_{i,\alpha})
\end{equation}
with $m_{\alpha}$ the alpha mass and $\textbf{v}_{i,\alpha}$ the velocity vector of the $i^{th}$ alpha particle considered. Using the Lorentz equation, this leads to 
\begin{equation}
\partial_t \mathcal{E}_{\alpha} = \textbf{J}_{\alpha}\cdot\textbf{E}
\end{equation}
where $\textbf{J}_{\alpha} = q_{\alpha}\sum_i^N\textbf{v}_{i,\alpha}$ is the alpha current, $\textbf{E}$ the electric field and $q_{\alpha}$ the alpha charge. The total energy of the system bulk + alpha particles is composed of the electromagnetic fields' energy, the bulk and the fast particles kinetic energy. The total energy density $\mathcal{E}$ reads
\begin{equation}
\mathcal{E} = \frac{\epsilon_0E^2}{2} + \frac{B^2}{2\mu_0} + \mathcal{E}_b + \mathcal{E}_{\alpha}
\end{equation}
with $\mathcal{E}_b$ the bulk kinetic energy. Since the total energy of the system is conserved, the quantity $\textbf{J}_{\alpha}\cdot\textbf{E}$ then stands for the power transferred from alpha particles to the rest of the plasma. When it is positive, it implies that alpha particles are taking energy away from the fields and giving it away when it is negative. \\ \\
In XTOR-K, the power exchange $\textbf{J}_{\alpha}\cdot\textbf{E}$ is computed by interpolating on the phase space grid $(E,\lambda,r)$ the instantaneous power exchange $w_i(\textbf{r}_{i,\alpha},\textbf{v}_{i,\alpha}) = q_{\alpha}\textbf{v}_{i,\alpha}\cdot\textbf{E}(r_{i,\alpha})$ of each  alpha particle; located in phase space at $(\textbf{r}_{i,\alpha},\textbf{v}_{i,\alpha})$, where $\textbf{r}_{i,\alpha}$ is its position vector. Due to the full f scheme used in XTOR-K, each kinetic particle carries the same weight, allowing for a somewhat simpler computation of the wave-particle power exchange than in delta f hybrid codes \cite{Briguglio2014}. The power exchange $\textbf{J}_{\alpha}\cdot\textbf{E}$ in then summed up on chosen time windows, in order to average $\textbf{J}_{\alpha}\cdot\textbf{E}$ on at least one mode rotation period.
\section{Nonlinear evolution equations of resonant particles}
The equations describing the nonlinear evolution of resonant particles can be obtained in the angle-action formalism (\cite{Brochard_PhD} chapter 2, \cite{Brochard2018}) from the system Hamiltonian
\begin{equation}
H(\textbf{J},\boldsymbol{\alpha},t) = H_{eq}(\textbf{J}) - \tilde{h}(t)\cos\Theta(t), \ \ \ \Theta(t) = \textbf{n}\cdot\boldsymbol{\alpha} - \int_0^t\omega(t')dt'
\end{equation}
where $H_{eq}$ stands for the equilibrium Hamiltonian, $\textbf{J}$ the actions describing a triplet of invariants, $\boldsymbol{\alpha}$ the angles linked to the particles characteristic frequencies, $\tilde{h}$ the amplitude of the perturbed Hamiltonian due to kinetic-MHD modes and $\Theta$ the phase of the perturbed Hamiltonian. The mode numbers $\textbf{n}$ depends on the considered wave-particle resonance. For the precessional resonance, $\textbf{n} = (0,0,1)$, yielding $\Theta_{prec} = \alpha_3 - \int_0^t\omega(t')dt'$ and for the passing resonance, $\textbf{n} = (0,-1,1)$ leading to $\Theta_{pass} = \alpha_3-\alpha_2-\int_0^t\omega(t')dt'$. $d\Theta/dt$ stands for the wave-particle resonance condition in equation \ref{rescond}. Using the Hamilton-Jacobi equations, the time evolution of the particles invariants of motion read
\begin{equation}
\frac{d P_{\varphi}}{dt} \equiv -\frac{\partial H}{\partial \alpha_3} = - n \tilde{h}(t)\sin\Theta
\end{equation}
\begin{equation}
\frac{dE}{dt} \equiv \frac{\partial H}{\partial t} = - \tilde{h}(t)\omega(t)\sin\Theta - \frac{d\tilde{h}}{dt}\cos\Theta
\end{equation}
where $n\equiv n_3$ the toroidal mode number. When the time evolution of the perturbed Hamiltonian $\partial_t \tilde{h}$ can be neglected, the transport of resonant particles $\dot{P}_{\varphi}$ can be linked to the wave-particle energy exchange $\dot{E}$ as 
\begin{equation}\label{NL_ev}
\dot{P}_{\varphi} = n \frac{\dot{E}}{\omega}
\end{equation}
Due to the non direct grid $(\psi,\theta,\varphi)$ chosen in XTOR-K, the sign convention for an instability such as the fishbone mode is $n=-m=-1$. In that case, it implies that resonant particles are transported ($\dot{P}_{\varphi} > 0$ with XTOR-K's convention) when they loose irreversibly energy to the mode ($\dot{E} < 0$). Such a transport in enhanced when the mode frequency chirps down.
\section{Macroscopic link between radial transport and energy exchange of resonant particles}
Eq. (\ref{NL_ev}) gives a link between the energy loss and the change of canonical toroidal momentum of a single charged particle. A macroscopic version can be derived that links the exchange of energy with the radial alpha particle flux in the early nonlinear fishbone phase. The current $\textbf{J}_{tot,\alpha}$ of alpha particles and the electric field $\textbf{E}_{tot}$  are split in unperturbed and perturbed parts 
\begin{eqnarray} 
    \nonumber
   \mathbf{J}_{tot,\alpha}=\mathbf{J}_{\alpha}+\mathbf{\tilde{J}}_{\alpha}
		\label{eq:fluid_velocity}\\
	  \nonumber\mathbf{E}_{tot}=\mathbf{E}+\mathbf{\tilde{E}}
	\label{eq:E_field}
\end{eqnarray}  
where the mean current and electric fields $\mathbf{J}_{\alpha}$ and $\mathbf{E}$ depend on $\psi$ and poloidal angle only.  The perturbed fields are assumed to be of the form
\begin{eqnarray} 
    \nonumber
 \mathbf{\tilde{J}}_{\alpha}(\mathbf{x},t)=
	\mathbf{J}_{\alpha,\omega}(\mathbf{x})e^{-i\omega t}+c.c.
		\label{eq:fluid_velocity_perturbed}\\
	  \nonumber	
	 \mathbf{\tilde{E}}(\mathbf{x},t)=
	\mathbf{E}_{\omega}(\mathbf{x})e^{-i\omega t}+c.c.
	\label{eq:E_field_perturbed}
\end{eqnarray}
where $(\mathbf{\tilde{J}}_{\alpha,\omega},\mathbf{\tilde{E}}_{\omega})$ are complex numbers and $\omega$ is assumed constant - hence  no frequency chirping. This is a reasonable assumption in the early nonlinear fishbone phase. The energy exchanged between the perturbed electromagnetic field and alphas per unit time integrated over a finite volume reads 
\begin{equation}
    \dot{W}=\int_{0}^{T}\frac{dt}{T}\int d^3\mathbf{x}\: \mathbf{\tilde{J}}_{\alpha}\cdot\mathbf{\tilde{E}}
	  \label{eq:power}
\end{equation}
where $T$ is a time longer than a period $2\pi/\omega$, but smaller that the time scales of interest. The perturbed electric field is related to the perturbed electric potential $\tilde{\phi}$ and the vector potential $\mathbf{\tilde{A}}$ via the relation 
\begin{equation}
    \mathbf{\tilde{E}}=-\frac{\partial \mathbf{\tilde{A}}}{\partial t}-\nabla \tilde{\phi}
	  \label{eq:electric_field_vs_potentials}
\end{equation}
Using the charge conservation equation
\begin{equation}
    \frac{\partial\tilde{\rho}}{\partial t}+\nabla\cdot\mathbf{\tilde{J}}_{\alpha}=0
	  \label{eq:charge_conservation}
\end{equation}
where $\tilde{\rho}$ is the charge density, one finds
\begin{equation}
    \dot{W}=\int_{0}^{T}\frac{dt}{T}\int d^3\mathbf{x}\: 
		\left(\tilde{\rho}\frac{\partial \tilde{\phi}}{\partial t}
		- \mathbf{\tilde{J}}_{\alpha}\cdot\frac{\partial  \mathbf{\tilde{A}}}{\partial t}\right)
	  \label{eq:power_bis}
\end{equation}
up to a surface term that is ignored here. It can be written as an integral in the phase space as
\begin{equation}
    \dot{W}=2\omega Im \int d^3\mathbf{x}d^3\mathbf{p}\: 
		F_\omega(\mathbf{x},\mathbf{p}) H_\omega^{\ast}(\mathbf{x},\mathbf{p}) 
	  \label{eq:power_bis}
\end{equation}
where $F_\omega$ is the perturbed distribution function and $H_\omega=e\tilde{\phi}_\omega-e\mathbf{v}\cdot\mathbf{\tilde{A}}_\omega$ is the perturbed Hamiltonian. Let us now compute the flux of particles. The Vlasov equation reads
\begin{equation}
    \frac{\partial \mathcal{F}}{\partial t}-\left\{\mathcal{H},\mathcal{F}\right\}=0
	  \label{eq:Vlasov}
\end{equation}
where $\mathcal{F}$  and $\mathcal{H}$ are the total distribution function and hamiltonian. We use here a set of action/angle variables noted $\left(\boldsymbol{\alpha},\mathbf{J}\right)$ and define the unperturbed distribution function $F$ as the integral of $\mathcal{F}$ over the angles $\boldsymbol{\alpha}$ and in time over the duration $T$. An exact conservation equation over $F$ is readily obtained by integrating the Vlasov equation over the angles, to find
\begin{equation}
    \frac{\partial F}{\partial t}+\frac{\partial \Gamma_i}{\partial J_i}=0
	  \label{eq:Vlasov_conservative}
\end{equation}
where the flux components $\Gamma_i$ read
\begin{equation}
    \Gamma_i=\int_{0}^{T}\frac{dt}{T}\int d^3\boldsymbol{\alpha}d^3\textbf{J}
		\left(-\frac{\partial \tilde{H}}{\partial \alpha_i}\right)\tilde{F}
	  \label{eq:flux_components}
\end{equation}
The two first actions are related to the energy $E$ and the pitch-angle variable $\lambda$. The details of this correspondence do not matter here. We are interested here in the third component $i=3$. The corresponding action is nothing else than the canonical toroidal momentum $J_3=P_\varphi$. Moreover the toroidal angle $\varphi$ is equal to the third angle $\alpha_3$ up to a periodic function of the two other angles $(\alpha_1,\alpha_2)$. Hence if the perturbed field has a single toroidal wave number , i.e. 
\begin{equation}
    \nonumber
    \mathbf{E}_{\omega} =\mathbf{E}_{n\omega}e^{in\varphi}
	  \label{eq:fourier_expansion_toroidal_angle}
\end{equation}
then the radial flux of particles reads
\begin{equation}
    \Gamma_3=2n Im \int d^3\boldsymbol{\alpha}d^3\textbf{J} F_{n\omega}H_{n\omega}^{\ast}
	  \label{eq:flux_components}
\end{equation}
If one defines an exchange of energy from alpha particles in the space of invariants of motion $(E,\lambda,P_\varphi)$, it then appears that
\begin{equation}\label{macro}
\Gamma_3 = \frac{n}{\omega} \ \dot{W}
\end{equation}
which is the kinetic version of Eq. (\ref{NL_ev}). In XTOR-K, the sign convention for a 1,1 instability such as the fishbone mode is $n=-m=-1$.
\section*{Aknowledgments}
G.B, R.D and X.G would like to thank J. Graves for helpful discussions about the effects of sheared plasma rotation onto the wave-particle resonance condition. G.B would also like to thank L. Chen for helpful discussions on the nonlinear evolution of Energetic Particle Modes. This work has been carried out within the framework of the EUROfusion Consortium and has received funding from the Euratom research and training programme 2014-2018 and 2019-2020 under grant agreement No 633053. The views and opinions expressed herein do not necessarily reflect those of the European Commission. We benefited from HPC resources of TGCC and CINES from GENCI (projects no. 0500198 and 0510813) and the PHYMATH meso-center at Ecole Polytechnique.
\bibliography{thesis_lib}{}
\end{document}